\documentclass[aps,pra,twocolumn,superscriptaddress,floatfix,longbibliography]{revtex4-2}
\usepackage{amsmath}
\usepackage{amssymb}
\usepackage{bm}
\usepackage{graphicx}
\usepackage{color}
\usepackage[normalem]{ulem}
\usepackage{soul}
\usepackage{url}

\usepackage{verbatim}

\usepackage{hyperref}
\hypersetup{pdftex,colorlinks=true,allcolors=blue,bookmarksnumbered=true}
\usepackage{hypcap}

\usepackage{booktabs}
\usepackage{multirow}
\usepackage{makecell}

\def\Tr{{\rm Tr}}
\def\Re{{\rm Re}}

\def\sgn{{\rm sgn}}
\def\a{{\hat a}}
\def\b{{\hat b}}
\def\c{{\hat c}}

\def\A{{\hat A}}
\def\B{{\hat B}}
\def\C{{\hat C}}
\def\X{{\hat X}}
\def\n{{\hat n}}
\def\o{{\hat O}}
\def\O{{\hat O}}
\def\N{{\hat N}}
\def\hrho{{\hat \rho}}
\def\Tr{{\rm Tr}}
\def\sig{{\sigma}}
\def\tsig{{\tilde \sigma}}
\def\P{{\hat P}}
\def\Q{{\hat Q}}
\def\tildeE{{\mathcal E_m(\N_A,\N_B)}}

\begin{document}
\author{Yaxing Zhang}
\affiliation{Department of Physics and Applied Physics, Yale University, New Haven, Connecticut 06511, USA \\
and Yale Quantum Institute, Yale University, New Haven, Connecticut 06511, USA}
\author{Jacob C. Curtis}
\affiliation{Department of Physics and Applied Physics, Yale University, New Haven, Connecticut 06511, USA \\
and Yale Quantum Institute, Yale University, New Haven, Connecticut 06511, USA}
\author{Christopher S. Wang}
\affiliation{Department of Physics and Applied Physics, Yale University, New Haven, Connecticut 06511, USA \\
and Yale Quantum Institute, Yale University, New Haven, Connecticut 06511, USA}
\author{R. J. Schoelkopf}
\affiliation{Department of Physics and Applied Physics, Yale University, New Haven, Connecticut 06511, USA \\
and Yale Quantum Institute, Yale University, New Haven, Connecticut 06511, USA}
\author{S. M. Girvin}
\affiliation{Department of Physics and Applied Physics, Yale University, New Haven, Connecticut 06511, USA \\
and Yale Quantum Institute, Yale University, New Haven, Connecticut 06511, USA}

\title{Drive-induced nonlinearities of cavity modes coupled to a transmon ancilla}
\date{\today}
\begin{abstract}
High-Q microwave cavity modes coupled to transmon ancillas provide a hardware-efficient platform for quantum computing. Due to their coupling, the cavity modes inherit finite nonlinearity from the transmons. In this work, we theoretically and experimentally investigate how an off-resonant drive on the transmon ancilla modifies the nonlinearities of the cavity modes in qualitatively different ways, depending on the interrelation among cavity-transmon detuning, drive-transmon detuning and transmon anharmonicity. For a cavity-transmon detuning that is smaller than or comparable to the drive-transmon detuning and transmon anharmonicity, the off-resonant transmon drive can induce multiphoton resonances among cavity and transmon excitations that strongly modify cavity nonlinearities as drive parameters vary. For a large cavity-transmon detuning, the drive induces cavity-photon-number-dependent ac Stark shifts of transmon levels that translate into effective cavity nonlinearities. In the regime of weak transmon-cavity coupling,  the cavity Kerr nonlinearity relates to the third-order nonlinear susceptibility function $\chi^{(3)}$ of the driven ancilla. This susceptibility function provides a numerically efficient way of computing the cavity Kerr particularly for systems with many cavity modes controlled by a single transmon.  It also serves as a diagnostic tool for identifying undesired drive-induced multiphoton resonance processes. Lastly, we show that by judiciously choosing the drive amplitude, a single off-resonant transmon drive can be used to cancel the cavity self-Kerr nonlinearity or the inter-cavity cross-Kerr. This provides a way of dynamically correcting the cavity Kerr nonlinearity during bosonic operations and quantum error correction protocols that rely on the cavity modes being linear.
\end{abstract}

\maketitle

\section{Introduction}

Modes of superconducting microwave cavities have emerged as a promising platform for quantum computing and quantum simulations due to their long lifetime and integrability with Josephson-junction-based quantum devices including superconducting qubits~\cite{reagor2016, chakram2020}. Compared to two-level systems, the large accessible Hilbert space of the cavity modes enables a hardware-efficient way to encode error-correctable logical qubits~\cite{ofek2016, hu2019, campagne-ibarcq2020, gertler2021a}. In order to manipulate the states of cavity modes, it is necessary to introduce a source of nonlinearity via coupling to a nonlinear ancillary system~\cite{ma2021}, such as a superconducting transmon~\cite{koch2007}. Due to this coupling, the cavity modes inherit finite nonlinearity from the ancillas which make their energy levels non-equidistant. Such static cavity nonlinearity limits the performance of bosonic error correction schemes, in particular for the type of encoding that involves a large number of cavity photons~\cite{albert2018}. It also lowers the fidelity of Gaussian bosonic operations such as beam-splitters that are essential ingredients for entangling operations between bosonic modes~\cite{gao2019}.  Recent experiments have shown that off-resonant drives on transmon ancillas may lead to significant modifications of the cavity Kerr nonlinearity~\cite{wang2020}.  This suggests a possibility to dynamically control cavity nonlinearities using off-resonant drives, and more importantly, motivates the development of a systematic theory to compute cavity nonlinearities in the presence of such drives.

In this work, we study the nonlinearities of cavity modes inherited from an off-resonantly driven transmon ancilla. Specifically, we investigate the dependence of the nonlinearity of the ``dressed" cavity modes on the drive parameters and its interrelation with transmon anharmonicity and cavity-transmon detuning. We consider that the cavity modes are linearly coupled to the same transmon and focus on the usual dispersive coupling regime, i.e., the coupling strengths are weak compared to cavity-transmon detunings. Coupling-induced hybridization between the cavity and transmon excitations results in finite nonlinearity of the dressed cavity modes.

Off-resonant transmon drives are useful in inducing controllable coupling between far-detuned cavity modes due to the four-wave frequency mixing capability of the transmon ancilla~\cite{wang2020}. However, they can also lead to undesired resonant or near-resonant hybridization between cavity excitations and transmon excitations due to multiphoton resonances. As we will show, such hybridization leads to a rather sensitive dependence of cavity nonlinearity strength on the drive parameters, which is further complicated by the drive-induced ac Stark shift of the transmon transition frequencies. By working in the basis of Floquet eigenstates of the driven Hamiltonian, we are able to capture these drive-induced effects non-perturbatively in the drive strength. 

Away from the drive-induced resonances, the cavity-transmon coupling results in cavity-photon-number-dependent dispersive shifts of transmon transition frequencies between lower levels~\cite{grimsmo2021}. This means that the ac Stark shift of transmon levels induced by the off-resonant drive depends parametrically on the cavity photon number, which translates into an effective cavity nonlinearity. This effect has recently been utilized to realize error-transparent gates~\cite{ma2020} and cavity Hamiltonian engineering~\cite{wang2021} through the use of multiple drives applied close to the transmon transition frequency between its lowest two levels with drive detunings smaller than or comparable to the transmon-cavity dispersive coupling strength.

Here, we explore the regime of large drive detuning, much larger than the transmon-cavity dispersive coupling strength such that the drive-induced effects are only weakly dependent on cavity photon numbers. Depending on the interrelation between the drive detuning and transmon anharmonicity, the drive-induced cavity nonlinearity displays qualitatively different behaviors. Interestingly, as we will show, by judiciously choosing the drive amplitude, the cavity nonlinearity induced by a single drive blue-detuned from the transmon is sufficient to cancel the static cavity Kerr nonlinearity. This simple Kerr cancellation scheme is to be contrasted with previous Kerr cancellation methods in which multiple resonant or near-resonant transmon drives (with drive detunings smaller or comparable to the cavity-transmon dispersive coupling strength) are applied to impart strongly photon-number-dependent phase shifts~\cite{krastanov2015, heeres2015, wang2021}.

In the presence of many cavity modes coupled to a transmon (cf. Ref.~\cite{chakram2020}), finding the nonlinearity of the dressed cavity modes and their dispersive coupling with each other and the transmon can be numerically daunting as it requires diagonalization of the full multi-mode Hamiltonian. In the absence of drive, approximate semiclassical method such as the so-called “black-box quantization” ~\cite{nigg2012,minev2021} can be used to find the cavity nonlinearity parameters perturbatively in the transmon anharmonicity. In the presence of a drive, however, we still need to deal with a large multi-mode Hilbert space. 

In the regime of weak transmon-cavity coupling, as we will show, computing cavity nonlinearities reduces to finding the nonlinear susceptibility functions of the transmon. This generalizes our previous results that connect transmon-induced linear properties of cavity modes such as frequency shift and linear decay rate with its linear susceptibility function~\cite{zhang2019}. Specifically, cavity self-Kerr and inter-cavity cross-Kerr are given by the third-order nonlinear susceptibility function of the transmon. The latter can be calculated rather efficiently as it only requires diagonalization of the Hamiltonian of the driven transmon. Importantly, although the transmon-cavity coupling is treated perturbatively in using the susceptibility function, the drive on the transmon is not (up to our fourth-order truncation of the cosine potential). Therefore, it allows us to capture and conveniently identify the undesired drive-induced multiphoton resonances previously mentioned. 

The rest of the paper is structured as follows. After describing the Hamiltonian of the system in Sec.\ \ref{sec:Hamiltonian}, we present a general formalism in Sec.\ \ref{sec:formal_theory} for computing nonlinearities of the dressed cavity modes inherited from an off-resonantly driven transmon. In Sec.\ \ref{sec:theory_no_drive}, we review how cavity nonlinearities arise in the absence of the drive. In Sec.\ \ref{sec:weak_coupling_limit}, we apply the formalism to the regime of weak transmon-cavity coupling in which the dominant cavity nonlinearity is the fourth-order Kerr nonlinearity. We identify the connection between the cavity Kerr nonlinearity with the third-order nonlinear susceptibility function of the transmon. Through the susceptibility function, we discuss the drive-induced cavity Kerr nonlinearity in the regime of small and large cavity-transmon detuning with respect to the transmon anharmonicity and drive-transmon detuning. In Sec.\ \ref{sec:large_cavity_detuning}, we describe analytical theories for the case of large cavity-transmon detuning where cavity modes are far detuned from drive-induced multiphoton resonances. The theories are validated through quantitative agreement with numerical and experimental results. In Sec.\ \ref{sec:Kerr_cancellation}, we discuss the optimal drive conditions to cancel cavity Kerr nonlinearity, and demonstrate that under these conditions, the phase correlation of a Schr\"odinger cat state can be extended far beyond the characteristic phase collapse time under Kerr nonlinearity. 

\begin{figure}
\includegraphics[width = 6 cm]{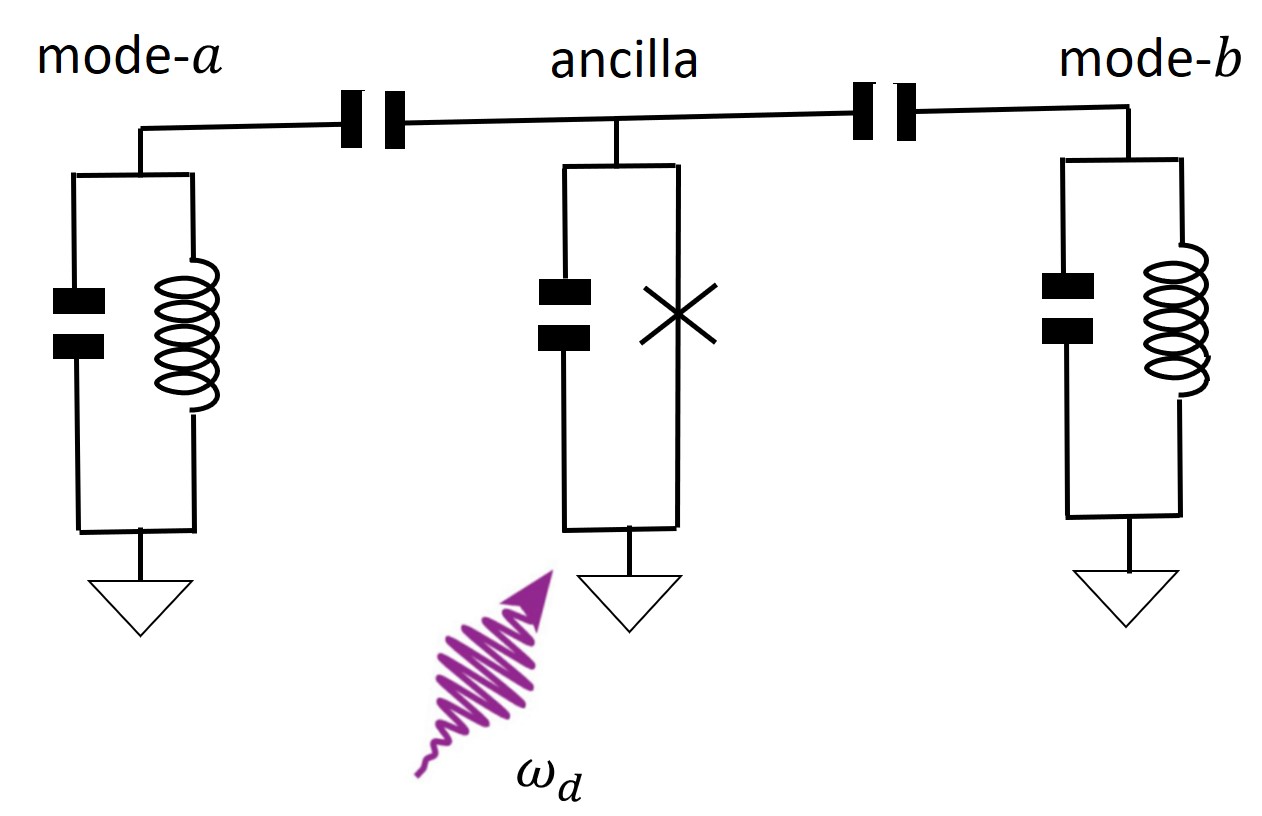}
\caption{A schematic showing two LC oscillators (cavity modes) coupled to a nonlinear LC oscillator (transmon ancilla).}
\label{fig:schematic}
\end{figure}

\section{The system Hamiltonian}
\label{sec:Hamiltonian}
We consider a system that consists of two linear cavity modes with frequency $\omega_a,\omega_b$ coupled to a driven nonlinear transmon ancilla; see Fig.~\ref{fig:schematic}. The transmon ancilla could serve the role of mediating coupling between the two cavity modes~\cite{wang2020} or assisting in the state preparation for a single cavity mode~\cite{heeres2017}. The linear modes could be modes of high-Q microwave cavities~\cite{reagor2016} or phonon cavities~\cite{hann2019a}. 
The system Hamiltonian reads, 
\begin{align}
\label{eq:full_Hamiltonian}
H&=H_{\rm cav}+H_{\rm anc}(t)+H_{\rm I},H_{\rm cav}=\hbar \omega_a \a^\dagger \a + \hbar \omega_b \b^\dagger \b,\nonumber \\
H_{\rm anc}(t) &= 4E_C \hat n^2 - E_J \cos \hat \phi - 2 e \n V_{\rm d}  \sin(\omega_{\rm d} t+\theta_d),\nonumber  \\
H_{\rm I} &= -2 e \n [ V_a (\a^\dagger  + \a) + V_b (\b^\dagger  + \b)],
\end{align}
where $H_{\rm cav}, H_{\rm anc}$ and $H_{\rm I}$ refer to the Hamiltonian of the linear cavity modes, the transmon ancilla and their interaction, respectively.  $V_a,V_b$ are the strengths of effective voltage fluctuations due to electric fields from the cavity modes $a,b$. 
The drive couples to the charge degree of freedom of the transmon ancilla with frequency $\omega_{\rm d}$ and phase $\theta_d$. 

For $E_C \ll E_J$ and a drive whose strength only populates the lower transmon levels, the transmon ancilla behaves as a weakly anharmonic oscillator. In this regime, one can expand the cosine potential and truncate to fourth order in $\hat \phi$, and then introduce the annihilation and creation operators $\c,\c^\dagger$, 
$\hat \phi = (8E_C/E_J)^{1/4}(c+c^\dagger )/\sqrt{2}, 
\n = -i(8E_C/E_J)^{-1/4}(c-c^\dagger )/\sqrt{2}.$ In terms of operators $\c, \c^\dagger$, the transmon Hamiltonian reads: 
\begin{align}
\label{eq:H_anc}
H_{\rm anc}(t)/\hbar &\approx \omega_c \c^\dagger \c -  \alpha (\c^\dagger +\c)^4/12 \nonumber \\
&+ (\c - \c^\dagger ) (e^{i\omega_d t} \Omega_d^* - e^{-i\omega_d t}\Omega_d),
\end{align}
where $\hbar\omega_c = \sqrt{8E_C E_J},\hbar\alpha = E_C, \hbar\Omega_d = e V_d \exp(-i\theta_d) (8E_C/E_J)^{-1/4}/\sqrt{2}$. The condition for neglecting higher order terms in the expansion can be found by comparing the sixth-order term with fourth-order term which leads to $\langle \hat \phi^2 \rangle \ll 1$, i.e. $(\alpha/\omega_c)\langle (\c^\dagger + \c )^2\rangle \ll 1$.

\subsection{The Rotating Wave Approximation}
To simplify the analysis, we switch to a frame that rotates at the drive frequency $\omega_d$ by making a unitary $U = \exp[-i (\a ^\dagger \a + \b ^\dagger \b+\c ^\dagger \c )\omega_d t]$. When the following conditions are satisfied, i.e., \[|\omega_{a,b,d}-\omega_c|,\alpha \langle ( \c^\dagger + \c)^2 \rangle \ll \omega_c,\] and \[|g_{a}|\sqrt{\langle \a \a^\dagger \rangle\langle \c\c^\dagger \rangle}, |g_{b}|\sqrt{\langle \b \b^\dagger\rangle \langle \c\c^\dagger \rangle}, |\Omega_d|\sqrt{\langle \c \c^\dagger \rangle}\ll \omega_c\] where $g_{a},g_b$ are defined below, one can apply the rotating wave approximation (RWA) and neglect terms that do not conserve the excitation number which leads to the following RWA Hamiltonian~\footnote{Leading-order corrections due to non-RWA terms in Eq.~(\ref{eq:H_anc}) and sixth-order terms from the expansion of the transmon cosine potential both scale as \unexpanded{$\alpha \langle \c^\dagger \c  \rangle/\omega_c $}. Thus if one were to include the non-RWA terms, one should also keep higher-order terms from the cosine expansion to be consistent.}:
\begin{align}
\label{eq:H_RWA}
H_{\rm RWA}= &  -\hbar\delta_{da} \a^\dagger \a - \hbar\delta_{db} \b^\dagger \b + H_{\rm anc}^{\rm RWA}+ H_I^{\rm RWA},  \\
 H_{\rm anc}^{\rm RWA}/\hbar = & -\delta_{dc}\c^\dagger \c - \frac {\alpha} {2}(\c^{\dagger } \c +1) \c^\dagger \c + \Omega_d  \c^\dagger + \Omega_d^* \c , \nonumber \\
 H_I ^{\rm RWA}/\hbar=& (g_a \a + g_b \b ) \c^\dagger + (g_a^* \a^\dagger + g_b^* \b^\dagger ) \c,\nonumber
\end{align}
in which  $\delta_{dx} = \omega_d - \omega_{x}, x \in \{a,b,c\}$ and $\hbar g_{a(b)} = -\sqrt{2} i eV_{a(b)}(8E_C/E_J)^{-1/4}$. Note that the transition frequency from the first excited state to the ground state of the transmon is $\omega_{10}=\omega_c - \alpha$~\footnote{In a classical sense, one can interpret $\omega_c$ as the frequency of the transmon nonlinear oscillator at zero energy, a notion that is convenient for semiclassical analysis.}. Since it is often convenient to speak of detunings of the drive and cavity modes from $\omega_{10}$, we define these detunings as follows: $$\delta_x \equiv \omega_x - \omega_{10}, \, x \in \{a,b,d\}.$$

\subsection{Parameter regimes of interest}
\label{sec:param_regime}
The RWA Hamiltonian in Eq.~(\ref{eq:H_RWA}) contains a total of six dimensionless parameters. Specifically, the static part of the system is controlled by two sets of parameters,  $g_{a(b)}/\delta_{a(b)}$ and $\delta_{a(b)}/\alpha$. The drive is controlled by two dimensionless drive parameters, $\Omega_d/\delta_d$ and $\delta_d/\alpha$. A central goal of this paper is to explore features of the cavity nonlinearities in different parameter regimes. As a guide to readers, Table~\ref{tb:parameter_regimes} summarizes the parameter regimes being explored in different sections. 

Throughout this paper, we focus on the regime where the ratio $g_{a(b)}/\delta_{a(b)}$ is much smaller than one. In the absence of the drive, this ratio controls the amount of hybridization between the cavity modes and the transmon. This regime is of particular interest when we use the cavity modes to store and encode quantum information. First, because the transmon ancilla is typically the lossier element, having a small ratio of $|g_{a(b)}/\delta_{a(b)}|$ helps reduce the amount of the transmon-cavity hybridization therefore reducing the ``inverse Purcell" decay of the cavity due to coupling to the transmon ``artificial atom"~\cite{reagor2016}. Second, as we will show in Sec.~\ref{sec:theory_no_drive}, reducing the ratio $|g_{a(b)}/\delta_{a(b)}|$ can suppress the strength of nonlinearities of the dressed cavity modes, making the cavity modes more suitable for bosonic operations or implementing bosonic error correction.

\begin{widetext}
\begin{table}[t]
  \centering
  \caption{Parameter regimes studied in different sections.}
    \begin{tabular}{|c|c|l|}
    \hline
    \multirow{2}{*}{Sec.~\ref{sec:theory_no_drive}} & \multirow{2}{*}{drive off} & Sec.~\ref{sec:4wm_no_drive}: large cavity-transmon detuning, $|\delta_{a(b)}|\gg \alpha$    \\  \cline{3-3}
     &  & Sec.~\ref{sec:TLS_no_drive}: small cavity-transmon detuning, $|\delta_{a(b)}|\ll \alpha$   \\
\hline    
\multirow{2}{*}{Sec.~\ref{sec:weak_coupling_limit}} &\multirow{2}{*}{\makecell{drive on, weak coupling, \\ perturbation in $g_a, g_b$}} & Sec.~\ref{sec:near_resonance}: small to intermediate cavity-transmon detuning, $|\delta_{a(b)}|\sim {\rm max}(\alpha,|\delta_d|)$ \\ \cline{3-3}
 & & Sec.~\ref{sec:asymptotic_regime}: large cavity-transmon detuning,  $|\delta_{a(b)}|\gg {\rm max}(\alpha,|\delta_d|)$ \\
    \hline
    \multirow{3}{*}{Sec.~\ref{sec:large_cavity_detuning}} & \multirow{3}{*}{ \makecell{drive on, \\ large cavity-transmon detuning, \\perturbation in $\delta_a^{-1}, \delta_b^{-1}$}} & Sec.~\ref{sec:weak_drive_limit}: weak drive, $|\Omega_d|\ll|\delta_d|$ \\ \cline{3-3}
  &   & Sec.~\ref{sec:small_drive_detuning}: small drive-transmon detuning, $|\delta_d|\ll \alpha$ \\ \cline{3-3}
  &  & Sec.~\ref{sec:semiclassical}: large drive-transmon detuning,  $|\delta_d|\gg \alpha$ \\
    \hline
    \end{tabular}%
\label{tb:parameter_regimes}%
\end{table}%
\end{widetext}


\section{Deriving dispersive Hamiltonian}
\label{sec:formal_theory} 
In the absence of the drive, it has been shown that the off-resonant cavity-transmon coupling generates a dispersive interaction between the so-called ``dressed" cavity modes and transmon~\cite{Blais2004b}. Also, the dressed cavity modes inherit finite nonlinearity from the transmon~\cite{kirchmair2013}. In this section, we present the formal theory of obtaining the dispersive Hamiltonian and cavity nonlinearities in the presence of the ancilla drive. 

In order to see how the dispersive Hamiltonian arises, let us first consider the limit of zero coupling, i.e. $H_I=0$. In this limit, eigenstates of the Hamiltonian $H_{\rm RWA}$ are product states. We label these states as $|\psi_m,N_a,N_b\rangle$, where $N_{a(b)}$ represents cavity Fock state with photon number $N_{a(b)}$ and $\psi_m$ represents an eigenstate of the ancilla Hamiltonian $H_{\rm anc}^{\rm RWA}$. It satisfies the stationary Schr\"odinger equation:
\begin{align}
\label{eq:H_anc_RWA}
H_{\rm anc}^{\rm RWA} \psi_m = \epsilon_m \psi_m.
\end{align}
Eigenstate $\psi_m$, stationary in the rotating frame, corresponds to a Floquet state of the driven transmon in the lab frame. Following the convention of previous work~\cite{zhang2019}, we label eigenstate $\psi_m$ as the state that adiabatically connects to Fock state $|m\rangle$ of the undriven transmon as the drive is ramped up or down. Eigenenergy $\epsilon_m$ in the rotating frame can be related to the $m$-th energy level $E_m$ of the transmon in the lab frame via the relation: $E_m = m\hbar \omega_d + \epsilon_m$. Because of the drive-induced ac Stark shift, $E_m$ is shifted from the bare energy level [$E_m^{\Omega_d =0} = m\hbar\omega_{10} - \hbar\alpha m(m-1)/2$] of the undriven transmon. 

Now let us consider turning on the transmon-cavity coupling $H_I$. Suppose that there is no degeneracy in the system eigenspectrum at $H_I=0$, then there is a unique state that adiabatically connects to the product state $|\psi_m,N_a,N_b\rangle$~\cite{Note-nonlinear-resonance}. We label this adiabatic state as  $|\overline{\psi_m,N_a,N_b}\rangle$. Written in the basis of the adiabatic eigenstates, the full RWA Hamiltonian in Eq.~(\ref{eq:H_RWA}) reads: 
\begin{align}
\label{eq:H_RWA_dressed_generic}
   H_{\rm RWA} =  \sum_{m,N_a,N_b} \mathcal E_m(N_a,N_b)  |\overline{\psi_m,N_a,N_b}\rangle \langle \overline {\psi_m,N_a,N_b}|,
\end{align}
where $\mathcal E_m(N_a,N_b)$ is the eigenenergy of the eigenstate $|\overline{\psi_m,N_a,N_b}\rangle$ and it can be thought of as an ancilla-state-dependent function of $N_a,N_b$. At zero coupling, $\mathcal E_m(N_a,N_b)$ is a simple sum of eigenenergies of the uncoupled cavities and driven ancilla [i.e.,~$\mathcal E_m^{H_I=0}(N_a,N_b) = -N_a \delta_{da}-N_b \delta_{db} + \epsilon_m$], while at finite coupling, $\mathcal E_m(N_a,N_b)$ is a more complicated function of $N_a,N_b$, as we will discuss in detail.


To gain further insight into the Hamiltonian in Eq.~(\ref{eq:H_RWA_dressed_generic}), we introduce operators $\A^\dagger,\A$ defined as $\A^{(\dagger)} = U_d \a^{(\dagger)} U_d^\dagger$, where $U_d$ is the unitary operator that diagonalizes the Hamiltonian $H_{\rm RWA}$. By construction, for each eigenstate, we have $|\overline{\psi_m,N_a,N_b}\rangle = U_d |\psi_m,N_a,N_b\rangle.$ Clearly, operators $\A^{\dagger},\A$ satisfy the bosonic commutation relation: $[\A,\A^\dagger]=1$. State $|\overline{\psi_m,N_a,N_b}\rangle$ is an eigenstate of $\A^\dagger\A$ with eigenvalue $N_a$: $\A^\dagger\A|\overline{\psi_m,N_a,N_b}\rangle = N_a |\overline{\psi_m,N_a,N_b}\rangle . $ One can think of $\A^\dagger,\A$ as creation and annihilation operators of a new ``dressed" mode whose excitation has overlap not just with that of the bare mode $a$ but also the transmon $c$ and mode $b$. We refer to this dressed mode as mode $A$. After defining operators $\B^\dagger,\B$ in a similar way and promoting $N_a,N_b$ in $\mathcal E_m(N_a,N_b)$ to operators, we rewrite Eq.~(\ref{eq:H_RWA_dressed_generic}) as follows:
%
\begin{align}
\label{eq:H_RWA_dressed_operator}
    H_{\rm RWA} &= \sum_m \mathcal E_m(\N_A,\N_B) \hat P_m, \\
\hat P_m &= \sum_{N_a,N_b}|\overline{\psi_m,N_a,N_b}\rangle \langle \overline {\psi_m,N_a,N_b}|, \nonumber
\end{align}
where $\N_A = \A^\dagger \A, \N_B = \B^\dagger \B$ are the occupation number operators of the dressed cavity modes, $\hat P_m$ is a projection operator that projects to the subspace $\{|\overline{\psi_m,N_a,N_b}\rangle, N_a,N_b=0,1,2...\}$.

One can interpret Eq.~(\ref{eq:H_RWA_dressed_operator}) as saying that the dynamics of the dressed cavity modes $A,B$ is controlled by an effective Hamiltonian $\mathcal E_m(\N_A,\N_B)$ conditioned on the coupled system being in the subspace $\{|\overline{\psi_m,N_a,N_b}\rangle,N_a,N_b = 0,1,2...\}$, or equivalently the transmon being in state $\psi_m$ in the limit of zero coupling. A more direct way to see this is to apply the unitary $U_d$ to the Hamiltonian $H_{\rm RWA}$ in Eq.~(\ref{eq:H_RWA_dressed_operator}), and one readily obtains that 
$U_d^\dagger H_{\rm RWA} U_d =  \sum_m \mathcal E_m(\N_a,\N_b) |\psi_m\rangle \langle \psi_m|,$
where $\N_a = \a^\dagger \a, \N_b = \b^\dagger \b$. As we will show in the next sections, while $\mathcal E_m(\N_A,\N_B)$ is linear in $\N_A,\N_B$ at $H_I=0$, it is generally nonlinear in $\N_A,\N_B$ at finite $H_I$ due to the nonlinearity of the transmon ancilla. This nonlinear dependence is the source of the nonlinearity of the dressed cavity modes.

\section{Cavity nonlinearities in the absence of a transmon drive}
\label{sec:theory_no_drive}
In this section, we review how the nonlinearities of the dressed cavity modes can be derived in the absence of an ancilla drive. An important dimensionless parameter here is the ratio between the ancilla anharmonicity and the cavity detuning from the ancilla $\alpha/|\delta_{a(b)}|$. In a way, this parameter controls the ``quantumness" of the dynamics of the coupled transmon-cavity system. The nonlinearities of the dressed cavity modes, as we show below, differ qualitatively in the two regimes of small and large $\alpha/|\delta_{a(b)}|$. 

\subsection{Transmon as a weakly anharmonic oscillator}
\label{sec:4wm_no_drive}
In the regime $\alpha/|\delta_{a(b)}|\ll 1$, the unequal spacing of the transmon levels (set by the transmon's anharmonicity $\alpha$) is masked by the large detuning between the the transmon and cavities. Therefore transitions between neighboring transmon states $|n+1\rangle$ and $|n\rangle$ are almost equally likely to excite the cavities. Put it differently, the transmon behaves almost like a linear oscillator when it interacts with the cavity modes.  

In this regime, a convenient way to solve the Hamiltonian in Eq.~(\ref{eq:H_RWA}) is to first find out the eigenmodes of the coupled system neglecting the transmon anharmonicity (the term $\propto \alpha$), and then treat the anharmonicity as a perturbation in the basis of the eigenmodes~\cite{nigg2012}. As we will show, the transmon anharmonicity is responsible for the nonlinearities of the eigenmodes. 

For small $g_{a(b)}/(\omega_{a(b)}-\omega_c)$, hybridizations between cavity and transmon modes are weak. Thus an eigenmode of the coupled system strongly overlaps with a certain bare mode. Specifically, the annihilation operator of the bare transmon mode expressed in terms of that of the eigenmodes can be written as the following:
\begin{align}
\c = \xi_A \A + \xi_B \B + \xi_C \C,
\end{align}
where $\xi_X$ is the linear participation ratio of the eigenmode $X\in \{A,B,C\}$ on the bare transmon mode $c$. To leading order in $g_{a(b)}/\delta_{a(b)}$, $\xi_{A(B)} \approx g_{a(b)}/(\omega_{a(b)}-\omega_c)\approx g_{a(b)}/\delta_{a(b)}, \xi_C \approx 1 + \mathcal O(\xi_A^2,\xi_B^2)$. We note that for an actual superconducting circuit, the eigenmodes and the participation ratios can be found using classical electromagnetic simulations~\cite{nigg2012,minev2021}.

In the rotating frame of mode $C$ and in the absence of an ancilla drive, the RWA Hamiltonian in Eq.~(\ref{eq:H_RWA}) expressed in terms of the ladder operators for the eigenmodes reads:
\begin{align}
\label{eq:H_RWA_eigenmodes_no_drive}
&H_{\rm RWA}/\hbar  =   \delta_{AC}\N_A +\delta_{BC}\N_B \nonumber \\  
& - \frac{\alpha}{2} \sum_{X_{1,2,3,4}\in \{A,B,C\}} \xi_{X_1}^*\xi_{X_2}\X_1^\dagger \X_2 (\xi_{X_3}^*\xi_{X_4}\X_3^\dagger \X_4 +1).
\end{align}
$\delta_{A(B)C}$ is the frequency difference between eigenmodes $A(B)$ and $C$: $\delta_{A(B)C} = \omega_{A(B)}-\omega_C\approx \delta_{a(b)}$.

Of primary interest to us are the quartic terms in Eq.~(\ref{eq:H_RWA_eigenmodes_no_drive}). It is straightforward to see that those terms that have unequal number of $\X^\dagger$ and $\X$ for any $X\in \{A,B,C\}$ are strongly off-resonant in the limit $\alpha \ll |\delta_{a(b)}|$. This can be most easily seen by making a unitary rotation $\hat U =\exp[-i(\N_A \delta_{AC} + \N_B \delta_{BC})t]$ such that the quartic terms generally oscillate at frequencies $\delta_{AC},\delta_{BC}$ or their linear combinations. In the lowest approximation, we can neglect the oscillating terms and obtain the following quartic Hamiltonian~\cite{nigg2012}:
\begin{align}
\label{eq:H_quar}
&H_{\rm quar}/\hbar \approx - \frac{1}{2}\sum_{X,X' \in\{A,B,C\}} \chi_{XX'} \N_X  \N_{X'}, \nonumber \\
&\chi_{XX} = \alpha |\xi_X|^4, \chi_{XX'} = 2 \alpha |\xi_X\xi_{X'}|^2.
\end{align}
The above equation readily shows that eigenmode $X$ has a self-Kerr nonlinearity of strength $\chi_{XX}$ and a cross-Kerr nonlinearity with another mode $X'$ of strength $\chi_{XX'}$. In the dispersive regime that we are considering where $|\xi_{A,B}|\ll 1, |\xi_C|\approx 1$, we have $\chi_{AA},\chi_{BB}, \chi_{AB}\ll \chi_{AC},\chi_{BC} \ll \chi_{CC}.$ 

In the regime of small $\alpha/|\delta_{a(b)}|$, the strengths of the nonlinearities of the cavity-like eigenmodes $A,B$ {\it weakly} depend on the states of the transmon-like eigenmode $C$. To zeroth order in $\alpha/|\delta_{a(b)}|$, the Kerr nonlinearities of the modes are independent of the state of the mode $C$ and have strengths $\chi_{AA},\chi_{BB},\chi_{AB}$, as shown in Eq.~(\ref{eq:H_quar}). To the next order, we find that there is a correction to the strength of cavity Kerr nonlinearities that is proportional to the transmon excitation number $\N_C$ and is suppressed by the small factor $\chi_{CC}/\delta_{A(B)C}$; see Appendix~\ref{app:correction_to_chi}. 

To the next order in $\alpha/\delta_{a(b)}$, there also emerge sixth-order nonlinearities for the cavity-like modes $A$ and $B$ whose strengths are smaller than Kerr nonlinearity by a factor of $\chi_{A(B)C}/\delta_{A(B)C}$; see Appendix~\ref{app:correction_to_chi}.  Written in terms of bare mode parameters, this factor becomes $\alpha|g_{a(b)}/\delta_{a(b)}|^2/\delta_{a(b)}$. We emphasize that this factor is simultaneously suppressed by two small parameters $\alpha/\delta_{a(b)}$ and $g_{a(b)}/\delta_{a(b)}$. This ensures that it is often a very good approximation to only keep Kerr nonlinearity for the cavity modes. We show in Sec.~\ref{sec:large_cavity_detuning}, this is not necessarily the case in the presence of a transmon drive. 

\subsection{Transmon as a two-level system}
\label{sec:TLS_no_drive}
In the regime $\alpha/|\delta_{a(b)}|\gg 1$, transmon transition frequency $\omega_{(n+1)n}$ from state $|n+1\rangle$ to $|n\rangle$ for any $n\geq 1$ is strongly off-resonant from cavity frequency $\omega_{a(b)}$, much stronger than the detuning of $\omega_{10}$ from $\omega_{a(b)}$. One can think of the cavity photons as being blocked from exciting the transmon to states $|n\geq 2\rangle$ and the transmon behaves like a strongly quantum two-level system when it interacts with the cavities. 

To leading order in $(\alpha/|\delta_{a(b)}|)^{-1} \ll 1$, one can replace $\c^\dagger$ and $\c$ in Eq.~(\ref{eq:H_RWA}) with $\sig^+$ and $\sig^-$ respectively, which are defined as $\sigma^+ = |1\rangle\langle 0|$ and $\sigma^- = |0\rangle\langle 1|$. Within the two-state manifold of the transmon, the RWA Hamiltonian reduces to the familiar Jaynes-Cummings Hamiltonian but with two cavity modes. This Hamiltonian can be unitarily transformed into a dispersive Hamiltonian~\cite{Carbonaro1979}. To fourth order in $g_{a(b)}/\delta_{a(b)}$ and switching to the rotating frame at frequency $\omega_{10}$, the dispersive Hamiltonian is found to be:
\begin{align}
\label{eq:H_TLS}
 & H_{\rm TLS}/\hbar = \delta_a \N_A + \delta_b \N_B - \Big(\frac{|g_a|^2}{\delta_a}+\frac{|g_b|^2}{\delta_b} \Big)\frac{\sigma_z}{2} \nonumber \\
& -\Big [\frac{|g_a|^2}{\delta_a} \N_A + \frac{|g_b|^2}{\delta_b}\N_B +\frac{|g_a|^4}{\delta_a^3} \N_A^2 + \frac{|g_b|^4}{\delta_b^3} \N_B^2   \nonumber \\
&+ \frac{2|g_ag_b|^2(\delta_a + \delta_b)}{\delta_a^2\delta_b^2} \N_A\N_B \Big] \sigma_z.
\end{align}
We use $H_{\rm TLS}$ to indicate that we have truncated the transmon to its first two levels. 

In contrast to the regime considered in the previous section, here, the strengths of the cavity nonlinearities  {\it strongly} depend on the state of the transmon. As shown in Eq.~(\ref{eq:H_TLS}), both the self-Kerr and cross-Kerr interaction strengths of the cavity modes have opposite sign when the transmon is in the ground and first excited state. When the transmon is in a higher level, the cavity Kerr strengths are much smaller, suppressed by small parameter $(\alpha/|\delta_{a(b)}|)^{-1}$. It is not hard to see that in this regime, sixth-order cavity nonlinearity is suppressed by the factor $|g_{a(b)}/\delta_{a(b)}|^2$ compared to the Kerr nonlinearity.


\section{Drive-induced change of cavity nonlinearities in the weak coupling regime}
\label{sec:weak_coupling_limit}
The drive on the transmon modifies its spectrum and eigenstates, which in turn, modifies the nonlinearities that the cavity modes inherit from the transmon. 
In general, calculating the cavity nonlinearities in the presence of drive requires diagonalizing exactly the coupled cavity-ancilla Hamiltonian in Eq.~(\ref{eq:H_RWA}) and finding the effective Hamiltonian $\tildeE$ in Eq.~(\ref{eq:H_RWA_dressed_operator}). The task of diagonalization can become numerically challenging in the case of large photon number in the cavity modes or a large number of cavity modes controlled by a single transmon. 

As mentioned in Sec.~\ref{sec:param_regime}, of primary interest to us is the regime where the cavity-transmon hybridization is weak which we will refer to as the weak coupling regime. In this regime, we can treat the cavity-transmon coupling $H_I$ as a perturbation to the uncoupled system, and compute cavity nonlinearities perturbatively in $H_I$. As we will show, this treatment allows us to alleviate the need of diagonalizing the full coupled system, and at the same time capture the non-perturbative modifications to the cavity nonlinearities due to the drive. 


\subsection{Scaling properties of  $\tildeE$}
\label{sec:E_m}
The goal of this section is to understand how the nonlinearities of the dressed cavity mode should scale with respect to the cavity-transmon coupling in the weak coupling regime. In the absence of degeneracies, the effective Hamiltonian $\tildeE$ of the dressed cavity modes is analytic in $\N_A,\N_B$. It is instructive to separate the coupling-induced part in $\tildeE$ and expand it with respect to $\N_A,\N_B$: 
\begin{align}
\label{eq:E_m_expansion}
&\delta \mathcal E_m(\N_A,\N_B) \equiv \mathcal E_m(\N_A,\N_B) - \mathcal E_m^{H_I = 0}(\N_A,\N_B),  \nonumber \\
&\delta \mathcal E_m(\N_A,\N_B)/\hbar = \sum_{n,n'=0}^\infty \frac{c_{nn',m}}{n!n'!}\N_A^n \N_B^{n'}, \nonumber \\
& c_{nn',m} = \frac{\partial^{n+n'}\delta \mathcal E_m(N_A,N_B)} {\hbar \partial{N_A^n}\partial{ N_B^{n'}}}\Big|_{N_A=N_B=0}.
\end{align}
One can identify the term proportional to $\N_{A(B)}^2$ as the self-Kerr nonlinearity of dressed mode $A(B)$ considered in Sec.~\ref{sec:theory_no_drive} and the term proportional to $\N_A\N_B$ as the cross-Kerr nonlinearity between modes $A$ and $B$. As we have seen in Sec.~\ref{sec:theory_no_drive}, the coefficients in front of these terms as well as higher-order terms in the expansion are finite due to the nonlinearity of the transmon and the finite cavity-transmon coupling.


In order to understand how the coefficient $c_{nn',m}$ in Eq.~(\ref{eq:E_m_expansion}) scales with the cavity-transmon coupling strengths in the weak coupling regime, we find $\delta \mathcal E_m(N_A,N_B)$ perturbatively in $H_I^{\rm RWA}$ of Eq.~(\ref{eq:H_RWA}). It is clear that in order to find terms proportional to $N_A^n N_B^{n'}$, we need to treat the perturbation at least to order $\sim \mathcal O(|g_a|^{2n} |g_b|^{2n'})$. It follows that to the leading order in the coupling strengths $g_a,g_b$: 
\begin{align}
\label{eq:c_scaling}
&c_{nn',m}  \sim  \mathcal O( |g_a|^{2n}|g_b|^{2n'}), \, n+n'>0,  \nonumber \\ 
&  c_{00,m}  \sim \mathcal O (|g_a|^2, |g_b|^2).
\end{align}
Since higher-order expansion coefficients $c_{nn',m}$ involve high-order perturbation in $g_a,g_b$, in the case where the perturbation theory applies, it suffices to consider the lowest-order cavity nonlinearity including the cavity self-Kerr and inter-cavity cross-Kerr; see Sec.~\ref{sec:Kerr_expression}. 

Often we are interested in the dynamics of the low-energy manifold of the dressed modes where there are only a few photons present. In this case, it is more convenient to express $\delta\mathcal E_m(\N_A,\N_B)$ in the normal ordered form:
\begin{align}
\label{eq:E_m_expansion_normal_order}
\delta \mathcal E_m(\N_A,\N_B)/\hbar = \sum_{n,n'=0}^\infty  \frac{\overline c_{nn',m}}{n!n'!}:\N_A^n \N_B^{n'}:,
\end{align}
where operators in between the two colons are normal ordered.  When parametrized in this form, the eigenenergy $\mathcal E_m(N_A,N_B)$ only depends on a finite set of coefficients $\overline c_{nn',m}$ with $n\leq N_A$ and $n'\leq N_B$. Specifically, we have the following relation: 
\begin{align}
\label{eq:E_m_expansion_normal_order_2}
   \delta \mathcal E_m(N_A,N_B)/\hbar =  \sum_{n=0}^{N_A}\sum_{n'=0}^{N_B}\frac{\overline c_{nn',m}}{n!n'!}\frac{N_A!}{(N_A-n)!}\frac{N_B!}{(N_B-n')!}.
\end{align}
Once we know $\delta \mathcal E_m(N_A,N_B)$ from $N_A = N_B = 0$ up to $N_A = n, N_B = n'$, coefficient $\overline c_{nn',m}$ can be be found by reverting the above relation.

Although the coefficients $\overline c_{nn',m}$ are generally not the same as $c_{nn',m}$, they become equal in the weak coupling limit. This can be seen as follows. By Wick's theorem, the operator $\N_A^n\N_B^{n'}$ can always be expressed as a normal-ordered operator :$\N_A^n\N_B^{n'}$: plus additional terms of normal ordered operators in which one or multiple pairs of $\A^\dagger,\A$ or $\B^\dagger,\B$ have contracted each other. This means that $\overline c_{nn',m}$ can be expressed as $c_{nn',m}$ plus an infinite sum of higher order coefficients $c_{n''n''',m}$ in which $n''\geq n, n'''\geq n'$ but they cannot take equal signs at the same time. Using the fact that $c_{nn',m}\sim \mathcal O(|g_a|^{2n}|g_b|^{2n'})$, we deduce the following relation that applies for $n\geq 2$ or $n'\geq 2$:
\begin{align}
    \overline c_{nn',m} = c_{nn',m}[1 + \mathcal O(|g_a|^{2},|g_b|^{2})]. 
\end{align}
For $n,n'<2$ we have $\overline c_{nn',m} =  c_{nn',m}$. We conclude that in the weak coupling regime, the coefficients $\overline c_{nn',m}$ also fall off polynomially in the coupling strength $g_a,g_b$.

\subsection{Expression for cavity Kerr nonlinearity}
\label{sec:Kerr_expression}
In the weak coupling regime, cavity nonlinearities are dominated by Kerr nonlinearities, i.e. terms quadratic in $\N_A,\N_B$ in Eq.~(\ref{eq:E_m_expansion_normal_order}). For clarity, we rewrite those terms below:
\begin{align*}
\delta\mathcal E_m(\N_A,\N_B)/\hbar \approx &  \frac{K_{A,m}}{2} :\N_A^2:  + \frac{K_{B,m}}{2} :\N_B^2:+ \nonumber\\ 
&+K_{AB,m}\N_A \N_B +...,
\end{align*} 
where $ K_{A,m}\equiv \overline c_{20,m},  K_{B,m}\equiv \overline c_{02,m},  K_{AB,m}\equiv \overline c_{11,m}$. $K_{A(B),m}$ and $K_{AB,m}$ represent ancilla-state-dependent cavity self-Kerr and cross-Kerr nonlinearities, respectively. We have chosen the normal ordered form of cavity self-Kerr as in Eq.~(\ref{eq:E_m_expansion_normal_order}). This choice is convenient for comparing with numerical diagonalization of the full coupled Hamiltonian. As has been noted before, coefficients $\overline c_{nn',m}$ are equal to $c_{nn',m}$ in Eq.~(\ref{eq:E_m_expansion}) in the weak coupling limit. 


As analyzed in the previous section, in the weak coupling regime, cavity nonlinearities can be computed perturbatively in transmon-cavity coupling $H_I^{\rm RWA}$ in Eq.~(\ref{eq:H_RWA}) using standard time-independent perturbation theory. To fourth-order in $H_I^{\rm RWA}$, we obtain terms in the eigenenergy $\mathcal E_m(N_A,N_B)$ that are second-order in $N_A,N_B$; see Eqs.~(\ref{eq:E_m_expansion},\ref{eq:c_scaling}). Identifying coefficients of those terms as cavity Kerr nonlinearities, we obtain their expressions to be as follows: 
\begin{widetext}
\begin{align}
\label{eq:self_Kerr}
\frac{K_{A,m}}{ 2 |g_{a}|^4} = \sum_{n=0}^\infty \sum_{j=\pm 1} \frac{|M_{a,nm}^{(j,j)}|^{2}}{\epsilon_{mn}/\hbar-2j\delta_{da}} +\sum_{n\neq m}\frac{|M_{a,nm}^{(+1,-1)}+M_{a,nm}^{(-1,+1)}|^{2}}{\epsilon_{mn/\hbar}}-[M_{a,mm}^{(+1,-1)}+M_{a,mm}^{(-1,+1)}][N_{a,mm}^{(+1,-1)}+N_{a,mm}^{(-1,+1)}],
\end{align}
\end{widetext}
\begin{widetext}
\begin{align}
\label{eq:cross_Kerr}
\frac{K_{AB,m}}{|g_{a}g_b|^2} = &\sum_{n=0}^\infty \sum_{j=\pm 1} \Big[\frac{|M_{a,nm}^{(j,j)}+M_{b,nm}^{(j,j)}|^{2}}{\epsilon_{mn}/\hbar-j(\delta_{da}+\delta_{db})} +\frac{|M_{a,nm}^{(j,-j)}+M_{b,nm}^{(-j,j)}|^{2}}{\epsilon_{mn}/\hbar+j(\delta_{da}-\delta_{db})} \Big] \nonumber\\ 
&+2{\rm Re}\sum_{n\neq m}  \frac{(M_{a,nm}^{(+1,-1)}+M_{a,nm}^{(-1,+1)})(M_{b,nm}^{(+1,-1)}+M_{b,nm}^{(-1,+1)})^*}{\epsilon_{mn}/\hbar} \nonumber \\ 
&- [M_{a,mm}^{(+1,-1)}+M_{a,mm}^{(-1,+1)}][N_{b,mm}^{(+1,-1)}+N_{b,mm}^{(-1,+1)}]-[M_{b,mm}^{(+1,-1)}+M_{b,mm}^{(-1,+1)}][N_{a,mm}^{(+1,-1)}+N_{a,mm}^{(-1,+1)}].
\end{align}
\end{widetext}
where $\epsilon_{mn}\equiv \epsilon_m-\epsilon_n$ and, the tensors $M,N$ in Eqs.~(\ref{eq:self_Kerr},\ref{eq:cross_Kerr}) above are defined as 
\begin{align}
\label{eq:M}
M_{a(b),nm}^{(i,j)}=&\sum_{m'}\frac{c_{nm'}^{(i)}c_{m'm}^{(j)}}{\epsilon_{mm'}/\hbar-j\delta_{da(b)}},  i,j = \pm 1, \nonumber \\
N_{a(b),nm}^{(i,j)}=&\sum_{m'}\frac{c_{nm'}^{(i)}c_{m'm}^{(j)}}{(\epsilon_{mm'}/\hbar-j\delta_{da(b)})^{2}}, i,j = \pm 1.
\end{align}
Here $c^{(\pm 1)}_{mn} $ represent the matrix elements of the operators $\c$ and $\c^\dagger$ between eigenstates $\psi_m$ and $\psi_{n}$ of the driven ancilla: $c^{(+1)}_{mn} = \langle \psi_m|\c ^{\dagger}|\psi_{n} \rangle$ ,$c^{(-1)}_{mn} = \langle \psi_m|\c|\psi_{n} \rangle$~\footnote{Note that the matrix elements $c^{(\pm 1)}_{mn}$ are to be distinguished from the expansion coefficients $c_{mn,k}$ introduced earlier in Eq.~(\ref{eq:E_m_expansion}).}. Self-Kerr $K_{B,m}$ of cavity-$B$ is given by the same expression as Eq.~(\ref{eq:self_Kerr}) with $a$ replaced by $b$ everywhere.
 
Equations~(\ref{eq:self_Kerr},\ref{eq:cross_Kerr}) comprise a major result of the paper. The result captures non-perturbatively the drive-induced change to the cavity Kerr nonlinearity through the matrix elements $c^{(\pm 1)}_{nm'}$ and eigenenergies $\epsilon_m$ of the driven ancilla. We emphasize that evaluating the matrix elements and eigenenergies only requires solving the ancilla Hamiltonian $H_{\rm anc}^{\rm RWA}$ in Eq.~(\ref{eq:H_RWA}). This greatly reduces the numerical complexity encountered in diagonalizing the full interacting transmon-cavity system. Moreover, being an explicit function of the cavity frequencies, the expressions for cavity Kerr allow us to efficiently explore the cavity frequency dependence of Kerr for given transmon and drive parameters; see Sec.~\ref{sec:drive_dependence_Kerr}.

In general, conditioned on the transmon being in different Floquet state $\psi_m$, the cavity Kerr $K_{A(B),m}$ and $K_{AB,m}$ are different. Of primary interest to us is the value of cavity Kerr when the driven transmon is in state $\psi_0$ that adiabatically connects to the vacumm state $|0\rangle$ of the undriven transmon as the drive is ramped up or down. 

In the absence of the transmon drive, analytical expressions for cavity Kerr can be readily obtained using Eqs.~(\ref{eq:self_Kerr},\ref{eq:cross_Kerr}). In this case, transmon driven eigenstates $\psi_m$ reduce to Fock states $|m\rangle$. The only non-zero matrix elements of the transmon ladder operators are those between neighboring Fock states. We find that the cavity self-Kerr and cross-Kerr when the transmon is in the vacuum state $|0\rangle$ read:
\begin{align}
\label{eq:Kerr_no_drive}
K_{A,0} &= -2|g_a|^4 \frac{\alpha}{\delta_{a}^{3}(2\delta_{a}+\alpha)}, \nonumber \\
K_{AB,0}&=-|g_{a}g_{b}|^2 \frac{2\alpha(\delta_{a}+\delta_{b})}{\delta_{a}^{2}\delta_{b}^{2}(\delta_{a}+\delta_{b}+\alpha)}.
\end{align}
It is straightforward to verify that the expressions for cavity Kerr in Eq.~(\ref{eq:Kerr_no_drive}) reduce to the results in Eq.~(\ref{eq:H_quar}) and Eq.~(\ref{eq:H_TLS}) in the limit of $\alpha \ll |\delta_{a(b)}|$ and $\alpha \gg |\delta_{a(b)}|$, respectively. We have also verified using Eq.~(\ref{eq:self_Kerr}) that the cavity Kerr when the transmon is in the first excited state $|1\rangle$ satisfies $K_{A,1} = K_{A,0}$ and $K_{A,1} = -K_{A,0}$ in the respective limit of small and large $\alpha/ |\delta_a|$, consistent with the discussion in Sec.~\ref{sec:theory_no_drive}. Similar results hold for the cross-Kerr.

\subsection{Connection between cavity nonlinearities and nonlinear susceptibility functions of the ancilla}
\label{sec:connection_to_chi}
An important feature of the expressions for cavity Kerr in Eqs.~(\ref{eq:self_Kerr},\ref{eq:cross_Kerr}) is that they are explicitly functions of the cavity frequencies. Once the matrix elements $c_{mn}^{(\dagger)}$ and quasienergies $\epsilon_m$ are computed by solving the ancilla Hamiltonian, the cavity Kerr nonlinearities as functions of cavity frequencies are uniquely determined. This allows us to efficiently explore the cavity-frequency-dependence of their Kerr nonlinearities for given ancilla and drive parameters. Before diving into details of this dependence in Sec.~\ref{sec:drive_dependence_Kerr}, we show that the cavity Kerr nonlinearities as  functions of the cavity frequencies are in fact related to the third-order nonlinear susceptibility functions of the transmon ancilla. 

We have previously shown that in the weak coupling regime, coupling-induced linear properties of the cavity modes such as cavity frequency shifts and linear (i.e., single-photon) decay rates can be calculated by treating the cavity operators $\a^\dagger +\a$ and $\b^\dagger + \b$ in the cavity-transmon coupling $H_I$ as weak classical drives (``probe" tones), and then computing the linear responses of relevant transmon dynamical variables to these weak probes~\cite{zhang2019}. Here, we generalize this method by considering transmon nonlinear responses to the probes, and show that the third-order nonlinear response (characterized by the third-order nonlinear susceptibility function) is directly proportional to the cavity Kerr nonlinearties. 

Let us consider a transmon-cavity interaction of the form: $H_I = -[\lambda_a (\a+\a^\dagger)  + \lambda_b (\b + \b^\dagger)]\O$, where $\O$ is some transmon operator and $\lambda_a,\lambda_b$ are the coupling strengths of the cavity fields to this operator. For the Hamiltonian considered in Eq.~(\ref{eq:full_Hamiltonian}), $\O$ is the transmon charge operator and $\lambda_{a(b)}\propto V_{a(b)}$. As in Ref.~\cite{zhang2019}, we switch to the interaction picture where operator $\a$ becomes $\a \exp(-i\omega_a t)$ and operator $\b$ becomes $\b \exp(-i\omega_b t)$. Then we treat the cavity operators as amplitudes of classical drives, and compute the expectation value of the response of the transmon operator $\O$ to the classical drives. Specifically, the third-order nonlinear response contains the following terms~(see Appendix~\ref{app:susceptibility}):

\begin{align}
\label{eq:nonlinear_response}
\langle \O^{(3)} \rangle_m & = \lambda_a^3 \a^\dagger \a^2 \chi_m^{(3)}(\omega_a,-\omega_a,\omega_a; \omega_a)e^{-i\omega_a t} \nonumber \\ &+ \lambda_a \lambda_b^2 \b^\dagger \b \a \chi_m^{(3)}(\omega_a,-\omega_b,\omega_b;\omega_a)e^{-i\omega_a t} + (a\leftrightarrow b)\nonumber \\
&+ \rm {H.c.}
\end{align}
When computing the transmon response, we have approximated the cavity operators $\a,\b$ as being constant in time because they are slowly varying on the time scale of the inverse cavity-transmon detunings. The third-order nonlinear susceptibility function $\chi^{(3)}_m(\omega_1,\omega_2,\omega_3;\omega_4)$ follows the standard definition in nonlinear optics in which the first three arguments represent the probe frequencies and the last argument represents the response frequency~\cite{boyd2008}; the subscript $m$ indicates that we are taking the expectation value with respect to ancilla eigenstate $\psi_m$. Without the ancilla drive at frequency $\omega_d$, the response frequency $\omega_4$ is equal to $\omega_1+\omega_2+\omega_3$; however, with the drive, $\omega_4$ can differ from that by integer multiples of $\omega_d$. Importantly, this susceptibility function is an intrinsic property of the driven ancilla and is not dependent on the ancilla-cavity coupling or the intrinsic cavity properties.  While there are also other terms in the third-order nonlinear response, we have only written explicitly terms that are related to the cavity Kerr nonlinearities, as we will show below. We also note that although we chose to write the cavity operators in a normal-ordered form on the right-hand side of Eq.~(\ref{eq:nonlinear_response}), the results for the cavity Kerr are not dependent on this choice to leading order in $\lambda_{a(b)}$.

In order to see how the susceptibility function $\chi^{(3)}$ relates to cavity Kerr, we look at the Heisenberg equations of motion for operators $\a,\b$ which read: 
\begin{align*}
\dot \a = i\hbar^{-1}\lambda_a \O e^{i\omega_a t}, \\
\dot \b = i\hbar^{-1}\lambda_b \O e^{i\omega_b t}.
\end{align*}
Upon the substitution of operator $\O$ in the above equations of motion with $\langle \O^{(3)}\rangle$ in Eq.~(\ref{eq:nonlinear_response}) and neglecting fast oscillating terms, we obtain that~\cite{Note4}
\begin{align*}
\dot \a = \, &i\hbar^{-1}\lambda_a^4 \a^\dagger \a^2 \chi_m^{(3)}(\omega_a,-\omega_a,\omega_a;\omega_a) \\ &+ i\hbar^{-1}\lambda_a^2\lambda_b^2 \b^\dagger \b \a \chi_m^{(3)}(\omega_a,-\omega_b,\omega_b;\omega_b)+... \\
\dot \b = \, & i\hbar^{-1}\lambda_b^4 \b^\dagger \b^2 \chi_m^{(3)}(\omega_b,-\omega_b,\omega_b;\omega_b) \\ & +i\hbar^{-1}\lambda_a^2\lambda_b^2 \a^\dagger \a \b \chi_m^{(3)}(\omega_b,-\omega_a,\omega_a;\omega_b)+...
\end{align*}

From the above equations of motion, we immediately identify the following relations between the cavity Kerr nonlinearities  
and transmon $\chi^{(3)}$:
\begin{align}
\label{eq:Kerr_chi_relation}
K_{A(B),m}= & -\hbar^{-1}\lambda_{a(b)}^4\Re\chi^{(3)}_m(\omega_{a(b)},\omega_{a(b)},-\omega_{a(b)};\omega_{a(b)}), \\
K_{AB,m}= & -\hbar^{-1}\lambda_a^2\lambda_b^2\Re\chi^{(3)}_m(\omega_a,\omega_b,-\omega_b;\omega_a).
\label{eq:Kerr_chi_relation_2}
\end{align}
%
In the absence of the ancilla decoherence, $\chi^{(3)}_m(\omega_{a(b)},\omega_{a(b)},-\omega_{a(b)};\omega_{a(b)})$ and $\chi^{(3)}_m(\omega_a,\omega_b,-\omega_b;\omega_a)$ are real functions. For the RWA Hamiltonian in Eq.~(\ref{eq:H_RWA}), we can substitute $\lambda_{a(b)}$ with $i\hbar g_{a(b)}$, and operator $\O$ with $-i(\c-\c^\dagger)$, and then the expressions for $\chi^{(3)}$ in Eqs.~(\ref{eq:Kerr_chi_relation},\ref{eq:Kerr_chi_relation_2}) under the RWA can be found from Eqs.~(\ref{eq:self_Kerr},\ref{eq:cross_Kerr}), respectively. 


More insight into the connection between the cavity Kerr in the weak coupling regime and the transmon $\chi^{(3)}$ can be gained as follows. In applying the leading-order perturbation theory to obtain the cavity Kerr in Eqs.~(\ref{eq:self_Kerr},\ref{eq:cross_Kerr}), the results are not sensitive to the difference between $\sqrt{N_{a(b)}}$ and $\sqrt{N_{a(b)}+m}$ ($m$ is some integer independent of $N_{a(b)}$) that come from the matrix elements of the bare cavity ladder operators $\a$ and $\b$. This means that to leading order in the perturbation theory, the cavity Kerr is not sensitive to the commutator between $\a^\dagger (\b^\dagger)$ and $\a (\b)$, thus justifying treating the quantized cavity modes as classical drives, as we did when computing the transmon response to the cavity fields in Eq.~(\ref{eq:nonlinear_response}).

\subsection{Dependence of cavity Kerr nonlinearities on the cavity-transmon detuning}
\label{sec:drive_dependence_Kerr}
In this section, we explore the dependence of the cavity Kerr nonlinearities on the cavity-transmon detuning using Eqs.~(\ref{eq:self_Kerr},\ref{eq:cross_Kerr}).
 
\begin{figure}[ht]
\includegraphics[width=8. cm]{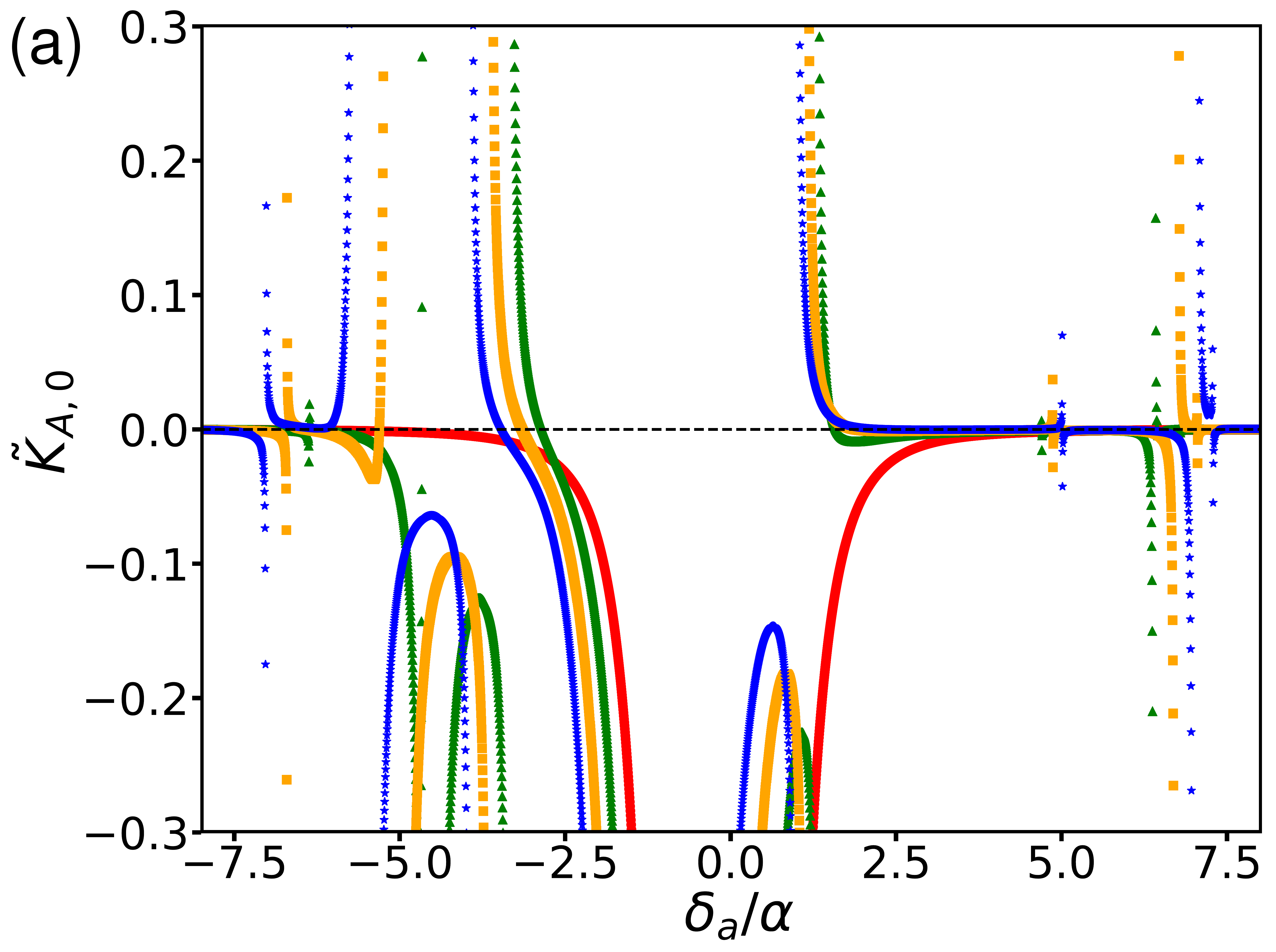} \hfill
\includegraphics[width=8.3 cm]{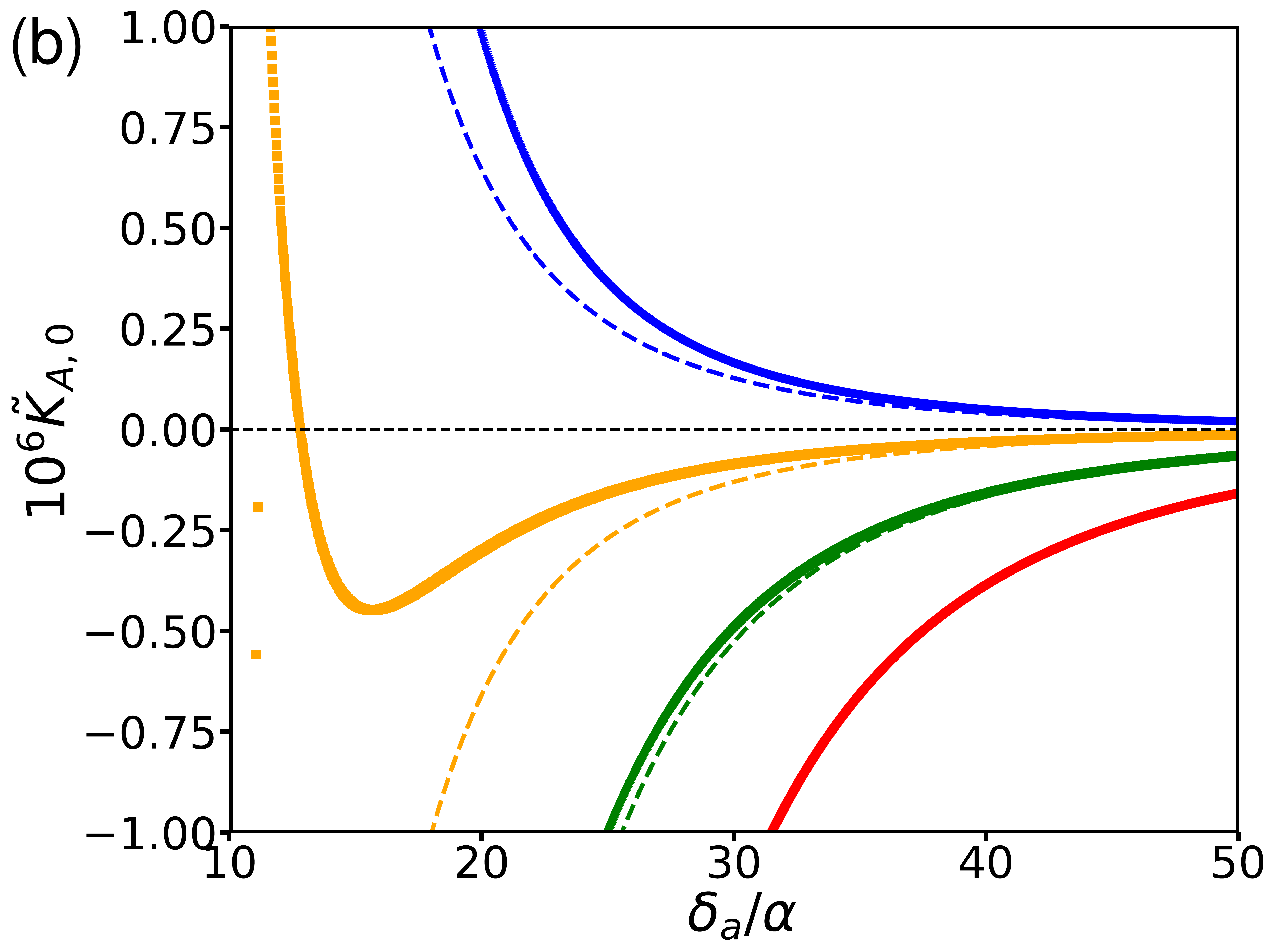} 
\caption{Cavity self-Kerr spectrum in the weak-coupling regime: (a) small-to-moderate cavity-transmon detuning; (b) large cavity-transmon detuning. The dimensionless cavity self-Kerr $\tilde K_{A,0}$ is defined as $\tilde K_{A,0}= \alpha^3 K_{A,0}/|g_a|^4$~\cite{Note-scaling}. $\delta_a$ is the detuning of cavity-a frequency from transmon transition frequency $\omega_{10}$. The spectrum is determined by two dimensionless drive parameters: $|\Omega_d/\delta_d|^2$ and $\delta_d/\alpha$. Their values are $|\Omega_d/\delta_d|^2 = 0$ (red dots), 0.3 (green triangles), 0.6 (orange squares), 0.9 (blue stars) and $\delta_d/\alpha = 3$ for all. The dashed lines in panel (b) refer to the large-$\delta_a$ asymptotic expression for cavity self-Kerr in Eq.~(\ref{eq:asymptotic_self_Kerr}). The two ``outlier" orange points in panel (b) are due to a sharp resonance near $\delta_a/\alpha = 11.$ The dashed red line (not visible) overlaps with the solid red curve. In both panels, the horizontal black dashed line indicates the zero of the cavity self-Kerr.}
\label{fig:self_Kerr_spec}
\end{figure}

Figure~\ref{fig:self_Kerr_spec} shows the cavity self-Kerr $K_{A,0}$ as a function of the cavity detuning $\delta_a$. As discussed in the previous section, this function is proportional to the transmon susceptibility function $\chi_0^{(3)}(\omega,-\omega,\omega;\omega)$. Colloquially, we shall refer it as the cavity self-Kerr spectrum. The spectrum can be qualitatively split into two regimes. The first regime [see Fig.~\ref{fig:self_Kerr_spec} (a)] is where the cavity detuning from the ancilla is of the order of ancilla anharmonicity and/or the drive detuning from the ancilla: $|\delta_{a}| \sim  \rm {max}( \alpha, |\delta_{d}|)$. In this regime, there is a rich dispersive structure in the cavity Kerr spectrum as a result of the drive-induced multiphoton resonances among cavity and transmon excitations. The second regime [see Fig.~\ref{fig:self_Kerr_spec} (b)] is where the cavity detuning from the ancilla is much larger than ancilla anharmonicity and drive detuning $|\delta_{a}| \gg \rm{max}(\alpha,|\delta_{d}|)$, so that the cavity is far away from any resonances. In this regime, the sharp dispersive structures associated with resonances become too weak to be visible and the cavity Kerr appears to be a much smoother function of the cavity detuning. We discuss features of these two regimes in more detail below. The cavity cross-Kerr $K_{AB,0}$ as a function of cavity detuning from the transmon shows similar features and is given in Appendix~\ref{app:cross_Kerr}.

\subsubsection{Small to moderate cavity-transmon detuning: near-resonant regime}
\label{sec:near_resonance}
 Because of the nonlinearity of the transmon, the drive can induce multiphoton resonances that result in resonant or near-resonant hybridization between the cavity excitations and transmon excitations even though the cavity modes are strongly off-resonant with the transmon in the absence of the drive. The resonance processes that give rise to the dispersive structures in the cavity self-Kerr (or cross-Kerr) spectrum are those that involve two photons at a time from the same cavity (or one from each cavity). As we will see below, these resonance processes occur in the regime where the cavity-transmon detuning is comparable to the maximum of transmon anharmonicity and the drive detuning: $|\delta_{a(b)}|\sim \rm{max}(\alpha,|\delta_{d}|)$.
 
The conditions for the resonant processes that affect cavity self-Kerr $K_{A(B),m}$ can be found by setting the denominator of the first term in Eqs.~(\ref{eq:self_Kerr}) to zero:
 \[ 
i) \quad 2j\omega_{a(b)} + (n-m-2j) \omega_d = \tilde \omega_{nm},\,j=\pm 1,
\]
Here $\tilde \omega_{nm}= (n-m)\omega_d + (\epsilon_n-\epsilon_m)/\hbar$ is the transmon transition frequency from the $n$-th to $m$-th level in the lab frame, with account taken of the drive-induced ac Stark shift. We use $\tilde\omega_{nm}$ to differentiate it from the un-Stark-shifted transition frequency $\omega_{nm}$. As a result of the use of the RWA, the resonance processes conserve the total excitation number.  Whenever the cavity frequency $\omega_a$ satisfies the above condition, the expression for the cavity self-Kerr in Eq.~(\ref{eq:self_Kerr}) diverges due to the perturbative treatment of the cavity-transmon coupling strength; these divergences can also be seen directly in Fig.~\ref{fig:self_Kerr_spec}(a). Similar conditions can be derived for cavity cross-Kerr $K_{AB,m}$; see Appendix~\ref{app:cross_Kerr}.


The pronounced dispersive structures near $\delta_a/\alpha = -3$ and $\delta_a/\alpha = 1.5$ for the green curve in Fig.~\ref{fig:self_Kerr_spec}(a) are due to processes $n = 3,m=0,j=1$ and $n=1,m=0,j=1$ in condition $i)$, respectively. The spectrum diverges when exactly on resonance (indicating the breakdown of the weak coupling/dispersive approximation). Generically, the cavity Kerr nonlinearity changes sign when the drive or system parameters are swept across the resonances. Notice that as the drive amplitude increases, the location of the divergences shift to the lower frequency as a result of the drive-induced ac Stark shift. Also the widths of the structures increase due to the increase in the resonance strengths.

In addition to the two-cavity-photon processes $i)$, single-cavity-photon processes also lead to sharp changes in the cavity self-Kerr as shown in Fig.~\ref{fig:self_Kerr_spec}(a). This is because single-cavity-photon processes affect the strengths of the intermediate virtual cavity-transmon transitions which in turn affect the size of the cavity Kerr, as can be seen from the expressions for the tensors $M,N$ in Eqs.~(\ref{eq:self_Kerr},\ref{eq:cross_Kerr}). The conditions for the drive-induced single-photon resonances that affect cavity self-Kerr $K_{A(B),m}$ are:
\[ 
ii) \quad j\omega_{a(b)} + (n-m-j) \omega_d = \tilde \omega_{nm},\, j=\pm 1.
\]
In Fig.~\ref{fig:self_Kerr_spec}(a), the divergence near $\delta_a/\alpha = -4$ for the green curve results from the process $n=2,m=0,j=1$ in condition $ii)$ and the divergence near $\delta_a/\alpha = 6$ corresponds to the process $n=1, m =0, j= -1$ in condition $ii)$. 

The modification to the cavity Kerr near a resonance shown in condition i) or ii) can be analyzed in the regimes of small and moderate anharmonicity, similar to the analysis in Secs.~\ref{sec:4wm_no_drive} and \ref{sec:TLS_no_drive}. It is instructive to explicitly write the resonance conditions in the limit of zero drive amplitude: condition i) becomes $2j\delta_{a(b)} + (k-2j)\delta_d = -\alpha k(2m+k-1)/2$ and condition ii) becomes $j\delta_{a(b)} + (k-j)\delta_d = -\alpha k(2m+k-1)/2$, where we have set $k=n-m$. 
Suppose that cavity $a$ or $b$ is near a specific resonance with given $k$ in condition $i)$. If $\alpha\ll |2j\delta_{a(b)} + (k-2j)\delta_d|$, then the cavity is simultaneously in near resonance with all processes with same $k$ but different $m$. In this case, the drive-induced change to $K_{A,m}$ weakly depends on $m$ similar to the situation in Sec.~\ref{sec:4wm_no_drive}. 
On the contrary, if $\alpha\gg |2j\delta_{a(b)} + (k-2j)\delta_d|$, it is possible to have $\delta_{a(b)}$ to be in near resonance with a particular transmon transition from state $m$ to $n$ but sufficiently far away from others. In this case, the near-resonant dynamics can be well described by restricting to the two level subspace (state $m$ and $n$) for the transmon. Hybridization of the cavity excitations with this subspace results in a characteristic change in cavity Kerr $K_{A(B),m}$ and $K_{A(B),n}$ similar to that described in Sec.~\ref{sec:TLS_no_drive}. 
An example of such near-resonant dynamics in which $\omega_a + \omega_d \approx \tilde \omega_{20}$ is analyzed in Appendix~\ref{app:near_resonance}.

In both cases, the modification to the cavity Kerr is accompanied by stronger hybridization between the cavity and transmon thus stronger decay the cavity inherits from the transmon via the Purcell effect. In Sec~\ref{sec:asymptotic_regime}, we shall focus on the regime of large cavity-transmon detuning where the cavity-transmon interaction remains strongly dispersive yet the cavity Kerr is subject to drive-induced modification. 

The divergent behavior of the cavity Kerr at the aforementioned multiphoton resonances indicates the breakdown of the perturbation theory that leads to Eqs.~(\ref{eq:self_Kerr},\ref{eq:cross_Kerr}). Generically, the perturbation theory becomes inaccurate when the distance of the cavity frequencies to one of the resonances become small or comparable to the coupling-induced frequency shifts of either cavity or relevant transmon transition frequency. Strengths of these frequency shifts are second-order in the transmon-cavity coupling; therefore they are typically stronger than the coupling-induced cavity Kerr nonlinearities. In Appendix~\ref{app:beyond_weak_coupling}, we discuss in more detail the breakdown of the perturbation theory by comparing it with the exact numerical diagonalization of the full system and show how to incorporate non-perturbative corrections to the weak-coupling expressions in Eqs.~(\ref{eq:self_Kerr},\ref{eq:cross_Kerr}).

\subsubsection{Large cavity-transmon detuning: asymptotic regime}
\label{sec:asymptotic_regime}
For a large cavity-transmon detuning [i.e., $|\delta_{a(b)}|\gg {\rm max}(\alpha,|\delta_{d}|)$], the cavity modes are far detuned from being in resonance with the driven transmon and thus  cavity-transmon coupling remains strongly off-resonant. This off-resonant coupling leads to a dispersive cross-Kerr interaction between the cavity-like eigenmodes and the transmon-like eigenmode as described in Sec.~\ref{sec:4wm_no_drive}. As a result of this coupling, transition frequencies of the transmon mode depend on the cavity photon numbers. In the presence of the transmon drive, drive-induced ac Stark shifts of transmon levels also depend on the cavity photon number, which translates into effective cavity nonlinearities. 

In contrast to the near-resonant regime discussed in the previous section, the cavity Kerr spectra do not develop sharp divergent features; see Fig.~\ref{fig:self_Kerr_spec}(b). Instead, it changes relatively smoothly as a function of the cavity-transmon detuning. For asymptotically large detuning, the cavity Kerr spectra decay as a power law in $1/\delta_a$ and $1/\delta_b$. This power law decay can be found by expanding the expressions for cavity Kerr in Eqs.~(\ref{eq:self_Kerr},\ref{eq:cross_Kerr}) with respect to $1/\delta_a$ and $1/\delta_b$. To the leading order in the expansion, we obtain that 
\begin{align}
\tilde K_{A,m} & \sim s_m \left (\frac{\delta_a}{\alpha}\right)^{-4},\label{eq:asymptotic_self_Kerr}
\\
 \tilde K_{AB,m} & \sim 2 c_m \left(\frac{\delta_a\delta_b}{\alpha^2}\right)^{-2},
\left|\frac{\delta_a}{\alpha}\right|,\left|\frac{\delta_b}{\alpha}\right| \rightarrow \infty.\label{eq:asymptotic_cross_Kerr}
\end{align}
As in Fig.~\ref{fig:self_Kerr_spec}, the dimensionless cavity self-Kerr and cross-Kerr are defined as $\tilde K_{A,m} = \alpha^3 K_{A,m}/|g_a|^4$, $\tilde K_{AB,m} = \alpha^3 K_{AB,m}/|g_ag_b|^2$. The asymptotic expression in Eq.~(\ref{eq:asymptotic_self_Kerr}) agrees well with the full expression as shown in Fig.~\ref{fig:self_Kerr_spec}(b). 

The expansion coefficients $s_m,c_m$ in Eqs.~(\ref{eq:asymptotic_self_Kerr},\ref{eq:asymptotic_cross_Kerr}) are controlled by two dimensionless drive parameters $\Omega_d/\delta_d$ and $\delta_d/\alpha$. In the absence of drive, we have $s_m=c_m=-1$  which follows from Eq.~(\ref{eq:Kerr_no_drive}).
At finite drive, analytical expressions for $s_m,c_m$ can be obtained by calculating the drive-induced ac Stark shift of transmon-like eigenmode with account taken of the cross-Kerr coupling between the transmon and cavity modes, and then expanding it with respect to the cavity photon numbers. Here, we discuss the implication of the results in the weak coupling regime, and describe the detailed calculation in Sec.~\ref{sec:large_cavity_detuning} which generally applies beyond the weak coupling regime. 

For weak drive, by evaluating the drive-induced ac Stark shift of transmon levels to second order in $\Omega_d$ (see Sec.~\ref{sec:weak_drive_limit}), we have:  
\begin{align}
\label{eq:sm_weakdrive}
    s_m  = & -1 + \Delta_m,\,c_m  = -1 + \Delta_m/2, \\
    \Delta_m \approx & 8\alpha |\Omega_d|^2\left[\frac{m+1}{(\delta_{d}+m\alpha)^3}-\frac{m}{[\delta_{d}+(m-1)\alpha]^3}\right]. \nonumber  
\end{align}
It is interesting to note that for $\delta_{d}>0$, the coefficients $s_0$ and $c_0$ become less negative as the drive power increases. At certain drive power, they even change sign as shown by Fig.~\ref{fig:self_Kerr_spec}(b). As we will discuss in more detail in Sec.~\ref{sec:Kerr_cancellation}, such dependence on the drive power provides a way to cancel the cavity Kerr nonlinearity. 

In order to go beyond the weak drive regime, one can consider two different limits. First, in the limit where the drive frequency is in near resonance with a specific transmon transition frequency but far detuned from others, $$|\omega_d - \omega_{(m_0+1)m_0}| \ll |\omega_d - \omega_{(m+1)m}|\sim \alpha,\,\rm{ for}\, m\neq m_0,$$i.e. $\delta_{d}/\alpha + m_0\ll 1$, the cavity Kerr nonlinearity is strongly altered by the drive only when the transmon is in state $m_0$ or $m_0+1$. Truncating to the subspace spanned by states $m_0$ and $m_0+1$ allows us to obtain an analytical expression for $\Delta_m$ beyond the weak-drive regime (see Sec.~\ref{sec:small_drive_detuning}):  
\begin{align}
\label{eq:Delta_m0}
    & \Delta_{m_0} = -\Delta_{m_0+1} \nonumber \\
    & = \frac{8(m_0+1)\sgn[\delta_{d}+m_0\alpha]\alpha|\Omega_d|^2} {[(\delta_{d}+m_0\alpha)^2+4(m_0+1)|\Omega_d|^2]^{3/2}}, \\
    &|\Delta_m| \ll |\Delta_{m_0}|,\,\rm{for}\,,m\neq m_0,m_0+1. \nonumber
\end{align} 
%
To leading order in $|\Omega_d|^2$, Eq.~(\ref{eq:Delta_m0}) can also be obtained from Eq.~(\ref{eq:sm_weakdrive}) in the limit $\delta_{d}/\alpha +m_0 \ll 1$. For stronger drive, $\Delta_{m_0}$ changes nonlinearly in the drive power. $|\Delta_{m_0}|$ reaches a maximum $(4/\sqrt{27})\alpha/|\delta_{dc}+(m_0+1)\alpha|$ at $|\Omega_d| = |\delta_{d}+m_0\alpha|/2$ and then decreases to zero in the limit $|\Omega_d| \gg |\delta_{d}+m_0\alpha|$. 

In the opposite limit where the drive is far away from any transmon transition frequency, $$|\omega_d - \omega_{(m+1)m}|\gg \alpha, \,\rm{for\,all}\,m,$$ namely, $\delta_{d} \gg \alpha$, dynamics of the driven ancilla becomes semiclassical.  One can solve the ancilla Hamiltonian perturbatively in the dimensionless parameter $\alpha/\delta_{dc}\ll 1$. To first order in $\alpha/\delta_{dc}$, we have (see Sec.~\ref{sec:semiclassical} for the detailed derivation):
\begin{align}
\label{eq:Delta_m_semiclassical}
    \Delta_m = &\frac{8Q_0^2}{3Q_0^2+1} - 12(m+\frac{1}{2})\frac{\alpha}{\delta_{dc}} \nonumber \\
    &\times\frac{Q_0^2(4+3Q_0^2)\sqrt{(3Q_0^2+1)(Q_0^2+1)}}{(1+3Q_0^2)^4},
\end{align}
where $Q_0$ is the solution to the cubic equation: $Q_0^3 + Q_0 = |\Omega_d|\sqrt{\alpha/\delta_{dc}^3}$. For weak drive, we have $\Delta_m \approx 8\alpha(|\Omega_d|^2/\delta_{dc}^3)[1-6(\alpha/\delta_{dc})(m+1/2)]$, which can also be obtained from Eq.~(\ref{eq:sm_weakdrive}) by taking the limit $\delta_{d}\gg \alpha$. For a strong drive, interestingly, we find that $\Delta_m$ saturates to a drive-independent value $\Delta_m = 8/3.$ 

An important qualitative difference between the small- and large-drive-detuning limit lies in 
the variation of the drive-induced change of cavity Kerr among different transmon states $\psi_m$. In the small-drive-detuning limit, the drive mainly couples to a two-level subspace of the transmon, resulting in a strong change of cavity Kerr conditioned on the transmon in this subspace. In the large-drive-detuning limit, however, the non-equidistance ($\sim \alpha$) of the transmon levels is masked by the relatively large drive detuning $\delta_{d}$. As a result, the drive-induced ac-Stark shifts of transmon levels are all close to each other (at least for lower levels). This results in a relatively weak dependence of cavity Kerr on the transmon levels. 

\begin{figure}[ht]
\includegraphics[width=8.5 cm]{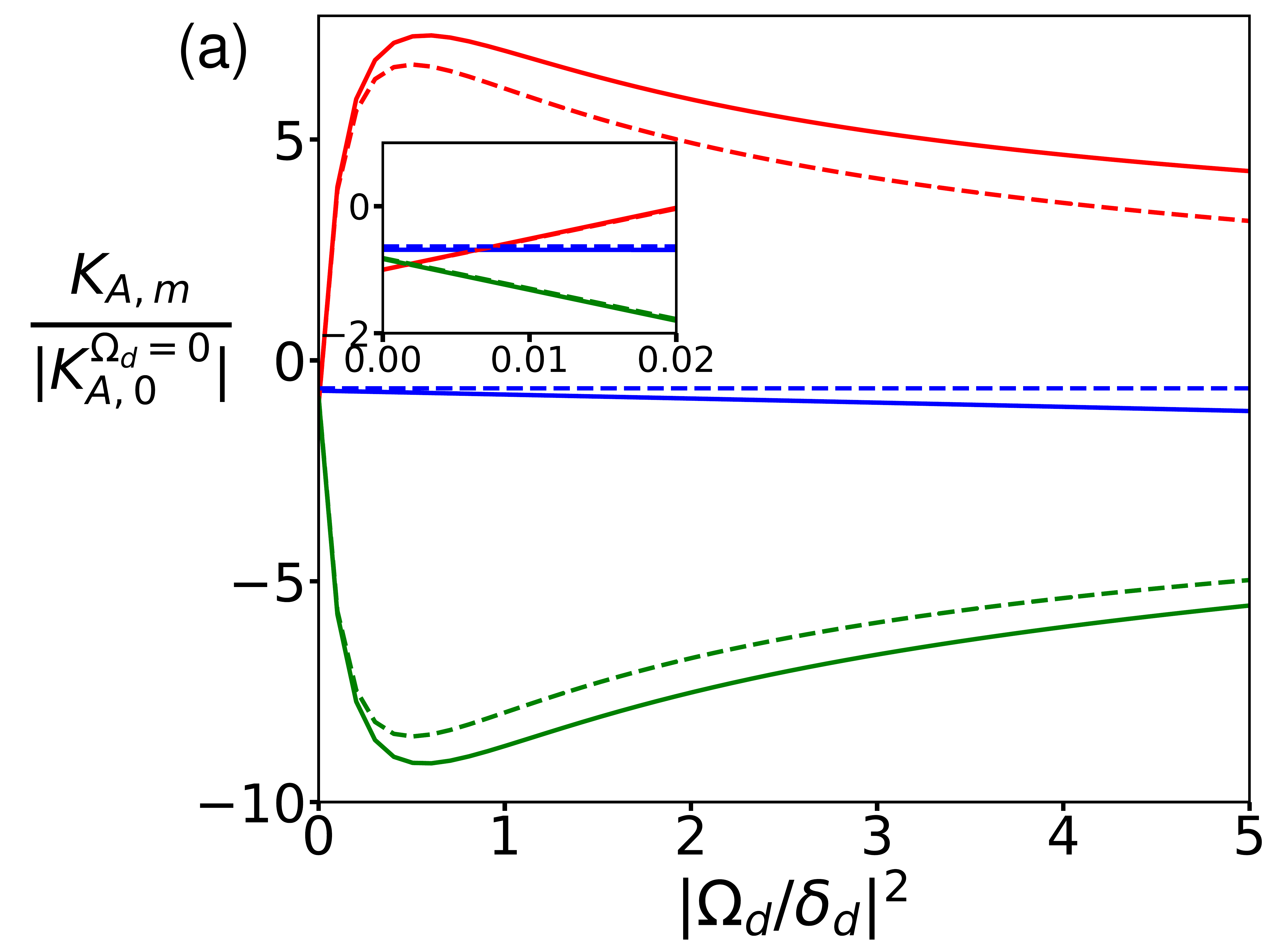} \hfill
\includegraphics[width=8.5 cm]{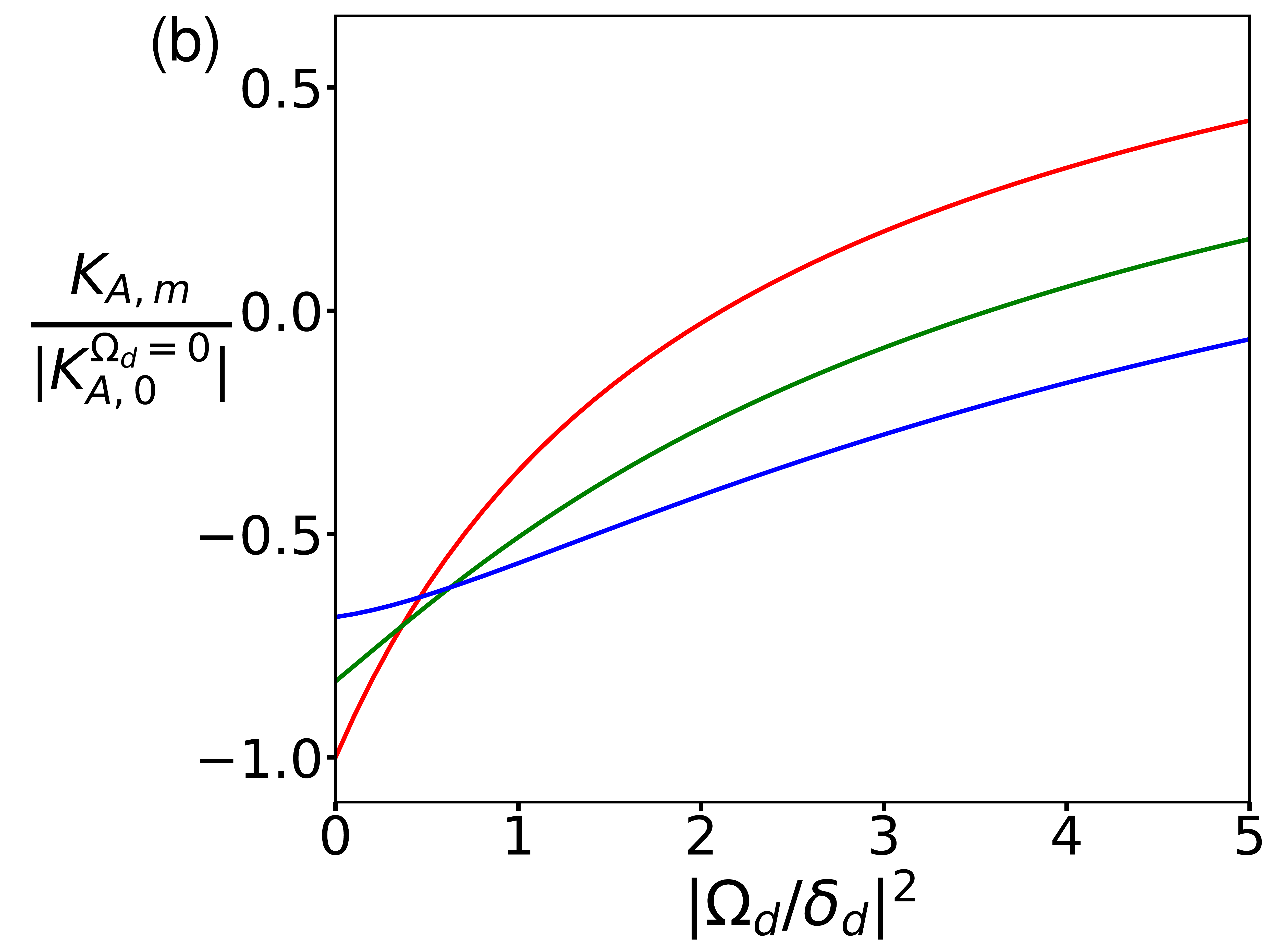} 
\caption{Variations of cavity self-Kerr $K_{A,m}$ among different transmon states $\psi_m$: (a) small drive-transmon detuning, $\delta_d = \alpha/10$; (b) large drive-transmon detuning, $\delta_d = 10\alpha.$ The red, green and blue curves refer to $m=0,1,2$, respectively. Solid lines are weak-coupling expression in Eq.~(\ref{eq:self_Kerr}), and dashed lines in panel (a) are analytical results in Eq.~(\ref{eq:Delta_m0}). Cavity-transmon detuning is $\delta_a = 50\alpha$ in both panels. Inset in panel (a) refers to the range $|\Omega_d/\delta_d|^2\leq 0.02$ zoomed in.}
\label{fig:self_Kerr_dispersion}
\end{figure}

To illustrate this difference, we show in Fig.~\ref{fig:self_Kerr_dispersion} the cavity self-Kerr as a function of the scaled drive power for the two different limits: $\delta_d\ll \alpha$ where the drive frequency is close to the transmon transition frequency $\omega_{10}$ and $\delta_d\gg\alpha$ where the drive is far away from all transmon transition frequencies. In the former case, consistent with  Eq.~(\ref{eq:Delta_m0}), the cavity Kerr $K_{A,0}$ and $K_{A,1}$ change in opposite direction and varies non-monotonically with respect to the drive amplitude. In contrast, the cavity self-Kerr $K_{A,m\neq 0,1}$ does not change much with respect to the drive amplitude. In the latter case, as predicted by Eq.~(\ref{eq:Delta_m_semiclassical}), cavity Kerr $K_{A,m}$ relatively weakly depends on $m$. Note that, for weak drive, we have $K_{A,m+1}> K_{A,m}$ [as expected from Eq.~(\ref{eq:H_quar_correction})]. At stronger drive, this hierarchy is flipped as predicted by the second term in  Eq.~(\ref{eq:Delta_m_semiclassical}). A comparison of the semiclassical result [Eq.~(\ref{eq:Delta_m_semiclassical})] with the full expression [Eq.~(\ref{eq:self_Kerr})] for cavity self-Kerr is given in Appendix~\ref{app:semiclassical_compare_weak_coupling}.

\section{Drive-induced cavity nonlinearities in the large cavity-transmon detuning regime}
\label{sec:large_cavity_detuning}
In this section, we focus on the parameter regime in which the individual cavity detuning from the transmon is much larger than the transmon anharmonicity: $|\delta_{a,b}|\gg \alpha $. This is a regime of significant experimental interest due to the relatively weak anharmonicity of the transmon. Further, we require that the cavity-transmon detunings are also much larger than the drive detuning, i.e., $|\delta_{a,b}|\gg |\delta_d |$, such that the cavity modes are far detuned from any drive-induced resonances. These two conditions allow us to go beyond the weak-coupling regime, in particular, to derive analytical expressions not just for the cavity Kerr nonlinearity but also higher-order nonlinearities in the presence of transmon drive. As we will show, convergence of higher-order cavity nonlinearities requires a more strict condition in the presence of the transmon drive than that without the drive. We compare the analytical results with experiments and numerics in Sec.~\ref{sec:experiment}.



As discussed in Sec.~\ref{sec:4wm_no_drive}, in the regime of $|\delta_{a(b)}|\gg \alpha$, it is convenient to express the RWA Hamiltonian in Eq.~(\ref{eq:H_RWA}) in terms of the ladder operators for the eigenmodes of the linear system:
\begin{align}
\label{eq:H_RWA_eigenmodes}
& H_{\rm RWA}/\hbar  =   \sum_{X\in\{ A,B,C\}}(-\delta_{dX}\N_X + \Omega_d^*\xi_X\X + \Omega_d\xi_X^*\X^\dagger) \nonumber \\  
& - \frac{\alpha}{2} \sum_{X_{1,2,3,4}\in \{A,B,C\}} \xi_{X_1}^*\xi_{X_2}\X_1^\dagger \X_2 (\xi_{X_3}^*\xi_{X_4}\X_3^\dagger \X_4 +1),
\end{align}
where $\delta_{dX} = \omega_d - \omega_X.$ Because of the hybridization between modes, the drive initially only acting on the bare transmon mode now acts on all the eigenmodes with a strength weighted by the participation factor $\xi_X$ introduced in Sec.~\ref{sec:4wm_no_drive}. Since we are primarily interested in the regime where $\xi_{A,B}\ll 1, \xi_C\approx 1$ and the drive is much closer to the transmon-like mode than the cavity-like modes, we will approximate $\xi_C\Omega_d$ as $\Omega_d$ and $\xi_{A(B)}\Omega_{d}$ as 0. 

In the absence of the drive, as we have shown in Sec.~\ref{sec:4wm_no_drive}, one can disregard non-dispersive terms in the second line of Eq.~(\ref{eq:H_RWA_eigenmodes}) to first order in $\alpha/|\delta_{a,b}|$. Such approximation does not necessarily apply in the presence of the drive since the drive can induce resonant or near-resonant interaction between cavities and transmon. This occurs when the drive detuning to the transmon is comparable to the cavity detunings to the transmon: $|\delta_d|\sim |\delta_{a(b)}|$. We give an example of how such drive-induced resonance can modify cavity nonlinearities in Appendix~\ref{app:near_resonance}. 

Under the condition $|\delta_{a,b}|\gg \rm{max}(|\delta_{d}|,\alpha)$, however, one can still neglect non-dispersive terms in the second line of Eq.~(\ref{eq:H_RWA_eigenmodes}). This leads to the following Hamiltonian:
\begin{align}
\label{eq:H_RWA_eigenmodes_dispersive}
   &  H_{\rm RWA} \approx H_C(\N_A,\N_B) + H_{AB}, \\
  &   H_C/\hbar =  -\hat \delta_{dC}(\N_A,\N_B) \N_C - \frac{\alpha}{2}\N_C (\N_C + 1) \nonumber \\ & +\Omega_d \C^\dagger + \Omega_d^* \C, \nonumber \\
  &  H_{AB}/\hbar =  -\sum_{X\in\{A,B\}}\delta_{dX}\N_X - \sum_{X,X'\in\{A,B\}} \chi_{XX'}\N_X\N_{X'},\nonumber\\
 & \hat \delta_{dC}(\N_A,\N_B)= \delta_{dC}+\chi_{AC}\N_A+\chi_{BC}\N_B. \nonumber
\end{align}
%
The definition of $\chi_{XX'}$ is below Sec.~\ref{sec:4wm_no_drive}. There is a shift in the frequency of eigenmodes $A,B$ due to the transmon anharmonicity which we have absorbed into $\delta_{dA(B)}$. We have approximated the anharmonicity $\chi_{CC}$ of eigenmode $C$ as the bare transmon anharmonicity $\alpha$ which differ by a factor of $|\xi_C|^4$. Note that the only coupling between mode $C$ and $A,B$ is the cross-Kerr coupling which we have absorbed into the definition of the drive detuning $\hat \delta_{dC}(\N_A,\N_B)$. 

The Hamiltonian $H_C(\N_A,\N_B)$ in Eq.~(\ref{eq:H_RWA_eigenmodes_dispersive}) can be interpreted as that of a driven transmon-like mode $C$. The drive detuning $\hat \delta_{dC}(\N_A,\N_B)$ depends ``parametrically" on the cavity photon number operators $\N_A,\N_B$ as a result of the cross-Kerr interaction between $A,B$ and $C$. Accordingly, the drive-induced ac Stark shift of the transmon levels depends parametrically on the cavity photon numbers. As we will show below, this dependence is generally nonlinear, which translates into effective nonlinearities of the cavity-like modes.

\subsection{Weak drive limit}
\label{sec:weak_drive_limit}
To leading order in the drive power, the drive-induced ac Stark shift to the $m$-th level of the transmon-like mode $C$ reads:
\begin{align}
\label{eq:ac_Stark_shift}
    & \hat {\delta\epsilon}_m^{\rm ac} (\N_A,\N_B) \approx  \frac{|\Omega_{d,m}|^2}{\hat \delta_{d,m}(\N_A,\N_B)} 
    - \frac{|\Omega_{d,m-1}|^2}{\hat \delta_{d,m-1}(\N_A,\N_B)}, \nonumber  \\
    &\hat \delta_{d,m}(\N_A,\N_B) \equiv \hat \delta_{dC}(\N_A,\N_B) +(m+1)\alpha,\nonumber  \\
    & \Omega_{d,m}\equiv \sqrt{m+1}\Omega_d.
\end{align}
$\hat \delta_{d,m}$ is the drive detuning from the transition frequency of the transmon-like mode between state $m+1$ and $m$. 
Expanding $\hat {\delta \epsilon}_m^{\rm ac} $ with respect to $\N_A,\N_B$, we obtain:
\begin{align}
\label{eq:Stark_shift}
\hat {\delta\epsilon}_m^{\rm ac} (\N_A,\N_B) =&\sum_{n=0}^{\infty}\left( \frac{|\Omega_{d,m-1}|^2}{(-\delta_{d,m-1})^{n+1}}-\frac{|\Omega_{d,m}|^2}{(-\delta_{d,m})^{n+1}}   \right)\nonumber \\ &\times  (\chi_{AC}\N_A+\chi_{BC}\N_B)^{n},
\end{align}
where $\delta_{d,m}\equiv \hat \delta_{d,m}(0,0) = \delta_{dC}+(m+1)\alpha$.
Identifying the coefficients in front of $\N_A^2$ and $\N_A\N_B$ as the drive-induced self-Kerr of mode-$A$ and cross-Kerr between modes $A$ and $B$, we reproduce Eq.~(\ref{eq:sm_weakdrive}) in the weak coupling limit. 

Higher-order terms in $\N_A,\N_B$ in Eq.~(\ref{eq:Stark_shift}) correspond to drive-induced change to higher-order cavity nonlinearities. To clearly see the condition of convergence of higher-order terms, let us take $m=0$. Eq.~(\ref{eq:Stark_shift}) simplifies to:
\begin{align}
\label{eq:Stark_shift_0}
\hat {\delta\epsilon}_0^{\rm ac} (\N_A,\N_B) =\frac{|\Omega_{d}|^2}{\delta_{d,0}} \sum_{n=0}^\infty  \left( -\frac {\chi_{AC}\N_A+\chi_{BC}\N_B}{\delta_{d,0}}\right)^{n}.
\end{align}
It follows that convergence of cavity nonlinearities when the transmon mode is in the lowest state requires that $\chi_{AC},\chi_{BC}\ll |\delta_{d,0}|. $ For general transmon state $m$, the condition becomes $\chi_{AC},\chi_{BC}\ll |\delta_{d,m}|,|\delta_{d,m-1}|$. 
Recall that in the absence of the drive, convergence of higher-order cavity nonlinearities requires a less strict condition, i.e., $\chi_{A(B)C}\ll |\delta_{A(B)C}|$; see Appendix~\ref{app:correction_to_chi}.

\subsection{Small drive-transmon detuning: two-level approximation}
\label{sec:small_drive_detuning}
In this section, we consider the regime $|\delta_{d,m_0}|\ll \alpha$ such that the drive frequency is close to a specific transmon transition frequency between states $m_0+1$ and $m_0$ and far from others. In this case, we can restrict the analysis to the Fock states $|m_0\rangle$ and $|m_0+1\rangle$ of the transmon-like mode $C$. Then Hamiltonian $H_C(\N_A,\N_B)$ in Eq.~(\ref{eq:H_RWA_eigenmodes_dispersive}) becomes: 
\begin{align}
\label{eq:H_RWA_small_drive_detuning}
&\frac{H_C(\N_A,\N_B)}{\hbar} \approx  -\hat \delta_{d,m_0}(\N_A,\N_B) \frac{\sig_z }{2}  + \Omega_{d,m_0} \sig_++\rm{H.c.} .
\end{align}
Here we have introduced $\sigma_z = |m_0+1\rangle\langle m_0+1| - |m_0\rangle\langle m_0|$, $\hat \sigma_+ = |m_0+1\rangle\langle m_0|$. 

Hamiltonian $H_C(\N_A,\N_B)$ in Eq.~(\ref{eq:H_RWA_small_drive_detuning}) can be  diagonalized: 
\begin{align}
\label{eq:tilde_delta_d_m0}
& \frac{H_C(\N_A,\N_B)}{\hbar} = -\hat {\tilde \delta}_{d,m_0}(\N_A,\N_B)\frac{\tsig_z}{2},  \nonumber \\
& \hat {\tilde \delta}_{d,m_0}(\N_A,\N_B) = \rm {sgn} (\hat \delta_{d,m_0})\sqrt{4|\Omega_{d,m_0} |^2 + \hat \delta_{d,m_0}^2(\N_A,\N_B)}.
\end{align}
Note that here eigenstates of $\tsig_z$ are rotated with respect to those of $\sig_z$ in Eq.~(\ref{eq:H_RWA_small_drive_detuning}); in the limit $\Omega_d \rightarrow 0$, the eigenstate of $\tsig_z$ with eigenvalue $+1(-1)$ continuously goes over to the eigenstate of $\sig_z$ in Eq.~(\ref{eq:H_RWA_small_drive_detuning}) with eigenvalue $+1 (-1)$.

As a result of the diagonalization, the transition frequency from transmon state $m_0+1$ to $m_0$ depends nonlinearly on the cavity photon number as manifested in $\hat {\tilde \delta}_{d,m_0}(\N_A,\N_B)$. Again, we expand $\hat {\tilde \delta}_{d,m_0}(\N_A,\N_B)$ with respect to $\N_A,\N_B$ and obtain:
\begin{align}
\label{eq:Delta_c_small_detuning}
     &\hat {\tilde \delta}_{d,m_0}(\N_A,\N_B)  = \sum_{n}\tilde \delta_{d,m_0}\left(\frac{\chi_{AC}\N_A+\chi_{BC}\N_B}{\tilde \delta_{d,m_0}}\right)^{n} \nonumber \\&\times \sum_{k= \rm {ceil}(n/2)}^{n}\frac{(-1)^{k+1}2^{k-n}(2k-3)!!}{(2k-n)!(n-k)!}\left(\frac{\delta_{d,m_0}}{\tilde \delta_{d,m_0}}\right)^{2k-n},
\end{align}
where $\tilde \delta_{d,m_0}= \rm {sgn}(\delta_{d,m_0})\sqrt{\delta_{d,m_0}^2+4|\Omega_{d,m_0}|^2}$. The summation over $k$ goes to zero when $\Omega_d = 0$ for $n\geq 2$. Equation~(\ref{eq:Delta_c_small_detuning}) shows that higher-order terms in the expansion are higher-order in $\chi_{A(B)C}/\tilde\delta_{d,m_0}$. Therefore,
the expansion series converges faster at stronger drive due to the increase in $|\tilde \delta_{d,m_0}|$ with the drive strength.  

Substituting Eq.~(\ref{eq:Delta_c_small_detuning}) into Eq.~(\ref{eq:tilde_delta_d_m0}) and combining with $H_{AB}$ in Eq.~(\ref{eq:H_RWA_small_drive_detuning}), we obtain that
\begin{widetext}
\begin{align}
\label{eq:H_RWA_TLS_expanded}
H_{\rm RWA}/\hbar = &-\left(\tilde \delta_{d,m_0}+\frac{\delta_{d,m_0}}{\tilde \delta_{d,m_0}}\sum_{X\in\{A,B\}}\chi_{XC}\N_X\right)\frac{\tilde \sig_z}{2}  -\sum_{X\in\{A,B\}}\delta_{dX}\N_X-\frac{1}{2}\sum_{X,X'\in\{A,B\}} (\chi_{XX'}+\Delta\chi_{XX'}\tilde \sig_z)\N_X \N_{X'} \nonumber \\
& + \left(\frac{\Delta\beta_A}{3!}\N_A^3 + \frac{\Delta\beta_B}{3!}\N_B^3+\frac{\Delta\beta_{AB}}{2!}\N_A^2\N_B +\frac{\Delta\beta_{BA}}{2!}\N_B^2\N_A + \frac{\Delta\sigma_{A}}{4!}\N_A^4 + \frac{\Delta\sigma_{B}}{4!}\N_B^4 + ...\right)\tilde\sigma_z, 
\end{align}
\end{widetext}
where ... represents the fourth- and higher-order terms in $N_A,N_B$ (excluding the $\N_A^4$ and $\N_B^4$ terms). The drive-induced nonlinearity parameters scaled by the static cavity Kerr parameters are given by the following: 
\begin{align}
\label{eq:Delta_chi_TLS}
\frac{\Delta\chi_{AA(BB)}}{\chi_{AA(BB)}} &= \frac{8 \alpha |\Omega_{d,m_0}|^2} {\tilde \delta_{d,m_0}^{3}},\nonumber\\
\frac{\Delta\chi_{AB}}{\chi_{AB}} &= \frac{4 \alpha |\Omega_{d,m_0}|^2 }{\tilde \delta_{d,m_0}^{3}}, \nonumber \\
\frac{\Delta\beta_{A(B)}}{\chi_{AA(BB)}} &= \frac{24\alpha|\Omega_{d,m_0}|^2}{{\tilde\delta_{d,m_0}^3}}\frac{\chi_{A(B)C}}{\tilde\delta_{d,m_0}}\frac{\delta_{d,m_0}}{\tilde\delta_{d,m_0}}, \nonumber \\
\frac{\Delta\beta_{AB(BA)}}{\chi_{AB}} &= \frac{12\alpha|\Omega_{d,m_0}|^2}{{\tilde\delta_{d,m_0}^3}}\frac{\chi_{A(B)C}}{\tilde\delta_{d,m_0}}\frac{\delta_{d,m_0}}{\tilde\delta_{d,m_0}}, \nonumber \\
\frac{\Delta\sigma_{A(B)}}{\chi_{AA(BB)}} & = - \frac{24\alpha|\Omega_{d,m_0}|^2}{\tilde\delta_{d,m_0}^3}\frac{\chi_{A(B)C}^2}{\tilde\delta_{d,m_0}^2}\frac{5\delta_{d,m_0}^2-\tilde\delta_{d,m_0}^2}{\tilde\delta_{d,m_0}^2}.
\end{align}

Equation~(\ref{eq:H_RWA_TLS_expanded}) immediately shows that the cavity Kerr and higher-order nonlinearity strengths when the transmon is in states $m_0$ and $m_0+1$ are modified by the drive, and the sign of the modification is opposite for the two states. We note that the above expressions for the drive-induced cavity nonlinearities go beyond the perturbation theory in $g_a,g_b$ laid out in Sec.~\ref{sec:weak_coupling_limit}. 
To fourth order in the coupling strengths $g_a,g_b$, the expressions for the drive-induced cavity Kerr nonlinearity $\Delta\chi_{XX'}$ reduce to the perturbative results in Eq.~(\ref{eq:Delta_m0}). 

At large drive strengths where $|\Omega_{d,m_0}| \gg |\delta_{d,m_0}|$, both the drive-induced change of the cavity fourth-order Kerr and higher-order nonlinearity strengths decay to zero with the increase of the drive amplitude. However, the former decays as $|\Omega_{d,m_0}/\delta_{d,m_0}|^{-1}$ while the latter decays as $|\Omega_{d,m_0}/\delta_{d,m_0}|^{-3}$ for the sixth-order nonlinearities (i.e., the $\Delta\beta$ terms) or $|\Omega_{d,m_0}/\delta_{d,m_0}|^{-5}$ for the eighth-order nonlinearities (i.e., the $\Delta\sigma$ terms).

\subsection{Large drive-transmon detuning: a semiclassical analysis}
\label{sec:semiclassical}
In this section, we consider that the drive is far detuned from any transmon transition frequency, i.e., $\delta_{dC} \gg \alpha$. In this regime, the driven transmon-like mode $C$ can be analyzed using a semiclassical approximation; see  Refs.~\cite{dykman1988a,dykman2012,zhang2019} which we follow here. 

For the purpose of semiclassical analysis, we introduce coordinate and momentum operators for mode $C$:
\begin{align}
    \Q = \sqrt{\frac{\lambda}{2}}(\C^\dagger +\C),\, \P = -i\sqrt{\frac{\lambda}{2}}(\C-\C^\dagger),
\end{align}
Operators $\Q,\P$ satisfy the commutation relation: $[\P,\Q]=-i\lambda$, where $\lambda$ can be thought of as an effective Planck constant and is to be specified. 

Substituting operators $\C,\C^\dagger$ with $\P,\Q$ in Eq.~(\ref{eq:H_RWA_eigenmodes_dispersive}), we obtain that 
\begin{align}
\label{eq:semiclassical_H}
    &H_C(\N_A,\N_B)/\hbar = \frac{\alpha}{2\lambda^2}\Big[- \frac{\lambda\hat \delta_{dC}}{\alpha}(\P^2+\Q^2) \nonumber\\
    &-\frac{1}{4}(\P^2+\Q^2)^2 +\frac{(2\lambda)^{3/2}\Omega_d}{\alpha}\Q\Big]+\frac{\hat\delta_{dC}}{2}+\frac{\alpha}{8}.
\end{align}

Now we define $\lambda$ to be 
\begin{align}
\label{eq:lambda}
\lambda = \frac{\alpha}{2|\delta_{dC}|}.
\end{align}
 It follows that Eq.~(\ref{eq:semiclassical_H}) becomes:
\begin{align}
\label{eq:H_C_semiclassical}
    H_C(\N_A,\N_B)/\hbar = \frac{2|\delta_{dC}|^2}{\alpha}\hat g + \frac{\hat\delta_{dC}}{2}+\frac{\alpha}{8},
\end{align}
where
\begin{align}
        \hat g \equiv & \hat g (\Q,\P,\N_A,\N_B) = -\frac{1}{2} \frac{\hat \delta_{dC}}{|\delta_{dC}|}(\P^2+\Q^2) \nonumber \\  &-\frac{1}{4}(\P^2+\Q^2)^2+\overline \Omega_d \Q, \, \overline \Omega_d = \frac{\sqrt{\alpha}\Omega_d}{|\delta_{dC}|^{3/2}}.  
\label{eq:Hamiltonian_g}
\end{align}
Without loss of generality, we assume $\Omega_d>0$.  

Hamiltonian $\hat g$ in Eq.~(\ref{eq:Hamiltonian_g}), a function of operators $\P,\Q,\N_A$ and $\N_B$, is a dimensionless Hamiltonian that controls the dynamics of the driven mode $C$. In the absence of the dispersive coupling to modes $A,B$, it is controlled by two parameters: the dimensionless drive amplitude $\overline \Omega_d$ and the scaled Planck constant $\lambda.$ 

In the regime of $\lambda\ll 1$, Hamiltonian $\hat g$ can be diagonalized perturbatively in the parameter $\lambda$. Note that it is already diagonalized in the Fock basis of modes $A,B$. We can  simplify the analysis by projecting onto any of their Fock states $|N_A,N_B\rangle$, or equivalently, replace operator $\N_A,\N_B$ with number $N_A,N_B$. The rest of the analysis follows that in Refs.~\cite{dykman2012,zhang2019}. First, we find out the extrema of function $g(Q,P,N_A,N_B)$ (where $Q,P$ are classical coordinate and momentum) with respect to $Q,P$ for fixed $N_A,N_B$. These extrema correspond to stable classical vibrational states of mode $C$ in the presence of a weak dissipation. Then we expand function $g$ about one of the extrema, and quantize the classical motion surrounding this point. Close to the extremum, this motion is just that of a harmonic oscillator, which we call ``auxiliary oscillator". The frequency of this oscillator is given by the curvature of function $g$ at the extremum. After these steps, we obtain, to order linear in $\lambda$, the following Hamiltonian:
\begin{align}
\label{eq:g_diag}
    &\hat g =  \hat g_0(\N_A,\N_B) - {\mathrm sgn}(\hat Q_0)\lambda \hat \nu_0 (\N_A,\N_B) (\C^\dagger_{\rm aux} \C_{\rm aux}+1/2), \\
    &\hat g_0(\N_A,\N_B) \equiv  \hat g[\hat Q_0,\hat P_0,\N_A,\N_B]. \nonumber
\end{align}
$\hat Q_0,\hat P_0$ is the location of a local extremum of the function $\hat g(Q,P,\N_A,\N_B)$ in the $Q-P$ plane if we think of $\N_A,\N_B$ as integers. $\hat Q_0,\hat P_0$ are generally functions of $\N_A,\N_B$ and can be found by solving the equations $\partial g (Q,P,\N_A,\N_B)/\partial P = \partial g (Q,P,\N_A,\N_B)/\partial Q = 0.$ Because $g$ is even in $P$, we always have $\hat P_0 = 0$ and $\hat Q_0$ satisfies the equation: $\hat Q_0 [\hat Q_0^2 + (\hat\delta_{dC}/|\delta_{dC}| )] = \overline \Omega_d.$  $\hat \nu_0$ is the frequency of the small oscillations about the extremum, and it is given by 

\begin{align}
\hat \nu_0(\N_A,\N_B) = &\sqrt{(\partial^2g/\partial Q^2)(\partial^2g/\partial P^2)|_{Q=\hat Q_0,P=P_0}} \nonumber\\
=&  \sqrt{((\hat \delta_{dC}/|\delta_{dC}|)+3\hat Q_0^2)((\hat \delta_{dC}/|\delta_{dC}|)+\hat Q_0^2)}.
\end{align}
$\C_{\rm aux}^\dagger, \C_{\rm aux}$ are the  creation and annihilation operators of the auxiliary mode. They are related to operators $\C^\dagger,\C$ via a squeezing and displacement transformation~\cite{dykman2012,zhang2019}. 

To find out the drive-induced change to the Kerr nonlinearity of modes $A,B$, we expand $g_0$ and $\nu_0$ in Eq.~(\ref{eq:g_diag}) to second order in $\N_A,\N_B$:
%
%
\begin{align}
\label{eq:g0_expansion}
\hat g_0(\N_A,\N_B) = \,& g_0 -\frac{Q_0^2}{2}\hat \eta +\frac{Q_0^2}{3Q_0^2+\sgn(\delta_{dC})}\frac{\hat \eta^2}{2},\\
\hat \nu_0(\N_A,\N_B) = \,& \nu_0 + \frac{\nu_0}{(\sgn(\delta_{dC})+3Q_0^2)^2}\hat \eta \nonumber \\ &+ \frac{3\nu_0 Q_0^2 (4\sgn(\delta_{dC})+3Q_0^2)}{(\sgn(\delta_{dC})+3Q_0^2)^4}\frac{\hat \eta^2}{2}, \\
\hat \eta = &\frac{\chi_{AC}\N_A+\chi_{BC}\N_B}{|\delta_{dC}|}.\nonumber
\end{align}
Here $g_0,\nu_0,Q_0$ without the hat are defined as the value of $\hat g_0,\hat \nu_0,\Q_0$ at $\N_A = \N_B = 0$, respectively. Substituting the expressions for $\hat g_0$ and $\hat \nu_0$ above into Eq.~(\ref{eq:g_diag}) and collecting terms quadratic in $\N_A$ or $\N_B$ or linear in $\N_A\N_B$, we obtain the Eq.~(\ref{eq:Delta_m_semiclassical}) in the weak coupling limit. A comparison between this semiclassical result with the full weak coupling calculation of cavity Kerr using Eq.~(\ref{eq:self_Kerr}) was shown in Appendix~\ref{app:semiclassical_compare_weak_coupling}. 

To third order in $\hat \eta$, we found that the correction to the right-hand side of Eq.~(\ref{eq:g0_expansion}) reads $-\hat \eta^3Q_0^2 (\sgn(\delta_{dC})+Q_0^2)/2(3Q_0^3+\sgn(\delta_{dC}))^3$. It is interesting to note that here while the coefficient of the quadratic in $\hat \eta$ term in Eq.~(\ref{eq:g0_expansion}) (which modifies cavity Kerr nonlinearity) saturates to a constant value at $Q_0\gg 1$, that of the cubic term decays as $1/Q_0^2$. This is in contrast to the regime of small drive-transmon detuning regime discussed in the previous section in which both the drive-induced cavity Kerr and sixth-order nonlinearity decays to zero at large drive.

\subsection{Comparison with experiment and numerical diagonalization}
\label{sec:experiment}
To confirm the analytic result, we perform numerical diagonalization of the full cavity-transmon Hamiltonian to find the strengths of the cavity nonlinearities. The theoretical results are further corroborated by experiments. 

For numerical diagonalization, we consider a model that consists of a single cavity mode-$a$ and a transmon described by Hamiltonian $H_{\rm RWA}$ in Eq.~(\ref{eq:H_RWA}) (with $g_b$ set to zero). We then diagonalize the Hamiltonian to find eigenstates $|\overline{\psi_m,N_a}\rangle$ and eigenenergies $\mathcal E_m(N_a)$; see Sec.~\ref{sec:formal_theory}. According to the analysis in Sec.~\ref{sec:E_m}, we parametrize the effective Hamiltonian $\mathcal E_m(\N_A)$ of dressed cavity mode-$A$ conditioned on the transmon in Floquet state $\psi_m$ in the following normal-order form: 
\begin{align}
\label{eq:single_mode_expansion}
 \mathcal E_m(\N_A) = & :\delta\omega_{A,m} \N_A + \frac{K_{A,m}}{2}\N_A^2 + \frac{\beta_{A,m}}{3!}\N_A^3 \nonumber \\ &+ \frac{\sigma_{A,m}}{4!}\N_A^4+...:.
\end{align}
Again, of primary interest to us is $m=0$ corresponding to the transmon in state $\psi_0$ that adiabatically connects to the transmon ground state as the drive is turned on/off.

The experiment was performed on a circuit QED setup that consists of a high-Q 3D microwave cavity coupled to a transmon ancilla; see Ref.~\cite{wang2020}. The experimental procedure to measure cavity nonlinearities is as follows. We first prepare a coherent state in the cavity mode and then turn on the pump on the transmon. We let the cavity coherent state evolve for some time and then measure the cavity Wigner function. We fit it to a Wigner function that is simulated using a Lindblad master equation with the Hamiltonian given in Eq.~(\ref{eq:single_mode_expansion}) and a single photon loss channel. This fitting allows us to extract the nonlinearity parameters in Eq.~(\ref{eq:single_mode_expansion}). Due to lack of sensitivity, we did not include the $\sigma_{A,0}$ term in the fit. More details on the experimental procedure can be found in Appendix~\ref{app:experiment}.

\begin{figure}[ht]
\includegraphics[width=9. cm]{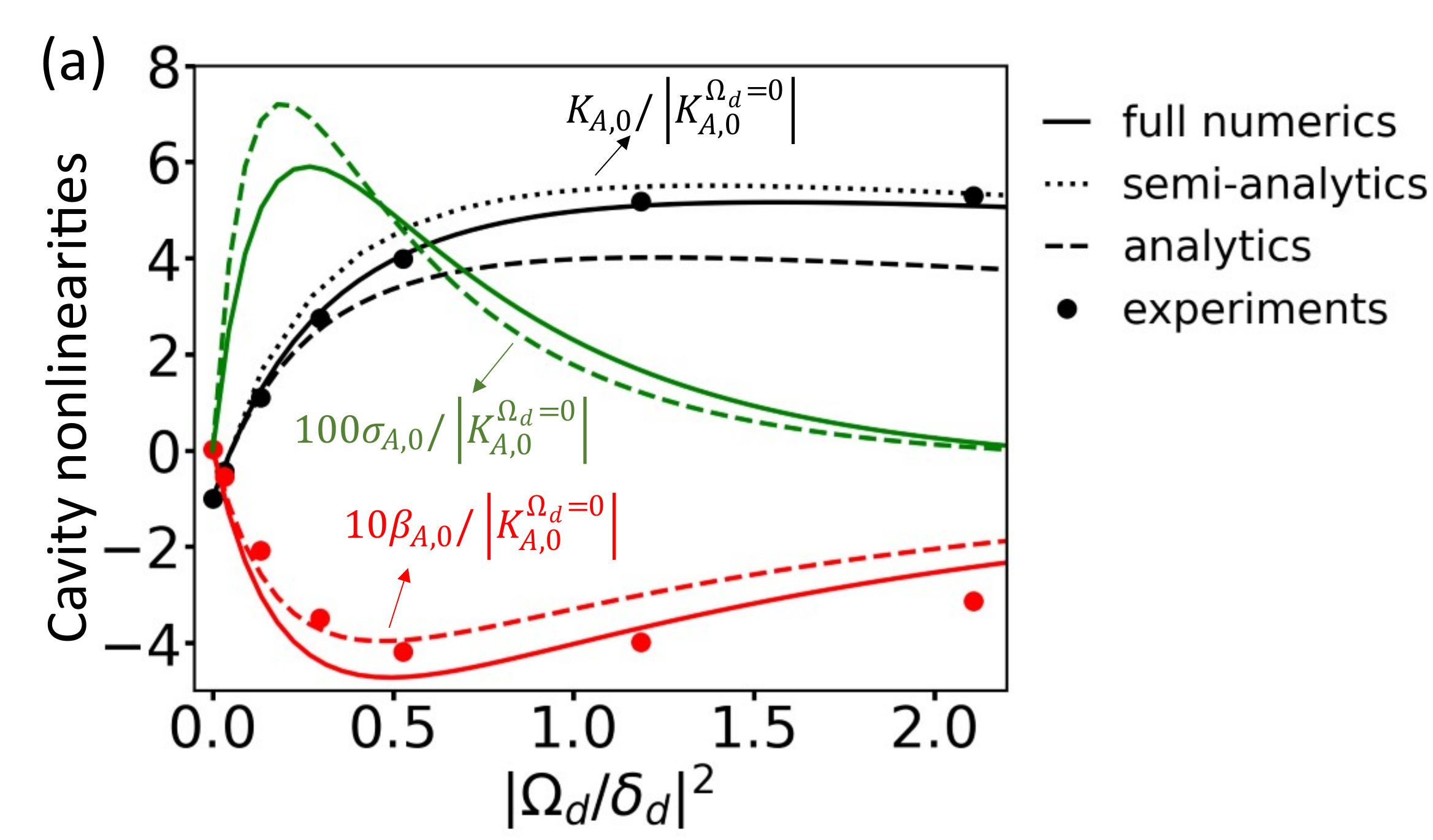} \\
\includegraphics[width=9. cm]{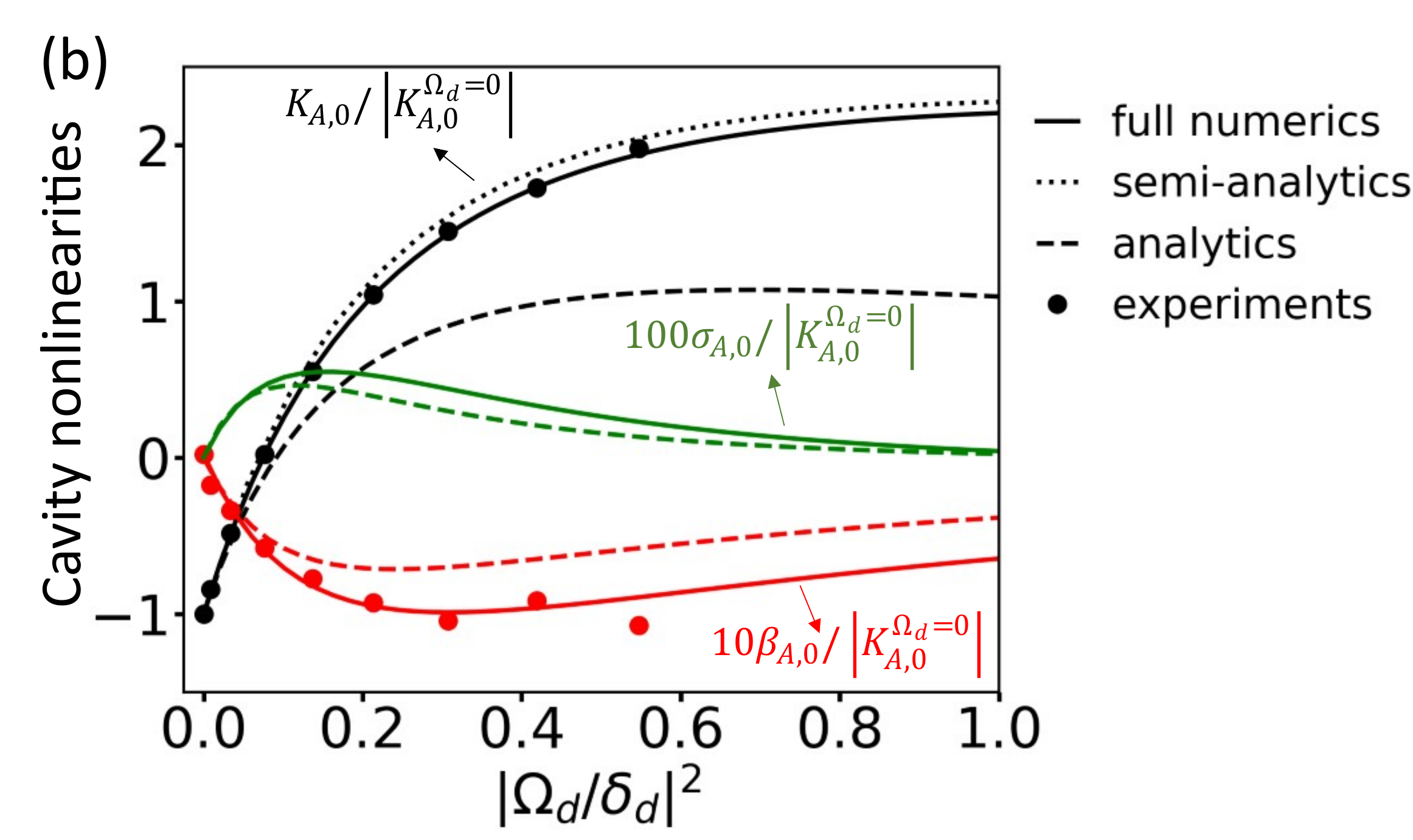} 
\caption{Cavity nonlinearities as a function of the drive power: Comparison between numerics (solid), semianalytics (dotted), analytics (dashed), and experiments (dots) for two different drive detunings, (a) $\delta_d/\alpha = 0.08$ and (b) $\delta_d/\alpha = 0.26$. The numerics is obtained through the diagonalization of the full cavity-transmon system. The semianalytics and analytics is obtained using Eq.~(\ref{eq:self_Kerr}) and Eq.~(\ref{eq:Delta_chi_TLS}), respectively. Lines or dots of different colors indicate different nonlinearity parameters as defined in Eq.~(\ref{eq:single_mode_expansion}). The cavity parameters are $\delta_a/\alpha = 9.64 ,g_a/\delta_a = 0.064$. The experimental values of bare transmon frequency $\omega_{10}/2\pi= 4.936$~GHz, anharmonicity $\alpha/2\pi = 0.168$~GHz. The theoretical curves are independent of the actual values of $\omega_{10}$ and $\alpha$. For these parameters, cavity self-Kerr in the absence of drive is $K_{A,0}^{\Omega_d = 0}/2\pi = -2.63$~kHz and cross-Kerr between cavity mode and transmon is $\chi_{AC}/2\pi = 1.25$~MHz.}
\label{fig:convergence}
\end{figure}

Figure~\ref{fig:convergence} shows the drive-power dependence of the nonlinearity parameters in Eq.~(\ref{eq:single_mode_expansion}) for $m=0$. While at zero drive the cavity nonlinearity is dominated by Kerr nonlinearity, there is a significant increase in higher-order cavity nonlinearities at finite drive amplitude. By comparing Fig.~\ref{fig:convergence}(a) with Fig.~\ref{fig:convergence}(b), we note that for the same amount of change in cavity Kerr, the change in higher-order nonlinearities is smaller for a larger drive-transmon detuning. In particular, at the drive power where $K_{A,0}$ crosses zero, the magnitude of $\beta_{A,0}$ approximately goes as $\delta_d^{-1}$ and $\sigma_{A,0}$ goes as $\delta_d^{-2}$, consistent with predictions of Eq.~(\ref{eq:Delta_chi_TLS}). This suggests that for the purpose of canceling cavity Kerr using an off-resonant transmon drive, it is preferable to use a larger drive-transmon detuning so the drive-induced higher-order cavity nonlinearities are suppressed while the Kerr is canceled; see Sec.~\ref{sec:Kerr_cancellation}.  
The results of the numerical diagonalization of the full cavity-transmon system match quite well with experimental results. The semianalytical results for the cavity Kerr nonlinearity based on Eq.~(\ref{eq:self_Kerr}) also match well with the full numerics and experiments. In obtaining the semianalytical results, we have used the dressed transmon frequency instead of the bare transmon frequency which produces a better agreement with the full numerics; see Appendix~\ref{app:beyond_weak_coupling}. Additionally, the analytical results shown in Fig.~\ref{fig:convergence} using Eq.~(\ref{eq:Delta_chi_TLS}) match well with both numerics and experiments at small $\delta_d/\alpha$, but deviate from them for larger $\delta_d/\alpha$.  This is because the two-level approximation used in obtaining Eq.~(\ref{eq:Delta_chi_TLS}) requires $|\delta_d|\ll \alpha$.

\begin{figure}[ht]
\includegraphics[width=8. cm]{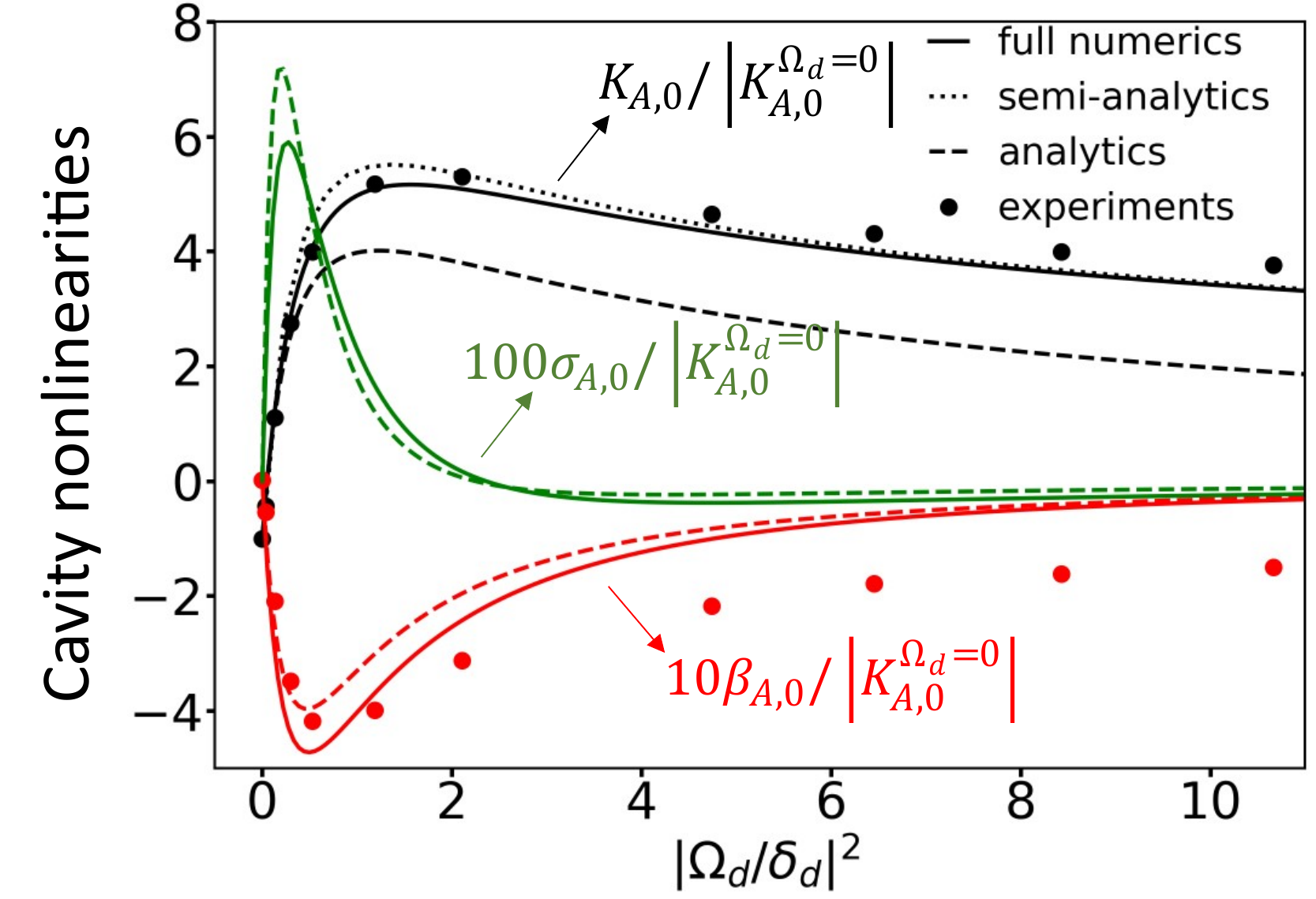}  
\caption{Cavity nonlinearities for a broader range of drive power. Same parameters as in Fig.~\ref{fig:convergence}(a).}
\label{fig:decay_nonlinearity}
\end{figure}

We show in Fig.~\ref{fig:decay_nonlinearity} the result with the same parameter as in Fig.~\ref{fig:convergence}(a) but for a broader range of drive powers. It shows that for large scaled drive powers, higher-order cavity nonlinearity decays faster than lower-order nonlinearity. Specifically, $K_{A,0}$ decays as $|\Omega_d/\delta_d|^{-1}$, $\beta_{A,0}$ decays as $|\Omega_d/\delta_d|^{-3}$ and $\sigma_{A,0}$ decays as $|\Omega_d/\delta_d|^{-5}$; see the text below Eq.~(\ref{eq:Delta_chi_TLS}). We believe the deviation of the experimental data from the theoretical result for $\beta_{A,0}$ at strong drives is partly due to the terms not included in the RWA Hamiltonian in Eq.~(\ref{eq:H_RWA}) such as the sixth-order terms from the cosine potential of the transmon.

\section{Cancellation of cavity Kerr nonlinearity}
\label{sec:Kerr_cancellation}
Finite nonlinearity results in non-equidistance of cavity energy levels. Classically, this leads to an energy-dependent cavity frequency. As a result, energy fluctuations of the cavity mode due to coupling to environment translates into frequency fluctuations or dephasing. Quantum mechanically, a somewhat similar situation occurs even without coupling to the environment. A cavity mode initially in a coherent state with mean photon number $\bar n$ will undergo deterministic phase ``scrambling" over the characteristic time scale $\tau_{\rm ph}\sim \pi/2K \sqrt{\bar n}$~\cite{harocheb}, where $K$ is the cavity self-Kerr. In contrast to the noise-induced pure dephasing, such phase scrambling is a unitary effect. A Schr\"odinger cat state displays similar behavior; see Fig.~\ref{fig:Wigner}. 

In the context of quantum error correction based on encoding information in the states of harmonic oscillators, logical states are typically designed to correct for photon loss. Because photon loss does not commute with unitary evolution under cavity nonlinearity, this leads to uncorrectable errors, which have been shown to be a leading factor limiting the performance of bosonic quantum error correction codes~\cite{ofek2016,hu2019,campagne-ibarcq2020}.


As analyzed in Secs.~\ref{sec:asymptotic_regime} and~\ref{sec:large_cavity_detuning}, in the regime of large cavity-transmon detuning, a single off-resonant drive can cancel cavity Kerr without inducing stronger cavity-transmon hybridization. In this section, we demonstrate numerically that such Kerr cancellation enables preserving the phase of a Schr\"odinger cat state stored in the cavity mode for a time that is much longer than the characteristic phase scrambling time $\tau_{\rm ph}$. The same method can be used to cancel the cross-Kerr between two cavity modes. 

\subsection{Numerical procedure}
\label{sec:numerical_procedure}
The numerical procedure to quantify the performance of the Kerr cancellation drive in preserving the Schr\"odinger cat state is as follows. We first construct an even Schr\"odinger cat state in the eigenbasis of the RWA Hamiltonian in Eq.~(\ref{eq:H_RWA}) (with $g_b = 0$): 
\begin{align}
\label{eq:cat_state}
    |\beta_+\rangle& = N^{-1/2} (|\beta\rangle + |-\beta\rangle), \\
    |\beta\rangle& = e^{-\frac{|\beta|^2}{2}}\sum_{N_a} \frac{\beta^{N_a}}{\sqrt{N_a!}} |\overline {\psi_0,N_a}\rangle, \nonumber 
\end{align}
where $N$ is a normalization factor equal to $2+2\exp(-2|\beta|^2)$. $\beta$ is the amplitude of the coherent state $|\beta\rangle$. Of interest to us is the regime where $|\beta|^2 \gtrsim 1$. Also we focus on the transmon being in state $\psi_0$ that adiabatically connects to the ground state as the drive is turned on or off. 

Then we let this state evolve under the full RWA Hamiltonian for some time $t$ and compute its overlap with an approximate state that evolves under a Kerr-free Hamiltonian with a simple linear frequency term:
\begin{align}
\label{eq:F(t)}
    F(t) &= |\langle \Psi_{\rm approx}(t)|\Psi(t)\rangle|^2 , \\
\Psi(t) & = e^{-iH_{\rm RWA}t}|\beta_+\rangle,\quad \Psi_{\rm approx}(t) = e^{-i \N_A \overline\omega t }|\beta_+\rangle. \nonumber
\end{align}
In practice, one can choose an $\overline \omega$ that maximizes the fidelity $F(t)$. Here we choose $\overline \omega$ to be the frequency at the mean photon number: $\overline \omega = (d\mathcal E_0 (N_A) /d N_A) |_{N_A = \langle \N_A\rangle}$, where $\langle \N_A \rangle \equiv \langle\beta_+| \N_A|\beta_+\rangle = |\beta|^2 \tanh (|\beta|^2)$. Since $N_A$ only takes discrete values, we further approximate the derivative as $\mathcal E_0 (\lceil \langle \N_A\rangle \rceil) - \mathcal E_0 (\lceil \langle \N_A\rangle \rceil-1)$ where $\lceil x \rceil$ is the ceiling function that maps $x$ to the least integer greater than or equal to $x$. 

In Sec.~\ref{sec:decay}, we will discuss the effects of transmon decoherence on the state fidelity. For here and in Secs.~\ref{sec:optimal_drive} and~\ref{sec:performance}, we focus on the coherent dynamics. 

\subsection{Optimal drive parameters and scaling of infidelity}
\label{sec:optimal_drive}
Before we show the performance of the Kerr cancellation drive, we discuss the optimal drive condition to maximize the fidelity $F(t)$. The aforementioned nonlinearity-induced phase scrambling of cavity coherent state or Schr\"odinger cat state results from finite variance of the cavity photon number distribution. To quantify this effect, we expand the eigenenergy $\mathcal E_0(N_A)$ with respect to $N_A$ about the mean photon number $N_A = \langle \N_A\rangle$:
\begin{align}
\label{eq:expansion}
\mathcal E_0(N_A) & =\sum_{n=0}^\infty \frac{1}{n!}\frac{d \overline  {\mathcal E}_0(N_A)^n}{d N_A^n}\Big |_{N_A = \langle \N_A\rangle} \delta N_A^n , \\ \quad \delta N_A & \equiv N_A - \langle \N_A\rangle. \nonumber 
\end{align}

In the absence of drive, the dominant cavity nonlinearity is the Kerr nonlinearity. Therefore, the second derivative of $\mathcal E_0(N_A)$ in Eq.~(\ref{eq:expansion}) is much larger than higher derivatives. A reasonable choice of drive parameters to minimize nonlinearity-induced phase scrambling is such that 
%
\begin{align}
\label{eq:condition}
\frac{d^2 \mathcal E_0(N_A)}{d N_A^2} \Big |_{N_A = \langle \N_A\rangle} = 0.
\end{align}
 Using the parametrization of  $\mathcal E_0(N_A)$ in Eq,~(\ref{eq:single_mode_expansion}) and keeping to $\beta_{A,0}$ term, we have that
\begin{align}
\label{eq:second_derivative}
 \frac{d^2 \mathcal E_0(N_A)}{d N_A^2}\Big |_{N_A = \langle \N_A\rangle} \approx  K_{A,0} + \beta_{A,0} (\langle \N_A\rangle - 1).        
\end{align}
We note that there is a contribution from higher-order nonlinearity $\beta_{A,0}$ to $(d^2 \mathcal E_0(N_A)/d N_A^2) |_{N_A = \langle \N_A\rangle}$, which can become significant for large $\langle \N_A\rangle$.

Neglecting higher-order cavity nonlinearities, the condition in Eq.~(\ref{eq:condition}) approximately becomes $K_{A,0} = 0$. To the lowest order in the drive amplitude, the condition to cancel cavity self-Kerr $K_{A,0}$ follows from Eq.~(\ref{eq:H_RWA_TLS_expanded}) to be: 
\begin{align}
\label{eq:cancellation_condition}
    8\alpha|\Omega_d|^2/\delta_{{d,0}}^3 = 1 
\end{align}
We remind the readers that here $\delta_{d,0}$ is the drive detuning from the transition frequency between the first two states of the transmon-like eigenmode $C$, which differs from transition frequency $\omega_{10}$ of the bare transmon by approximately $|g_a|^2/\delta_a$; i.e., $\delta_d \approx \delta_{d,0} - |g_a|^2/\delta_a. $ In the case $\delta_d$ is comparable to $|g_a|^2/\delta_a$, it is important to take into account this frequency shift. The condition in Eq.~(\ref{eq:cancellation_condition}) can be rewritten as $2\alpha|\Omega_d|^2/\delta_{{d,0}}^2 = \delta_{d,0}/4$, where the left-hand side is approximately the drive-induced ac Stark shift of the transition frequency of transmon mode $C$. The precise drive power required to cancel the cavity Kerr is higher than that set by this condition because for $\delta_d>0$, the cavity Kerr becomes sublinear in the drive power as the drive power increases as can be seen from Fig.~\ref{fig:convergence}.

Upon the cancellation of the $\delta N_A^2$ term in Eq.~(\ref{eq:expansion}), the major contribution to the remaining infidelity $1-F$ comes from the $\delta N_A^3$ term  whose coefficient is proportional to $\beta_{A,0}$. Since the infidelity should be independent of the sign of $\beta_{A,0}$, we have to leading order in $\beta_{A,0}$: $1-F\propto \beta_{A,0}^2$. As discussed in Sec.~\ref{sec:large_cavity_detuning} [cf. Eq.~(\ref{eq:H_RWA_TLS_expanded})], the magnitude of $\beta_{A,0}$ scales as $\chi_{AC}/\delta_{d,0}$ at the Kerr cancellation point. It follows that the infidelity should scale as: 
\begin{align}
\label{eq:scaling}
1-F\propto (\chi_{AC}/\delta_{d,0})^2.    
\end{align}


\subsection{Numerical results}
\label{sec:performance}
\begin{figure}[t]
\includegraphics[width=9.0 cm]{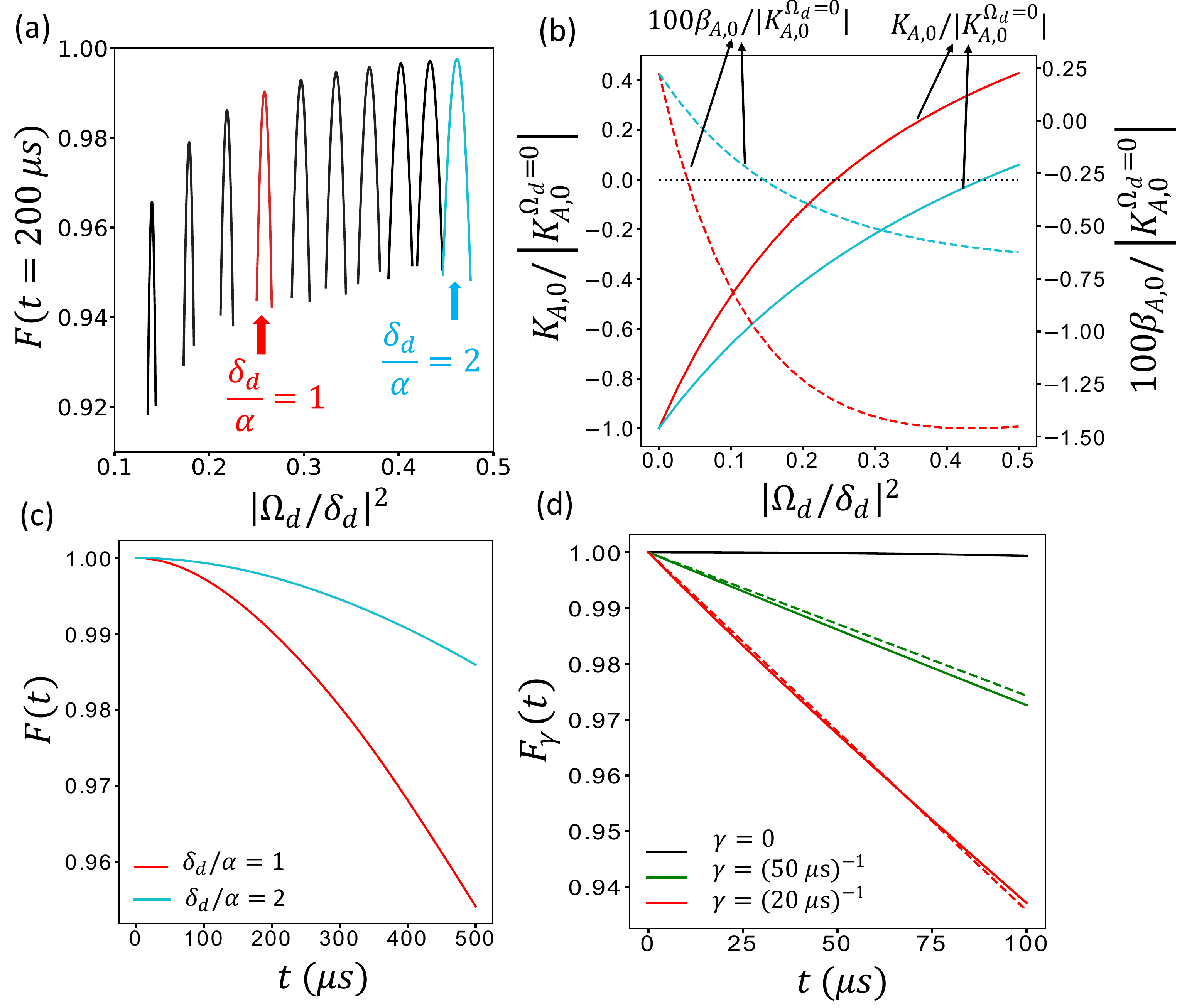}
\caption{Performance of the Kerr cancellation drive. Hamiltonian parameters of the static cavity-transmon system are the same as in Fig.~\ref{fig:convergence}. (a)~Fidelity F(t) at $t= 200\,\mu s$ as a function of scaled drive power for a set of equally-spaced drive detunings. Each ``peak" corresponds to a fixed drive detuning. From left to right, the scaled drive detuning $\delta_d/\alpha$ increases from 0.5 to 2. (b) The cavity Kerr nonlinearity $K_{A,0}$ (solid lines) and the leading higher-order nonlinearity $\beta_{A,0}$ (dashed lines) as a function of the scaled drive power for drive detuning $\delta_d/\alpha = $ 1 (red) and 2 (cyan). The same color encoding is used in panel (a). The horizontal black dashed line indicates where zero is for the cavity Kerr nonlinearity. (c)~The fidelity $F(t)$ as a function time for $\delta_d/\alpha = 2 $(cyan) and 1 (red) at the optimal drive amplitude $\Omega_d = \Omega_d^{\rm opt}$. The fidelity decreases quadratically in time $t$ at short time. (d) The fidelity $F_\gamma(t)$ for various transmon decay rates at drive detuning $\delta_d/\alpha=2.$ The solid lines show the results of the master equation in Eq.~(\ref{eq:master_equation}). The dashed lines show the results of Eq.~(\ref{eq:scaling_3}) where $\kappa_\gamma$ and $W_{\psi_0\rightarrow \psi_m}$ are given by Eqs.~(\ref{eq:kappa_gamma}) and~(\ref{eq:escape_rate}), respectively; the cavity inverse Purcell decay (the term $\propto \kappa_r$ in Eq.~(\ref{eq:scaling_3})) accounts for the majority ($\approx 83\%$) of the infidleity $1-F_\gamma(t)$, while the incoherent excitation from transmon state $\psi_0$ to $\psi_1$ (the term $\propto W_{\psi_0\rightarrow \psi_1}$ in Eq.~(\ref{eq:scaling_3})) accounts for the rest. }
\label{fig:Kerr_cancellation}
\end{figure}

Parameters of the cavity-transmon system are chosen to be the same as in Fig.~\ref{fig:convergence}. For these parameters, the cavity Kerr in the absence of a transmon drive $K_{A,0}^{\Omega_d=0}/2\pi = -2.63$~kHz. We choose the size of the Schr\"odinger cat state to be $\beta = \sqrt{3}$. This means that after $\tau_{\rm ph} = \pi/2\sqrt{\langle \N_A \rangle} |K_{A,0}^{\Omega_d=0}| \approx 55~\mu s$, the state dephases and the fidelity $F$ drops to close to zero in the absence of a transmon drive. Applying an off-resonant transmon drive significantly increases the fidelity $F(t)$. For a drive detuning $\delta_d/\alpha =2$, the fidelity $F(t)$ remains above 98.5\% for as long as 500 $\mu s$; see Figure~\ref{fig:Kerr_cancellation}(c).

Figure~\ref{fig:Kerr_cancellation}(a) shows that given a drive detuning, there exists an optimal drive amplitude $\Omega_d^{\rm opt}$ that maximizes the fidelity. 
The scaled optimal drive power $|\Omega_d^{\rm opt}/\delta_d|^2$ approximately increases linearly with the drive detuning, as predicted by Eq.~(\ref{eq:cancellation_condition}). We have verified that the value of the optimal drive power matches that given by Eqs.~(\ref{eq:condition},\ref{eq:second_derivative}). 
$\Omega_d^{\rm opt}$ is slightly larger than the drive amplitude at the Kerr cancellation point due to due to finite $\beta_{A,0}$; see Fig~\ref{fig:Kerr_cancellation}(b).

Figure~\ref{fig:Kerr_cancellation}(a,b) demonstrates two advantages of using a large drive detuning. First, the infidelity $1-F$ at the optimal drive amplitude $\Omega_d^{\rm opt}$ decreases quadratically with the increase of the drive detuning, consistent with our analysis around Eq.~(\ref{eq:scaling}). Second, away from the optimal drive power, the fidelity drops slower at larger drive detuning, i.e. it is less sensitive to deviation from the optimal drive power. This is related to the fact that the slope of cavity Kerr at the zero crossing point decreases with the increase of the scaled drive detuning $\delta_d/\alpha$; see Fig.~\ref{fig:Kerr_cancellation}(b) and Eq.~(\ref{eq:cancellation_condition}). 
Approximating the optimal drive amplitude as that given by the condition in Eq.~(\ref{eq:cancellation_condition}), one can show that the deviation from the maximal fidelity scales with respect to the distance to the optimal drive power as follows:
\begin{align}
\label{eq:scaling_2}
    F^{\rm opt}(t) - F(t) \propto  \left(\frac{\alpha}{\delta_{d,0}}K_{A,0}^{\Omega_d = 0}t\right)^2 \left[\left(\frac{\Omega_d^{\rm opt}}{\delta_{d,0}}\right)^2-\left(\frac{\Omega_d}{\delta_{d,0}}\right)^2\right]^2.
\end{align}
The ``width" of $F$ as a function of drive power curve in Fig.~\ref{fig:Kerr_cancellation}(a) decreases as $1/t$. 

\begin{figure}[t]
\includegraphics[width=4.1 cm]{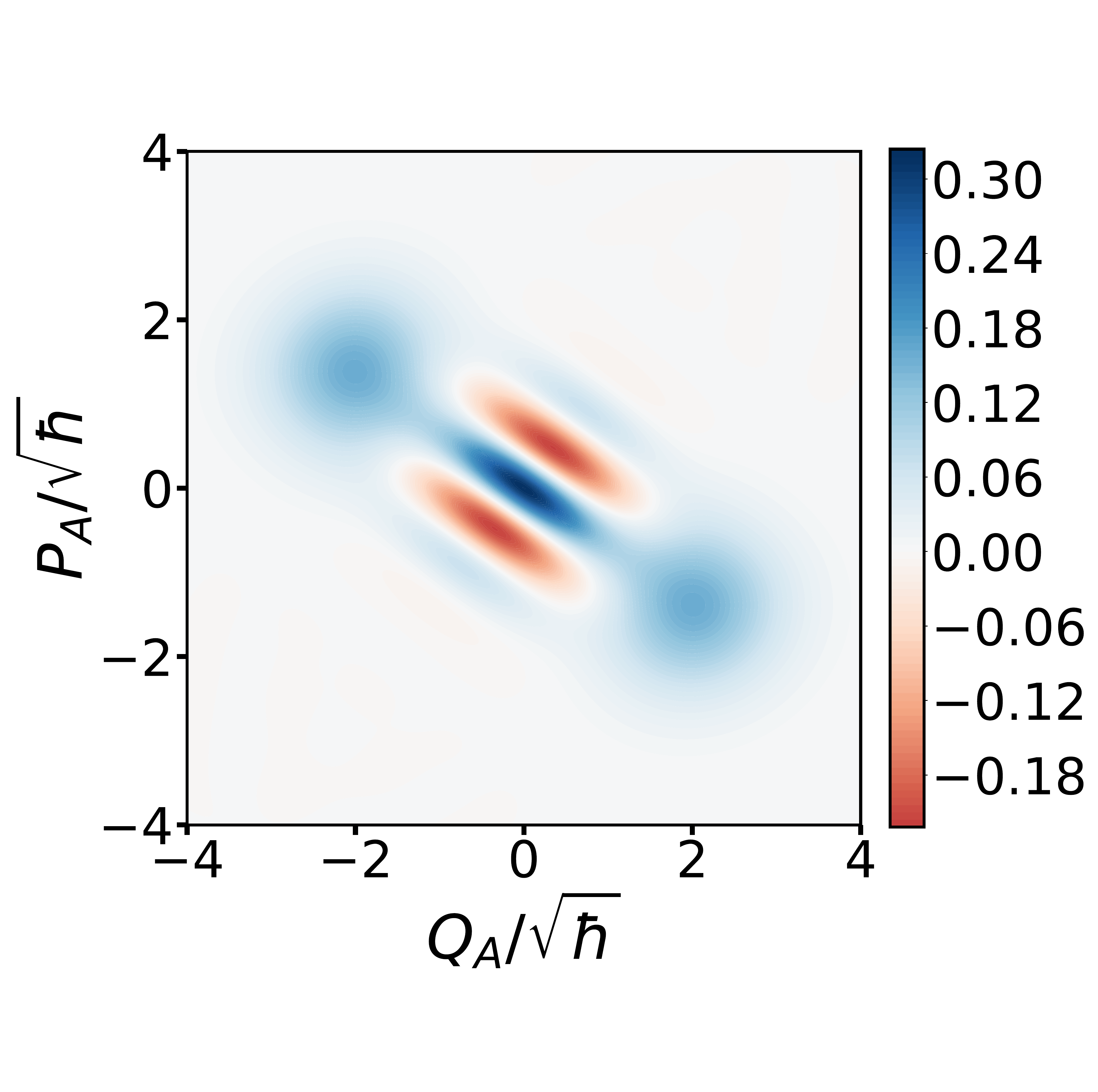}\quad
\includegraphics[width = 4.1 cm]{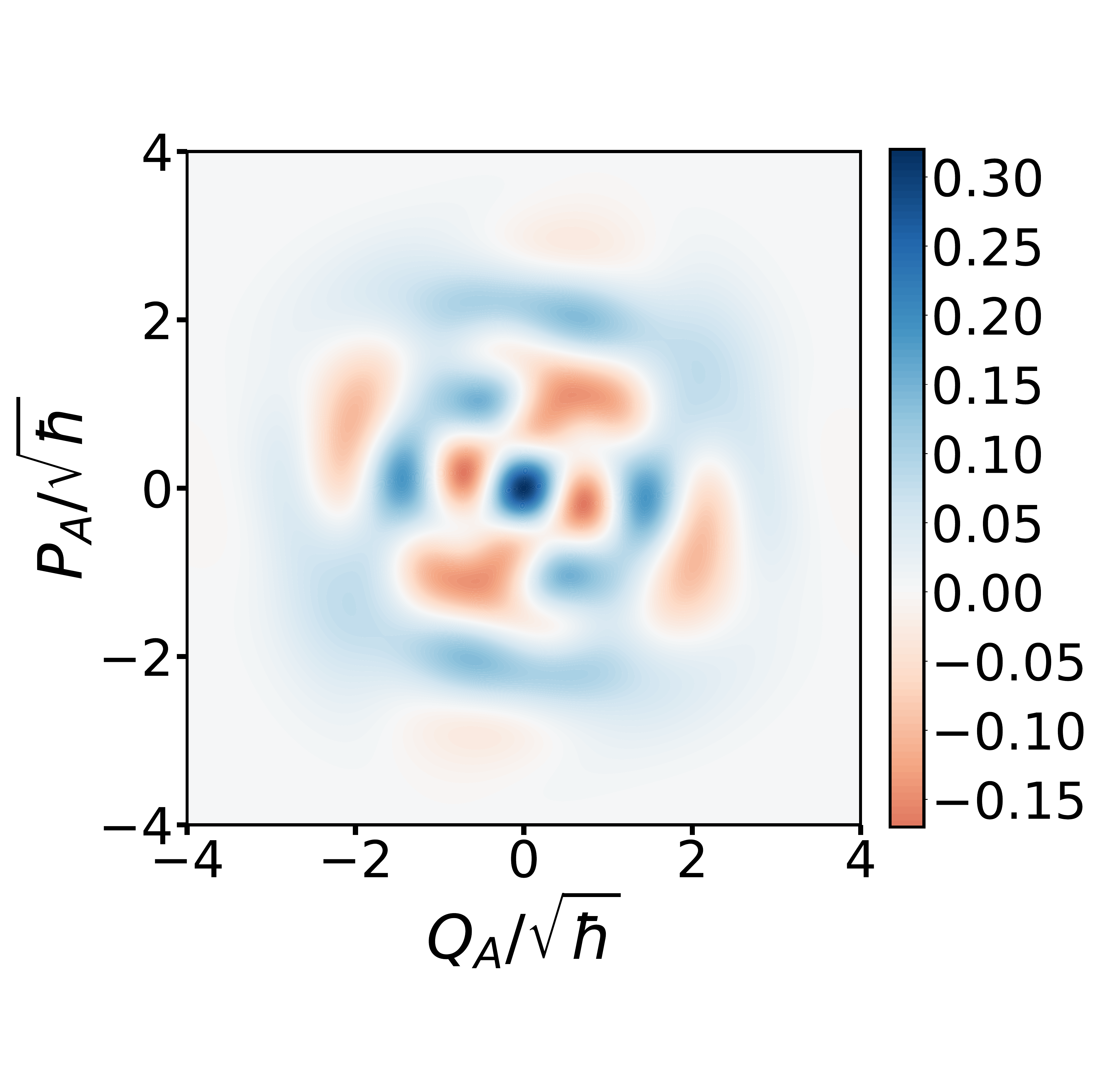}
\caption{Wigner functions of state $\Psi(t)$ defined in Eq.~(\ref{eq:Wigner_def}) with (left) and without (right) the Kerr cancellation drive at the characteristic phase scrambling time $t = 55~\mathrm{\mu s} \approx \tau_{\rm ph}$. The drive amplitude is chosen to be at $\Omega_d = \Omega_d^{\rm opt}$ that maximizes the fidelity $F(t)$ in Fig.~\ref{fig:Kerr_cancellation}(a) and the drive detuning is $\delta_d/\alpha = 2$.  As shown in Fig.~\ref{fig:Kerr_cancellation}(c), with the 
Kerr cancellation drive, overlap of state $\Psi(t)$ with Kerr-free evolution remains above 98.5\% up to 500~$\mathrm{\mu s}$; without Kerr cancellation drive, this overlap drops to zero near $t = \tau_{\rm ph}$. } 
\label{fig:Wigner}
\end{figure}

Figure~\ref{fig:Wigner} shows examples of the Wigner functions of state $\Psi(t)$. At $t \approx \tau_{\rm ph}$, the cat state maintains its phase coherence in the presence of the Kerr cancellation drive, but completely dephases without the drive. The Wigner function is defined in a standard way for the dressed cavity mode as follows: 

\begin{align}
\label{eq:Wigner_def}
W(Q_A,P_A) = &\frac{1}{\pi\hbar} \int_{-\infty}^\infty \Psi^*(Q_A+Q_A')\Psi(Q_A-Q_A')\nonumber \\&\times e^{2iP_AQ_A'/\hbar}dQ_A',
\end{align}
where $\Psi(Q_A) \equiv \langle Q_A|\Psi(t)\rangle$. $|Q_A\rangle$ is eigenstate of coordinate operator of the dressed cavity mode $A$: $\hat Q_A = \sqrt{\hbar/2}(\A^\dagger +\A).$  



\subsection{Effects of transmon decay}
\label{sec:decay}
We have discussed how the Schr\"odinger cat state $|\beta_+\rangle$ undergoes phase scrambling under unitary evolution and how such a process can be suppressed by canceling cavity Kerr nonlinearity. In a circuit QED setup that consists of high-Q 3D microwave cavities coupled to a transmon ancilla, the transmon is typically the most lossy element. In this section, we consider the effect of transmon decay on the state fidelity. 

To model transmon decay, we consider a model where the transmon dynamical operator $\hat c^\dagger + \hat c$ is linearly coupled to some bath variable whose spectral density is assumed to be smooth over the characteristic frequency scale of $H_{\rm RWA}$ in Eq.~(\ref{eq:H_RWA}). The bath is further assumed to be in thermal equilibrium at zero temperature. Upon a Markov approximation and in the rotating frame of the drive, we obtain the following Lindblad master equation: 
\begin{align}
\label{eq:master_equation}
    \dot \rho = -\frac{i}{\hbar}[H_{\rm RWA},\rho] - D[\sqrt{\gamma}\hat c]\rho, \, D[\hat c] \equiv \{\rho,\c^\dagger \c\}/2  - \c \rho\c^\dagger. 
\end{align}
We then compute the overlap of the density matrix with the approximate Kerr-free state: $F_\gamma(t) = \langle \Psi_{\rm approx}(t)|\rho(t)|\Psi_{\rm approx}(t)\rangle$, with $\rho(0) = |\beta_+\rangle\langle \beta_+|$. To differentiate this fidelity from the fidelity without transmon decay, we use a subscript $\gamma$.
The result for $F_\gamma(t)$ is shown in Fig.~\ref{fig:Kerr_cancellation}(d). For the chosen parameters, the infidelity $1-F_\gamma$ is dominated by transmon decay, and the reduction in $F_{\gamma}$ can be approximated as purely coming from incoherent processes, i.e. $F_{\gamma}\approx \langle\Psi(t)|\rho(t)|\Psi(t) \rangle.$

Transmon decay leads to a linear in time reduction of the fidelity $F_\gamma(t)$ at a short time scale (short compared to the characteristic decoherence time of the dressed cavity mode and transmon). Such reduction in fidelity has two origins. First, the dressed cavity mode inherits finite decay from the lossy transmon, an effect sometimes referred to as ``inverse Purcell decay"~\cite{reagor2016}. To leading order in the cavity-transmon coupling strength $g_a$, the rate of this inherited decay is linear in the cavity photon number, i.e. $|\overline {\psi_0,N_A}\rangle$ decays into $|\overline {\psi_0,N_A-1}\rangle$ with a rate given by $N_A\kappa_\gamma$ where $\kappa_\gamma \propto \gamma$. To leading order in $\kappa_{\gamma} t$, the decrease in the fidelity $F_\gamma(t)$ due to the cavity inverse Purcell decay comes from the no jump evolution that maps $\rho(0)$ to $\exp(-\kappa_\gamma \N_A t/2)\rho(0) \exp(-\kappa_\gamma \N_A t/2)$. It follows that at short time $\kappa_\gamma t\ll1$, we have that  $(1-F_\gamma(t))_{\kappa_\gamma}=\kappa_\gamma t \langle \N_A \rangle$. 

Second, the transmon undergoes transitions from state $\psi_0$ to some other state $\psi_{m\neq 0}$, which occurs even when the bath is at zero temperature. These transitions result in the dressed cavity mode seeing a different effective Hamiltonian $ \mathcal E_{m\neq 0}(\N_A)$ which has a different effective cavity frequency $\delta\omega_{A,m}$; see Eq.~(\ref{eq:single_mode_expansion}). One can think of that once the transmon escapes state $\psi_0$, the Schr\"odinger cat state will rotate in the phase space with a speed that is different from $\bar \omega$ by the amount of $\delta\omega_{A,m\neq0}-\delta\omega_{A,0}$ that is approximately equal to $-m\chi_{AC}$. After the transition, the state overlap with $\Psi_{\rm approx}(t)$ will therefore rapidly oscillate at a rate set by $\chi_{AC}$. Since the time of escape is random, these oscillations will be averaged with  respect to the escape time from time $t= 0$ to the observation time $t$. For $\gamma t\ll 1$, the resulting infidelity $1-F_\gamma(t)$ due to escaping from state $\psi_0$ can be expressed as follows: 
\begin{align*}
&    (1-F_{\gamma}(t))_{\rm esc} = \sum_m W_{\psi_0\rightarrow \psi_m} \int_0^t dt_{\rm esc} \nonumber \\  &\times (1-|\langle \beta_+ |e^{-i(\delta\omega_{A,m} - \delta\omega_{A,0})\N_A (t-t_{\rm esc})}| \beta_+\rangle |^2). 
\end{align*}
where $W_{\psi_0\rightarrow \psi_m}\propto \gamma$ is the rate of transition from state $\psi_0$ to $\psi_m$ and $\delta\omega_{A,m}-\delta\omega_{A,0}\approx -m \chi_{AC}$. On a time scale that is much larger than $1/\chi_{AC}$, the oscillations of the state overlap in the integrand will be averaged to a constant value between 0 and 1 that depends on the size $\beta$ of the Schr\"odinger cat state $|\beta_+\rangle$.  

Combining the infidelity due to the cavity inverse Purcell decay and transmon escaping state $\psi_0$, we obtain the overall infidelity $1-F_\gamma(t)$ to be (on the time scale shorter than the decoherence time of the dressed cavity and transmon but longer than $1/\chi_{AC}$):
\begin{align}
\label{eq:scaling_3}
    1- F_\gamma (t) = (\kappa_\gamma \langle \N_A\rangle +   C \sum_m W_{\psi_0\rightarrow \psi_m} ) t + \mathcal O(t^2), 
\end{align}
where $\langle \N_A\rangle= |\beta|^2 \tanh(|\beta|^2)$, and coefficient $C = 1-[J_0(2|\beta|^2)+J_0(2i|\beta|^2)]/2\cosh^2(|\beta|^2)\in [0,1]$. For $|\beta|^2=3$ shown in Fig.~\ref{fig:Kerr_cancellation}, we have $C\approx 0.67$ and $\langle \N_A\rangle\approx 3$. Using the Fermi's golden rule expressions for $\kappa_\gamma$ and $W_{\psi_0\rightarrow \psi_m}$ below, we verify that Eq.~(\ref{eq:scaling_3}) matches well with the master equation simulation using Eq.~(\ref{eq:master_equation}); see the dashed lines in Fig.~\ref{fig:Kerr_cancellation}(d).


\begin{figure}[t]
\includegraphics[width=4. cm]{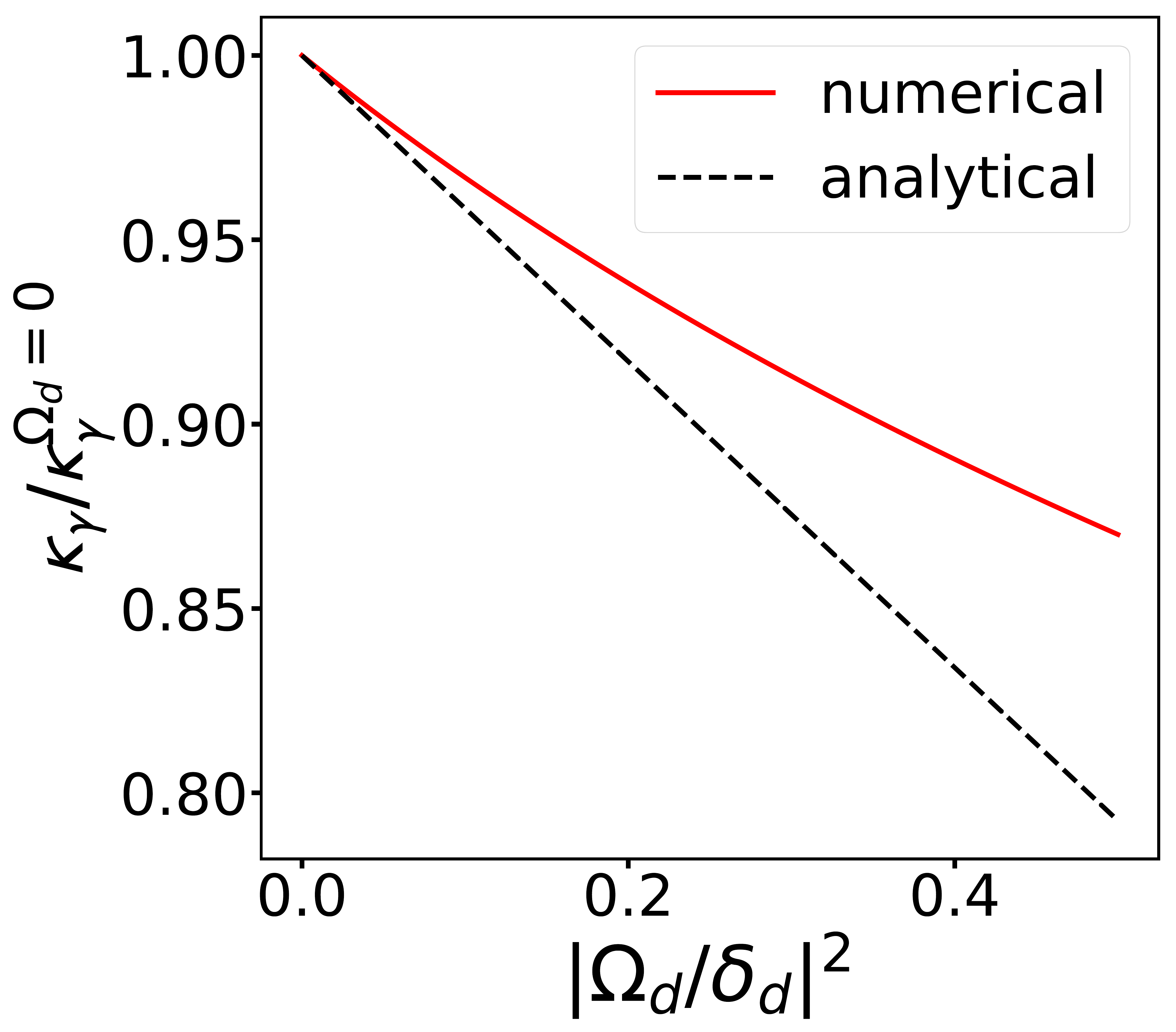} \quad
\includegraphics[width = 4. cm]{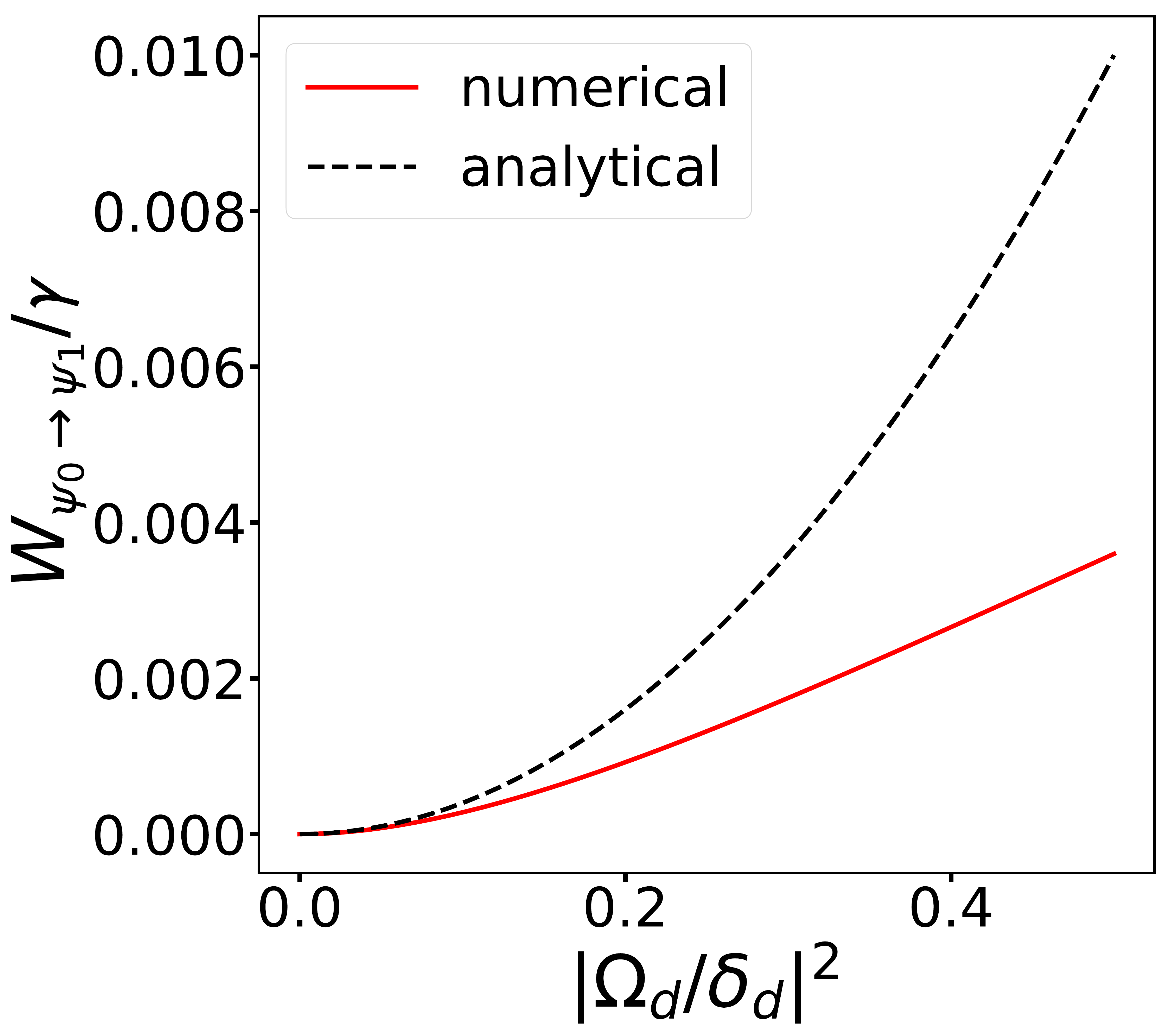}
\caption{Left: Inverse Purcell decay rate of the dressed cavity mode. Right: Incoherent transition rate from transmon Floquet state $\psi_0$ to $\psi_1$. Hamiltonian parameters of the static cavity-transmon system are the same as in Fig.~\ref{fig:convergence}. The scaled drive detuning is $\delta_d/\alpha = 2.$ For this drive detuning, the drive amplitude at Kerr cancellation is $|\Omega_d/\delta_d|^2\approx 0.45$; see Fig.~\ref{fig:Kerr_cancellation}(b). The numerical results are obtained using Fermi' s golden rule formula in Eqs.~(\ref{eq:kappa_gamma},\ref{eq:escape_rate}). The analytical results correspond to the expressions in Eqs.~(\ref{eq:kappa_gamma_2},\ref{eq:escape_rate_2}) which are derived using the Fermi's golden rule formula but are perturbative in the drive power. The deviation of the analytical from the numerical results at large drive strengths is partly due to the fact that the drive pushes the transmon frequency $\omega_{10}$ away from the drive frequency through ac Stark shift which makes the effective drive detuning larger. For the chosen drive parameters, transition rate $W_{\psi_0\rightarrow \psi_{m>1}}$ from transmon Floquet state $\psi_0$ to $\psi_{m> 1}$ is negligible compared to $W_{\psi_0\rightarrow \psi_1}.$} 
\label{fig:decoherence_rate}
\end{figure}

The inverse Purcell decay rate $\kappa_\gamma$ in Eq.~(\ref{eq:scaling_3}) can be calculated using Fermi's golden rule: 
\begin{align}
\label{eq:kappa_gamma}
\kappa_\gamma = |\langle \overline {\psi_0,0}| \hat c|\overline {\psi_0,1}\rangle|^2 \gamma. 
\end{align}
To leading order in the drive amplitude and the transmon-cavity coupling $g_a$, we found it to be: 
\begin{align}
\label{eq:kappa_gamma_2}
   & \kappa_\gamma  \approx  \left|\frac{g_a}{\delta_a}\right|^2 \left(1-\frac{4\alpha|\Omega_d|^2}{\delta_d^3}\frac{\delta_d}{\delta_a}\right) \gamma.
\end{align}
We also only kept leading order term in $\delta_d/\delta_a$ and $\alpha/\delta_a$ which are much smaller than one in the considered large cavity-transmon detuning regime. Part of the drive-induced inverse Purcell decay rate (the term proportional to $|\Omega_d|^2$) simply comes from the drive-induced ac Stark shift pushing the transmon frequency $\omega_{10}$ away or closer to the cavity frequency depending on the sign of $\delta_a$. Notably, at the Kerr cancellation point [see Eq.~(\ref{eq:cancellation_condition})], this term is suppressed by the small ratio of $\delta_d/\delta_a$ compared to the static inverse Purcell decay. In general, the drive-induced inverse Purcell decay can surpass the static value in particular when the cavity is in the vicinity of the drive-induced cavity-transmon reasonance that we discussed in Sec.~\ref{sec:near_resonance}. Detailed discussions of this scenario can be found in Ref.~\cite{zhang2019}.

The Fermi's golden rule expression for $W_{\psi_0\rightarrow \psi_m}$ in Eq.~(\ref{eq:scaling_3}) is given by 
\begin{align}
\label{eq:escape_rate}
    W_{\psi_0\rightarrow \psi_m} = |\langle \psi_m|\hat c|\psi_0\rangle|^2 \gamma.
\end{align}
To leading order in the drive amplitude, the transition rate from $\psi_0$ to $\psi_1$ is much larger than transition rates to other states and was found to be~\cite{zhang2019}: 
\begin{align}
\label{eq:escape_rate_2}
       & W_{\psi_0\rightarrow \psi_1}  \approx  \left| \frac{\alpha \Omega_d^2}{\delta_d^2(2\delta_d+\alpha)}\right|^2\gamma. 
\end{align}
A comparison between the perturbative analytical results in Eqs.~(\ref{eq:kappa_gamma_2},\ref{eq:escape_rate_2}) and numerics shows good agreement; see Fig.~\ref{fig:decoherence_rate}.


On a time scale that is much larger than transmon decoherence time ($\gamma t\gg1$), the transmon has undergone many incoherent transitions among its Floquet states untill observation time $t$. Because of the transmon-state-dependent cavity frequency, these random transitions dephase the dressed cavity. Taking into account transitions between states $\psi_0$ and $\psi_1$, this pure dephasing rate scales as $\kappa_{\rm ph}\sim W_{\psi_0\rightarrow \psi_1}(\chi_{AC}/\gamma)^2$. The results shown in Fig.~\ref{fig:Kerr_cancellation}(d) refers to the intermediate regime $\gamma t \lesssim 1$ and the infidelity $1-F_\gamma$ scales approximately linearly in $\gamma$.

In summary, in the presence of the Kerr cancellation drive, the fidelity of the cavity Schr\"odinger cat state is no longer limited by the coherent phase scrambling, but rather limited by the decoherence processes which include both transmon decoherence and the intrinsic decoherence of the cavity modes. Using realistic experimental parameters (see the caption of Fig.~\ref{fig:convergence}), we found that the cavity state infidelity due to transmon dissipation is about $3\%$ at $t = 100~\mu s$ for a cat state of size $\beta = \sqrt{3}$ and transmon decay rate $\gamma = (50~\mu s)^{-1}$ [see Fig.~\ref{fig:Kerr_cancellation}(d)]. Through semi-analytic analysis, we found that the inverse Purcell effect accounts for the majority ($\approx 83\%$) of this infidelity; the rest ($\approx 17\%$) comes from the incoherent excitation of the transmon from state $\psi_0$ to $\psi_1$, which occurs even at zero temperature due to the finite drive. For the parameters we used, the inverse Purcell effect limits the cavity lifetime to about 5 ms, which is comparable to the intrinsic lifetime of the state-of-art 3D microwave cavities~\cite{chakram2020}. 

\section{Conclusions}
We have studied the nonlinearities of cavity modes inherited from an off-resonantly driven superconducting transmon. These nonlinearities can be tuned in situ by the drive. In different regimes, the form of this tunability is qualitatively different.

First, for a small-to-moderate cavity-transmon detuning that is comparable to drive-transmon detuning and/or transmon anharmonicity, the drive can induce multi-photon resonances among the cavity and transmon excitations.
In the vicinity of these resonances, cavity nonlinearity parameters experience sharp changes as a function of the drive parameters. Second, for large cavity-transmon detuning where the cavity is far away from these resonances, off-resonant cavity-transmon interaction leads to a cavity-photon-number-dependent dispersive shift in transmon transition frequencies. This results in drive-induced ac Stark shifts of the transmon levels also depending on the cavity photon number which translates into an effective cavity nonlinearity. Depending on the interrelation between the drive-transmon detuning and transmon anharmonicity, this ac Stark shift shows qualitatively different behavior ranging from strongly quantum to semiclassical, which in turn, leads to different features in the cavity nonlinearities. 

For large cavity-transmon detuning, cavity nonlinearity induced by a single drive blue-detuned from the transmon can be used to cancel the cavity Kerr nonlinearity that is the dominant nonlinearity without the drive. In the case of multiple cavity modes coupled to the same transmon, the drive can also cancel the cross-Kerr interaction between cavity modes. Compared to previous Kerr cancellation methods~\cite{krastanov2015, heeres2015,wang2021}, this simple scheme only requires one drive, and does not require populating the transmon excited states therefore suppressing the susceptibility to transmon loss. We demonstrate numerically the performance of Kerr cancellation by extending the phase correlation of a cavity Schr\"odinger cat state well beyond the characteristic phase collapse time under Kerr nonlinearity. This Kerr-cancellation method is particularly suitable to the recently realized grid state encoding using a microwave cavity mode that involves a large number of cavity photons~\cite{campagne-ibarcq2020}.  

In the limit of weak transmon-cavity coupling, computing cavity nonlinearity reduces to calculating the nonlinear susceptibility function of the driven transmon. For systems with a large number of cavity modes coupled to the same transmon (cf.~\cite{chakram2020}), computing the susceptibility function is numerically more efficient than solving the full transmon-cavity Hamiltonian as the former only requires diagonalizing the Hamiltonian of the driven transmon. This method based on susceptibility function can be useful for characterizing multi-mode microwave cavities or acoustic cavities controlled by transmon ancillas. 

For future research, it would be interesting to investigate whether one can exploit the aforementioned drive-induced multiphoton resonances between cavity and transmon excitations as a way to dynamically and robustly control cavity nonlinearity. Although the cavity modes may inherit unfavorable decoherence properties from the typically lossier transmon ancilla, its nonlinearity strength can be tuned over a great range for a relatively small change in the drive amplitude or drive detuning due to the resonance. This tunable nonlinearity can be useful in many ways, including cavity state preparation~\cite{vrajitoarea2020c} and quantum simulations of many-body systems such as those described by the Bose-Hubbard model~\cite{carusotto2020}. 

\acknowledgments
We thank Luke Burkhart, Ben Chapman, Sal Elder, Connor Hann, and Yao Lu for useful discussions. This work was supported by ARO W911NF-18-1-0212 and by the Yale Quantum Institute. R.J.S. is a cofounder of, and equity shareholder in, and S.M.G. is a consultant for, Quantum Circuits, Inc. 

\appendix

\section{Higher-order corrections to Eq.~(\ref{eq:H_quar})}
\label{app:correction_to_chi}
The effect of the non-dispersive terms in Eq.~(\ref{eq:H_RWA_eigenmodes_no_drive}) can be captured by going to higher-order in the perturbation theory in terms of the small parameter $\alpha/\delta_{a(b)}$. Specifically, one can make a unitary transformation (Schrieffer-Wolff transformation) $U_S=\exp(iS)$ to eliminate the non-dispersive terms order by order in $\alpha/\delta_{a(b)}$. To first order in $\alpha/\delta_{a(b)}$, we have $S = i\sum_n H_n/\Delta_n$ (cf. Ref.~\cite{bukov2015a}), where $H_n$ represents a non-dispersive term, and it would become proportional to $\exp(i\Delta_nt)$ upon a unitary transformation $\hat U =\exp[-i(\delta_{AC}\N_a +\delta_{BC}\N_b) t]$. It follows that the Hamiltonian after the unitary reads:
\begin{align}
\label{eq:H_quar_correction}
& U_S^\dagger H_{\rm RWA} U_S\approx  H_0 + \sum_n \frac{[H_n,H_{n}^\dagger]}{2\Delta_n} = \tilde \delta_{AC} \N_A+\tilde \delta_{BC}\N_B  \nonumber \\ 
&- \frac{1}{2}\sum_{X,X' \in \{A,B,C\}} \chi_{XX'}\N_X\N_{X'} (1-\epsilon_{XX'}\N_C) \nonumber\\
 & + \sum_{X\in\{A,B\}}\frac{1}{3!}\beta_X \N_X^3 + \frac{1}{2}\left( \beta_{AB}\N_A^2\N_B + \beta_{BA}\N_B^2\N_A \right), 
\end{align}
where
\begin{align}
&\tilde\delta_{A(B)C} = \delta_{A(B)C} + \alpha(|\xi_C|^2-|\xi_{A(B)}|^2)/2,\nonumber \\ 
& \epsilon_{AA(BB)} = 9\frac{\chi_{CC}}{\delta_{A(B)C}},\quad \epsilon_{CC} = -\frac{\chi_{AC}}{\delta_{AC}}-\frac{\chi_{BC}}{\delta_{BC}}, \nonumber \\
& \epsilon_{A(B)C} =  \frac{3\chi_{CC}}{2\delta_{A(B)C}}, \nonumber \\
&\epsilon_{AB} = 2\chi_{CC}\left(\frac{1}{\delta_{AC}+\delta_{BC}}+\frac{2}{\delta_{AC}}+\frac{2}{\delta_{BC}}\right),\nonumber \\
&\beta_{A(B)} = \frac{3\chi_{AA(BB)}\chi_{A(B)C}}{2\delta_{A(B)C}}, \nonumber \\
&\beta_{AB(BA)} = \chi_{AA(BB)}\chi_{B(A)C} \bigg( \frac{1}{\delta_{A(B)C}}+\frac{4}{\delta_{B(A)C}} \nonumber\\
 &+\frac{1}{2\delta_{A(B)C}-\delta_{B(A)C}}\bigg).
\end{align}
The above equation shows that the fractional difference in the nonlinearity strength $\chi_{XX'}$ among different transmon states varies linearly in the transmon mode excitation number $N_C$ with the proportionality constant being $\epsilon_{XX'}\sim \chi_{CC}/\delta_{A(B)C}\ll 1$, for $X,X'\in\{A,B\}$. Additionally, there emerges sixth-order nonlinearity for the cavity-like modes. Their strengths characterized by coefficients $\beta_{A(B)}$ and $\beta_{AB(BA)}$ are smaller than the fourth-order Kerr nonlinearity by a factor of $\sim \chi_{A(B)C}/\delta_{A(B)C}\ll 1$.


4
4

\section{Linear and nonlinear susceptibilities of the driven ancilla}
\label{app:susceptibility}
Due to the weak coupling between the cavity modes and the transmon ancilla, one can treat the cavity fields as weak probe tones that act on the transmon ancilla. As we have previously shown in Ref.~\cite{zhang2019}, the linear and nonlinear susceptibilities of the periodically driven ancilla to additional weak classical probe tones are related to the ancilla-induced linear and nonlinear properties of the cavity modes. In the absence of ancilla decoherence, the calculation of the susceptibility functions is equivalent to the perturbative calculation in the weak-coupling regime presented in Sec.~\ref{sec:weak_coupling_limit}. In this section, we derive the expressions for the linear and nonlinear susceptibilities of the driven transmon ancilla. The procedure to derive susceptibilities follows the standard linear and nonlinear response theory~\cite{Kubo1957a, boyd2008}. To make the formalism general, we work in the lab frame and derive formulas that go beyond the RWA. 



We start with the equation of motion for the density matrix of the driven ancilla:
\begin{align}
\label{eq:density_matrix_eom}
\dot \hrho &= -i[H_{\rm anc + bath}(t)+H_{\rm probe}(t),\hrho]/\hbar,  \\
H_{\rm probe}(t) &= -f(t) \hat O. \nonumber
\end{align}
$H_{\rm anc + bath}$ is the total Hamiltonian of the ancilla under the periodic drive and the bath degrees of freedom that couple to the ancilla. $H_{\rm probe}$ describes a weak time-dependent classical ``force" $f(t)$ (the probe) coupled to an ancilla operator $\hat O$. The probe $f(t)$ plays the role of the cavity fields that act upon the transmon. Since the cavity modes couple to the charge degree of freedom of the transmon, later we will consider the case where $\hat O$ is the charge operator of the transmon ancilla, here we keep the formalism general.

Equation~(\ref{eq:density_matrix_eom}) has a formal solution:
\begin{align}
\label{eq:formal_solution}
\hrho(t) = \hat L_{\rightarrow}(t,0) \hrho(0) - \frac{i}{\hbar}\int_0^t dt_1 \hat L_{\rightarrow}(t,t_1) [H_{\rm probe}(t_1),\hrho(t_1)],
\end{align}
where $\hat L_{\rightarrow}$ is a superoperator that acts on all operators to its right according to: $\hat L_{\rightarrow}(t,t_1) \hat O = \hat U(t,t_1) \hat O U^\dagger (t,t_1), \hat U(t,t_1) = T \exp[-i \int_{t_1}^t H_{\rm anc + bath}(t')]$. 

With the help of the formal solution~(\ref{eq:formal_solution}), the density matrix can be found perturbatively in the probe strength $f(t)$ via iteration:
\begin{align}
\label{eq:iteration}
\hrho &= \sum_{m=0}^\infty \hrho ^{(m)},\hrho^{(m)}\sim \mathcal O(f^m), \nonumber \\
\hrho^{(m)}(t) &= - \frac{i}{\hbar}\int_0^t dt_1 \hat L_{\rightarrow}(t,t_1) [H_{\rm probe}(t_1),\hrho^{(m-1)}(t_1)], \nonumber \\
\hrho^{(0)}(t) &= \hat L_{\rightarrow}(t,0) \hrho(0).
\end{align}

Of interest to us is the response of the ancilla operator $\hat O$ to the coupling to the probe. This response is manifested as the change in the expectation value of the operator which formally reads:
\begin{align*}
\langle \o(t) \rangle= \sum_{m=0}^\infty \langle \o^{(m)}(t) \rangle,
\end{align*}
where $\o^{(m)}(t) \sim \mathcal O(f^{m})$ is the operator $\o$ solved to $m$-th order in $H_{\rm probe}$ in the Heisenberg picture. For instance, the unperturbed operator $\o^{(0)}(t) = \hat U^\dagger  (t,0) \o  \hat U(t,0)$. In the Schr\"odinger picture, we have that $\langle \o^{(m)}(t) \rangle = \Tr [\hat O \hrho^{(m)}(t)]$. 

Using Eq.~(\ref{eq:iteration}) and the cyclic property of the trace operation, we find the $m$-th order response of the ancilla operator $\o$ to the probe to be 
\begin{widetext}
\begin{align}
\label{eq:formal_response}
&\langle \o^{(m)}(t) \rangle =  \int_0^t d t_m\int_0^{t_{m}}dt_{m-1}...\int_0^{t_2}dt_1 \chi^{(m)}(t,t_m,...,t_1)f(t_m)...f(t_1), m\geq 1, \nonumber \\
&\chi^{(m)}(t,t_m,...,t_1)=\left(\frac{i}{\hbar}\right)^m\langle [...[[\o^{(0)}(t), \o^{(0)}(t_m)],\o^{(0)}(t_{m-1})]...,\o^{(0)}(t_1)]\rangle.
\end{align}
\end{widetext}
where $\langle...\rangle   \equiv  \Tr (...\hrho(0) )$. $\chi^{(m)}(t,t_m,...,t_1)$ is an $m$-th order response function in the time domain associated with operator $\o$ and is an intrinsic property of the driven ancilla. 
 
\subsection{Time-domain structure of the response function}
In the absence of coupling to the bath, some generic properties of the response function $\chi^{(m)}(t,t_1,...,t_m)$ can be inferred by inserting a complete set of ancilla Floquet states between adjacent operators. The Floquet states are eigenstates of the periodically driven ancilla: \[ \psi_m(t) = e^{-i\epsilon_m t} u_m(t), \]
where $\epsilon_m$ is the quasienergy, and $u_m (t)$ is called Floquet mode and has the same periodicity as the drive, i.e., $u_m(t+2\pi/\omega_d) = u_m(t)$. The Floquet modes form a complete set of states at any instant of time: $\sum_m |u_m (t)\rangle \langle u_m(t)| = \hat I$.  Using the relation $\hat U(t,0) \psi_m(0) = \psi_m (t)$, we immediately obtain that for $\rho(0) = |u_{l_0}\rangle \langle u_{l_0}\rangle$:
\begin{align}
\label{eq:response_function}
&\chi^{(m)}(t,t_m,...,t_1) =\left(\frac{i}{\hbar}\right)^m\sum_{\vec l_m,\vec K_m}\prod_{m'=0}^m O_{l_{m'+1}l_{m'},K_{m'+1}} \nonumber \\
&\times \exp[-i\epsilon_{l_{m'}}(t_{m'+1}-t_{m'})+iK_{m'+1}\omega_d t_{m'+1}]+...
\end{align}
where $ O_{mn,K}$ is the $K$-th Fourier component of the matrix element $\langle u_m(t)| \hat O |u_n(t) \rangle$: 
$O_{mn,K} = (2\pi / \omega_d)^{-1} \int_0^{2\pi/\omega_d} dt \langle u_m(t)| \o |u_n(t) \rangle \exp(-iK\omega_d t).$
To make the notation compact, we have defined $t_{m+1}\equiv t$ and introduced $\vec l_m = \{ l_1,l_2..,,l_m \},\vec K_m = \{ K_1,K_2...,K_{m+1} \}$. Also due to the trace operation, in Eq.~(\ref{eq:response_function}), we identify $m+1$ with $0$ such that $l_{m+1} \equiv l_0, t_{m+1} \equiv t_0 \equiv t.$ Note that Eq.~(\ref{eq:response_function}) only shows one term from the commutators in $\chi^{(m)}$; other terms represented by ... in Eq.~(\ref{eq:response_function}) will have similar structure but with pairs of $t_m$ and $t_n$ switching order. 

Equation~(\ref{eq:response_function}) shows that in the absence of external drive (i.e. all $K_{m'}$' s are zero), the response function $\chi^{(m)}(t,t_m,...,t_1)$ is a function of $m+1$ time differences (difference in the arguments) out of which only $m$ time differences are independent. In other words, $\chi^{(m)}$ is independent of initial time. In the presence of periodic drive, $\chi^{(m)}(t,t_m,...,t_1)$ is periodically modulated as a function of all $m+1$ arguments in addition to the dependence on $m$ time differences.  This property of the response function $\chi^{(m)}$ holds when the ancilla is coupled to a Markovian bath and $\rho(0)$ is the steady state. If $\rho(0)$ is a transient state, then in general $\chi^{(m)}$ will depend on all the arguments even without periodic drive. 

\subsection{Linear response to harmonic probes}
Now let us consider the response to a specific form of the probe that consists of multiple harmonic drives: \[ f(t) = \sum_\omega f_\omega e^{-i\omega t},\] where $\omega$ can be positive and negative and $f_{-\omega}=f_\omega^*.$ In the case where operator $\o$ is the ancilla charge operator, the probes can represent the fields from the cavity modes considered in the main text, except that here the probes are classical fields. 

Let us study first the property of the linear response $\langle \o^{(1)}(t) \rangle$ in Eq.~(\ref{eq:formal_response}). First, at the linear response level, the response $\langle \o^{(1)}(t) \rangle$ to the probe is a sum of response to each probe at the respective probe frequency. 

Second, as a generic property, upon integration of the right-hand side of Eq.~(\ref{eq:formal_response}) and in the absence of ancilla decoherence, the lower integration limit will yield terms that oscillate as a function of $t$ at the transition frequencies of the driven ancilla, whereas the upper limit will yield terms that oscillate at the probe frequency plus integer multiples of the drive frequency $\omega+K\omega_d$. This property can be readily seen from Eq.~(\ref{eq:response_function}). In the presence of ancilla dissipation, the terms from the lower limit will decay to zero in the long time limit $t\rightarrow \infty$, whereas those from the upper limit will remain oscillating at frequency $\omega+K\omega_d$, even in the long time limit. 

As mentioned in the main text, of interest to us is the dispersive regime where the cavity (probe) frequency is far away from any ancilla transition frequency. Thus, the terms from the lower integration limit can be neglected when we consider their back action on the cavity since they are strongly off-resonant with cavity frequency. This approximation holds even in the transient regime where the size of those terms is comparable to terms from the upper integration limit. 

Under the above considerations, we obtain that 
\begin{align}
&\langle \o^{(1)}(t) \rangle = \sum_{\omega,K} f_\omega  \chi(\omega;\omega+K\omega_d) e^{-i(\omega + K\omega_d) t}, \nonumber \\
 & \chi(-\omega;-\omega-K\omega_d) =  \chi^*(\omega;\omega+K\omega_d).
\end{align}
The linear susceptibility $\chi(\omega;\omega')$ in frequency domain is a function of both the probe frequency $\omega$ and response frequency $\omega'$; they do not need to be the same for a driven system. 
The susceptibility $\chi$ is given by the Fourier transform of the response function in the time domain: 
\begin{align}
\label{eq:chi_general}
\chi(\omega;\omega+K\omega_d) =&\int ^t_0 dt_1 \chi^{(1)}(t,t_1) \nonumber \\
 &  \times \exp[i\omega (t-t_1) + i K\omega_d t].
\end{align}
As discussed above, only the smooth terms (non-rotating terms from the upper integration limit) should be kept in the calculation of $\chi$. In the absence of ancilla dissipation, $\chi$ is time-independent; otherwise, $\chi$ slowly changes in time on the scale of ancilla relaxation time in the transient regime. 
As discussed in Ref.~\cite{zhang2019}, real and imaginary parts
 of $\chi(\omega;\omega)$ correspond to the ancilla-induced cavity frequency shift and inverse Purcell decay. For $K\neq 0$, $\chi(\omega;\omega+K\omega_d)$ relates to ancilla-mediated beam-splitter coupling (or two-mode squeezing coupling if $\omega$ and $\omega+K\omega_d$ have opposite signs) between one cavity at frequency $|\omega|$ and another at frequency $|\omega+K\omega_d|$. 

In the absence of ancilla decoherence, and taking initial density matrix to be $\rho(0) = |u_m\rangle \langle u_m|$ and denoting the corresponding $\chi$ as $\chi_m$, we obtain from Eq.~(\ref{eq:chi_general}) the following result:
\begin{align}
\label{eq:chi_explicit}
\chi_m(\omega;\omega+K\omega_d) = &\sum_{n,K'}\left(\frac{O_{mn,K'-K}O_{nm,-K'}}{-\hbar\omega-K'\hbar\omega_d+\epsilon_{nm}} \right. \nonumber \\ 
&\left. - \frac{O_{nm,K'-K}O_{mn,-K'}}{-\hbar\omega-K'\hbar\omega_d+\epsilon_{mn}}\right),
\end{align}
where $\epsilon_{nm}\equiv \epsilon_n-\epsilon_m$.  We have used a subscript $m$ for the susceptibility $\chi$ to indicate that this is calculated with respect to $\rho(0)=|u_m\rangle\langle u_m|$. We emphasize that the expressions for the susceptibility in Eqs.~(\ref{eq:chi_general},\ref{eq:chi_explicit}) apply for the probe-ancilla coupling of the general form in Eq.~(\ref{eq:density_matrix_eom}). They are applicable for any nonlinear ancilla with a time-periodic Hamiltonian not limited to a voltage-driven transmon, and apply beyond the RWA. 

Now we consider that the operator $\o$ that the probe field is coupled to is proportional to transmon charge operator: $\o = i(\c^\dagger - \c)$. In the case where the transmon ancilla can be well approximated as a weakly nonlinear oscillator and both the drive and probes are relatively close in frequency to the ancilla $|\omega-\omega_c|,|\omega_d-\omega_c|\ll\omega_c$, one can apply RWA and neglect terms that do not preserve excitation number in the Hamiltonian; see Sec.~\ref{sec:Hamiltonian} for the conditions of RWA. Under RWA, the frequencies at which the linear response oscillate are limited to the first harmonic $K = 0$ and $K = -2 \sgn(\omega)$. The expression for the susceptibility in Eq.~(\ref{eq:chi_general}) is simplified to (for $\omega>0$)
\begin{align}
&\chi(\omega;\omega) \approx \left(\frac{i}{\hbar}\right) \int ^t_0 dt_1 \langle [\c^{(0)}(t),\c^{\dagger (0)}(t_1)]\rangle e^{i\omega (t-t_1)}, \nonumber \\
&\chi(\omega;\omega - 2\omega_d) \approx \left(\frac{i}{\hbar}\right) \int ^t_0 dt_1 \langle [i\c^{\dagger (0)}(t),i\c^{\dagger (0)}(t_1)] \rangle \nonumber \\ 
& \times e^{i\omega (t-t_1) - 2i\omega_d t},
\end{align}
which coincide with the results obtained in Ref.~\cite{zhang2019}. In the absence of ancilla decoherence, explicit expression for $\chi$ under RWA can be obtained from Eq.~(\ref{eq:chi_explicit}) by only keeping the term $K'=-1$ in the summation (for $\omega>0$) and using that the non-vanishing matrix elements of $\o$ under RWA are approximately given by $O_{mn,\pm 1} \approx \pm ic^{(\pm 1)}_{mn}$, where $c^{(\pm 1)}_{mn}$ is the matrix element of operator $\c$ or $\c^{\dagger}$ between ancilla RWA eigenstates $\psi_m$ and $\psi_n$ defined in Eq.~(\ref{eq:H_anc_RWA}) [see the text below Eq.~(\ref{eq:M})].


\subsection{Third-order nonlinear response} 

Of primary interest to us in this work is the third-order nonlinear response of the drive ancilla to the probe. Due to the nonlinearity of the driven ancilla and the beating between different probe frequencies, third-order nonlinear response of the ancilla can in general oscillate at any frequency combination of three probe frequencies, 
\begin{widetext}
\begin{align}
\label{eq:third_order_response}
&\langle \o^{(3)}(t) \rangle = \sum_{\omega,\omega',\omega'',K} f_\omega f_{\omega'}f_{\omega''} \chi^{(3)}(\omega,\omega',\omega'';\omega+\omega'+\omega''+K\omega_d) e^{-i(\omega+\omega'+\omega''+K\omega_d)t} \nonumber \\
&\chi^{(3)}(\omega,\omega',\omega'';\omega+\omega'+\omega''+K\omega_d)  =  \int ^t_0 dt_3 \int ^{t_3}_0 dt_2  \int ^{t_2}_0 dt_1  \chi^{(3)}(t,t_3,t_2,t_1)  \nonumber \\
& \times \mathcal P \exp{[i\omega(t-t_3)+i\omega'(t-t_2)+i\omega''(t-t_1)+iK\omega_d t]},
\end{align}
\end{widetext}
where $\mathcal P$ indicates a summation over terms that are invariant with respect to permuting $\omega,\omega',\omega''$. Similar to the linear susceptibility, only smooth terms in $\chi^{(3)}$ need to be kept. As in Eq.~(\ref{eq:chi_explicit}), explicit expression for $\chi^{(3)}$ can be obtained in the absence of ancilla decoherence from Eq.~(\ref{eq:third_order_response}). 

As discussed in Sec.~\ref{sec:connection_to_chi} of the main text, $\Re \chi^{(3)}(\omega,\omega,-\omega;\omega)$ characterizes the ancilla-induced self-Kerr of a cavity at frequency $\omega$ whereas $\Re \chi^{(3)}(\omega,\omega',-\omega';\omega)$ characterizes the ancilla-induced cross-Kerr between two cavities with frequency $\omega$ and $\omega'$. Taking the initial state to be $\rho(0) = |u_m\rangle \langle u_m|$ and denoting the corresponding susceptibility as $\chi^{(3)}_m$, we obtain from Eq.~(\ref{eq:third_order_response}) that in the absence of ancilla decoherence, the nonlinear susceptibilities $\chi_m^{(3)}(\omega,-\omega,\omega;\omega)$ and $\chi_m^{(3)}(\omega,-\omega',\omega';\omega)$ read:
\begin{widetext}
\begin{align}
\label{eq:self_Kerr_nonRWA}
\chi_m^{(3)}(\omega,-\omega,\omega;\omega) & = -\sum_{n,K}\left[\sum_{j=\pm1}\frac{|\tilde{M}_{nm,K}^{(j)}(\omega)|^{2}}{\epsilon_{mn}+2j\hbar\omega-K\hbar\omega_{d}}+\frac{|\tilde{M}_{nm,K}^{(-1)}(\omega)+\tilde{M}_{nm,K}^{(+1)}(\omega)|^{2}}{\epsilon_{mn}-K\hbar\omega_{d}}\right]\nonumber \\
 & +[\tilde{M}_{mm,0}^{(-1)}(\omega)+\tilde{M}_{mm,0}^{(+1)}(\omega)][\tilde{N}_{mm,0}^{(-1)}(\omega)+\tilde{N}_{mm,0}^{(+1)}(\omega)],
\end{align}
\end{widetext}
\begin{widetext}
\begin{align}
\label{eq:cross_Kerr_nonRWA}
\chi_m^{(3)}(\omega,-\omega',\omega';\omega) &=-\sum_{n,K}\left[\sum_{j=\pm1}\frac{|\tilde{M}_{nm,K}^{(j)}(\omega)+\tilde{M}_{nm,K}^{(j)}(\omega')|^{2}}{\epsilon_{mn}+j\hbar(\omega+\omega')-K\hbar\omega_d}+\sum_{j=\pm1}\frac{|\tilde{M}_{nm,K}^{(j)}(\omega)+\tilde{M}_{nm,K}^{(-j)}(\omega')|^{2}}{\epsilon_{mn}+j\hbar(\omega-\omega')-K\hbar\omega_d}\right.\nonumber \\
 & \left.+2{\rm Re}\frac{(\tilde{M}_{nm,K}^{(+1)}(\omega)+\tilde{M}_{nm,K}^{(-1)}(\omega))(\tilde{M}_{nm,K}^{(+1)}(\omega')+\tilde{M}_{nm,K}^{(-1)}(\omega'))^{*}}{\epsilon_{mn}-K\hbar\omega_d}\right]\nonumber \\
 & +\left\{[\tilde{M}_{mm,0}^{(-1)}(\omega)+\tilde{M}_{mm,0}^{(+1)}(\omega)][\tilde{N}_{mm,0}^{(-1)}(\omega')+\tilde{N}_{mm,0}^{(+1)}(\omega')]+(\omega \leftrightarrow \omega')\right\},
\end{align}
\end{widetext}
where tensors $\tilde{M}_{mn,K}^{(j)}$ and $\tilde{N}_{nm,K}^{(j)}$
read:
\begin{equation}
\tilde{M}_{mn,K}^{(j)}(\omega)=\sum_{n'K'}\frac{O_{mn',K-K'}O_{n'n,K'}}{\epsilon_{nn'}-K'\hbar\omega_{d}+j\hbar\omega},
\end{equation}
\begin{equation}
\tilde{N}_{mn,K}^{(j)}(\omega)=\sum_{n'K'}\frac{O_{mn',K-K'}O_{n'n,K'}}{(\epsilon_{nn'}-K'\hbar\omega_{d}+j\hbar\omega)^{2}}.
\end{equation}
Similar to the expression for the linear susceptibility in Eq.~(\ref{eq:chi_explicit}), Eqs.~(\ref{eq:self_Kerr_nonRWA},\ref{eq:cross_Kerr_nonRWA}) for the nonlinear susceptibility work for any periodically-driven ancilla with an ancilla-probe couplng of the form in Eq.~(\ref{eq:density_matrix_eom}) and apply beyond the RWA. 

For a transmon ancilla capacitively coupled to cavity modes, we substitute operator $\o$ with $i(\c^\dagger -\c)$ and under the RWA as discussed in Sec.~\ref{sec:Hamiltonian}, the expressions for $\chi^{(3)}(\omega,\omega,-\omega;\omega)$ and $\chi^{(3)}(\omega,\omega',-\omega';\omega)$ follow from Eq.~(\ref{eq:third_order_response}) and read:
\begin{widetext}
\begin{align}
\label{eq:chi_self_Kerr}
\chi^{(3)}(\omega,\omega,-\omega;\omega) \approx  \left(\frac{i}{\hbar}\right)^3 &\int ^t_0 dt_3 \int ^{t_3}_0 dt_2  \int ^{t_2}_0 dt_1  \Bigg\{ \langle  [[[\c^{(0)}(t),\c^{\dagger (0)}(t_3)],\c^{\dagger (0)}(t_2)],\c^{(0)}(t_1)]\rangle \exp{[-i\omega (t_3 + t_2 - t_1)]} \nonumber \\
+& \langle [[[\c^{(0)}(t),\c^{\dagger (0)}(t_3)],\c^{ (0)}(t_2)],\c^{\dagger (0)}(t_1)] \rangle \exp{[-i\omega (t_3 - t_2 + t_1)]}  \nonumber \\
+ &\langle [[[\c^{(0)}(t),\c^{(0)}(t_3)],\c^{\dagger (0)}(t_2)],\c^{\dagger (0)}(t_1)]  \rangle \exp{[-i\omega ( - t_3 + t_2 + t_1)]} \Bigg\}  \exp(i\omega t),
\end{align}
\end{widetext}
\begin{widetext}
\begin{align}
\label{eq:chi_cross_Kerr}
\chi^{(3)}(\omega,\omega',-\omega';\omega) \approx  \left(\frac{i}{\hbar}\right)^3 &\int ^t_0 dt_3 \int ^{t_3}_0 dt_2  \int ^{t_2}_0 dt_1  \Bigg\{ \langle  [[[\c^{(0)}(t),\c^{\dagger (0)}(t_3)],\c^{\dagger (0)}(t_2)],\c^{(0)}(t_1)]\rangle \exp{(-i\omega t_3 -i\omega' t_2 + i\omega' t_1)} \nonumber \\
+& \langle [[[\c^{(0)}(t),\c^{\dagger (0)}(t_3)],\c^{ (0)}(t_2)],\c^{\dagger (0)}(t_1)] \rangle \exp{(-i\omega t_3 - i\omega' t_1 +i\omega' t_2)}  \nonumber \\
+ &\langle [[[\c^{(0)}(t),\c^{(0)}(t_3)],\c^{\dagger (0)}(t_2)],\c^{\dagger (0)}(t_1)]  \rangle \exp{(-i\omega' t_2  - i\omega t_1 +i\omega' t_3)} \Bigg\}  \exp(i\omega t) \nonumber \\
&+ (\omega \leftrightarrow \omega').
\end{align}
\end{widetext}
%
Using Eqs.~(\ref{eq:Kerr_chi_relation},\ref{eq:Kerr_chi_relation_2}) in the main text and the expressions for $\chi^{(3)}$ above, we obtain the same results for the cavity self-Kerr and cross-Kerr as in Eqs.(\ref{eq:self_Kerr},\ref{eq:cross_Kerr}).

It immediately follows from Eq.~(\ref{eq:chi_self_Kerr}) that for large $|\omega-\omega_c|$, $\Re \chi^{(3)}(\omega,\omega,-\omega;\omega)\sim \mathcal O((\omega-\omega_c)^{-4}).$ The terms $\sim O((\omega-\omega_c)^{-3})$ would come from taking all $\c^{(0)} (t')$ in the integrand to be $\c ^{(0)}(0) \exp(-i\omega_c t')$, which would necessarily vanish due to the commutators. This is consistent with our analysis in Sec.~\ref{sec:asymptotic_regime}.

\section{Inter-cavity cross-Kerr nonlinearity.}
\label{app:cross_Kerr}
In this section, we briefly discuss cavity cross-Kerr $K_{AB,m}$ as a function of cavity frequencies [i.e., the susceptibility function $\chi^{(3)}(\omega,-\omega',\omega';\omega)$ in Eq.~(\ref{eq:Kerr_chi_relation_2})], which we refer to as cavity cross-Kerr spectrum. 

\begin{figure}[ht]
\includegraphics[width=8. cm]{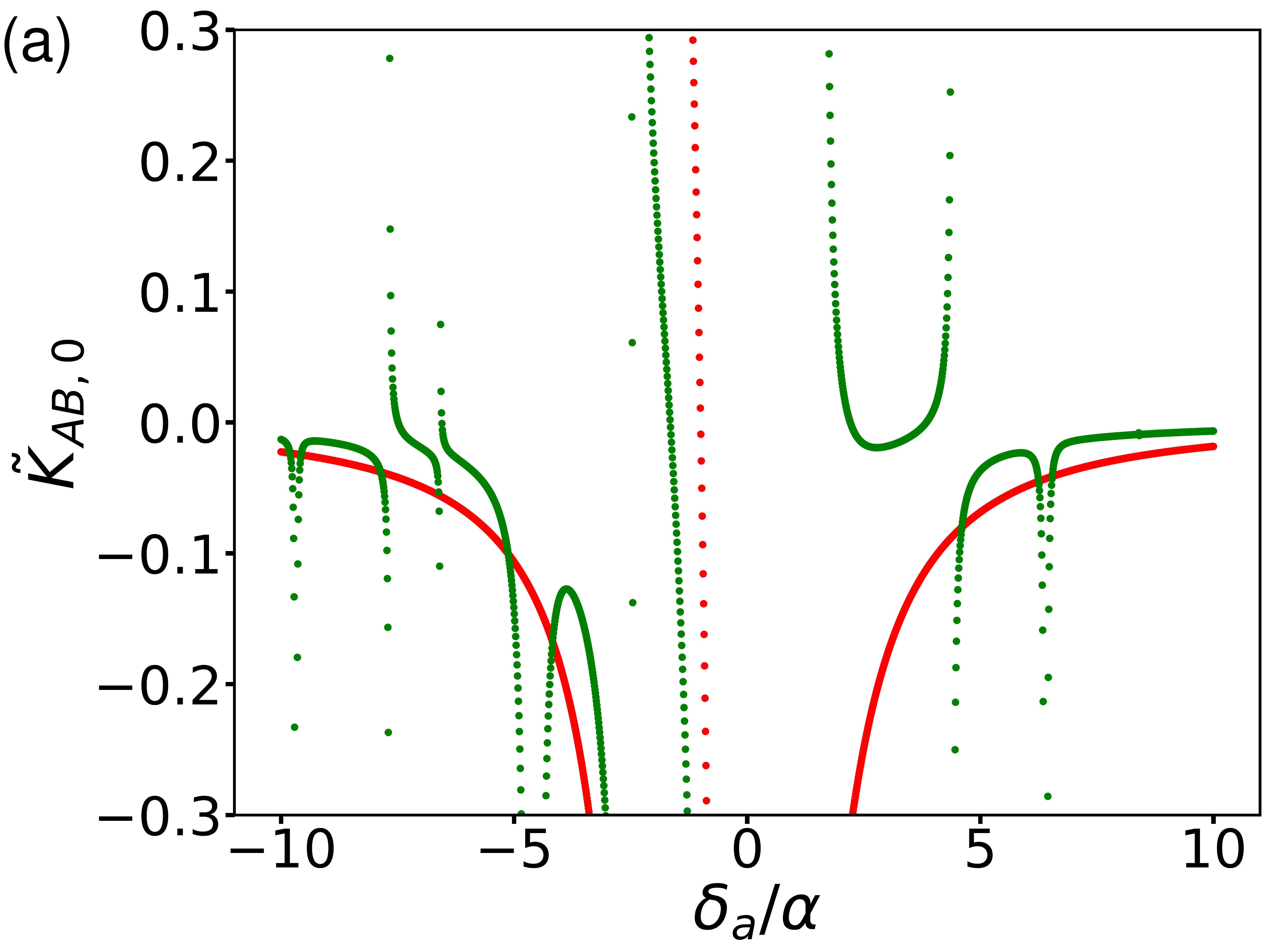} \hfill
\includegraphics[width=8. cm]{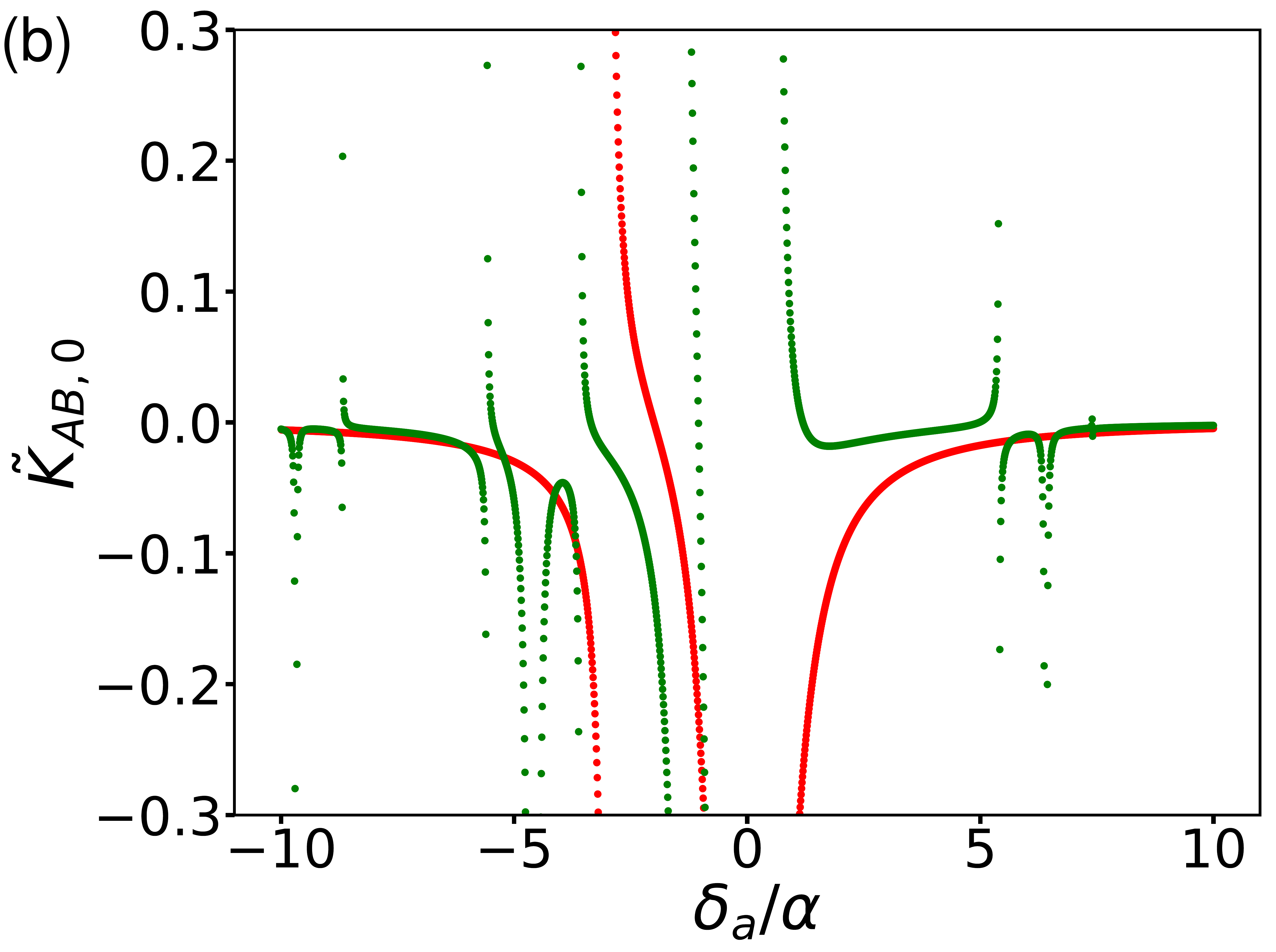}
\caption{Cavity cross-Kerr spectrum: (a) $\delta_b/\alpha = 1$; (b) $\delta_b/\alpha = 2.$ The dimensionless cavity cross-Kerr $\tilde K_{AB,0}$ is defined as: $\tilde K_{AB,0}= \alpha^3 K_{AB,0}/|g_ag_b|^2.$ For fixed $\delta_b/\alpha$, the spectrum is controlled by two dimensionless drive parameters $\delta_d/\alpha = 3$ and $|\Omega_d/\delta_d|^2 = 0$ (red dots), 0.3 (green dots). Same for both panels. }
\label{fig:cross_Kerr}
\end{figure}

Figure~\ref{fig:cross_Kerr} shows the cavity cross-Kerr $K_{AB,0}$ as a function of cavity-$a$ detuning $\delta_a$ for fixed cavity-$b$ detuning $\delta_b$. In the presence of the transmon drive, the cross-Kerr spectrum shows rich dispersive structures as a result of the drive-induced multiphoton resonance processes. The locations and strengths of these structures depend sensitively on the value of $\delta_b$ as can be seen going from the upper to the lower panel of Fig.~\ref{fig:cross_Kerr}. 

The resonance processes that are responsible for the strong dispersive structures in $K_{AB,m}$ as a function of $\delta_a,\delta_b$ are as follows: 
\[
iii) \,(n-m-2j)\omega_d + j(\omega_a+\omega_b) = \tilde \omega_{nm},\, j = \pm 1,
\]
\[
iv) \, (n-m)\omega_d + j(\omega_b-\omega_a) = \tilde \omega_{nm},\, j = \pm 1.
\]
In addition to the above processes that involve photons from both cavity modes, the processes that involve only one cavity mode [see resonance condition ii) in Sec.~\ref{sec:near_resonance}] also affect the cross-Kerr $K_{AB,m}$.

\section{Cavity nonlinearities in the vicinity of a drive-induced cavity-transmon resonance}
\label{app:near_resonance}
While the cavities are off-resonant with the transmon in the absence of the drive, turning on the drive can bring them into near resonance with certain transition between transmon states. Cavity nonlinearities are modified as a result of stronger hybridization with the transmon. To illustrate this point,  we consider as an example that cavity $a$ is in the vicinity of a resonance: $\omega_a + \omega_d \approx \tilde \omega_{20}$.

We consider the regime $|\delta_a|\gg \alpha$ which is convenient for analytical analysis and also of experimental interest. As in Sec.~\ref{sec:large_cavity_detuning}, we start from the Hamiltonian in Eq.~(\ref{eq:H_RWA_eigenmodes}) that is expressed in terms of the eigenmodes of the linear part of the system described in Sec.~\ref{sec:4wm_no_drive}. The condition $\omega_a + \omega_d \approx \tilde \omega_{20}$ implies that the drive is also far- detuned from the transmon-like mode $C$ with a detuning much larger than its anharmonicity, i.e., $|\delta_{dC}|\gg \alpha$. As discussed in Sec.~\ref{sec:semiclassical}, for such a large-detuning drive, the leading-order effect (leading-order in $\alpha/\delta_{dC}$) of the drive is to induce a classical displacement on mode $C$. This displacement can be found non-perturbatively in the drive strength with account taken of the finite nonlinearity of the mode by solving the classical equation of motion. We go to the displaced frame for mode $C$ by performing the standard displacement transformation $\hat D = \exp[d_C \C^\dagger - d_C^* \C]$ such that $\hat D^\dagger \C \hat D = \C+ d_C$, where $d_C \equiv Q_0/\sqrt{2\lambda}$. $Q_0$ is the classical displacement [see text below Eq.~(\ref{eq:g_diag})] and $\lambda$ is the scaled Plack' s constant [see Eq.~(\ref{eq:lambda})].


After the displacement transformation, the quartic terms in Eq.~(\ref{eq:H_RWA_eigenmodes}) now capture various four-wave mixing processes involving the drive. We keep the near-resonant term that corresponds to the process ($\omega_a+\omega_d\leftrightarrow \tilde \omega_{20}$) and disregard non-resonant terms. After switching to a rotating frame where mode-$A$ has zero frequency, we arrive at the following Hamiltonian: 
\begin{align}
\label{eq:TLS_2}
&H_{\rm RWA} \approx H_{\vec \sigma} -\sum_{X\in\{A,B\}} \frac{\hbar\chi_{XX'}}{2} \N_X \N_{X'},  \\
& H_{\vec\sigma}/\hbar = \hat \delta_{\rm eff}(\N_A,\N_B) \frac{\sig_z }{2} +  \Omega_{\rm eff} \sig_+ \A + \Omega_{\rm eff}^* \sig_- \A^\dagger,  \nonumber \\
&\hat \delta_{\rm eff}(\N_A,\N_B) = \tilde \omega_{20} - \tilde \omega_A - \omega_d - 2\chi_{AC}\N_A-2\chi_{BC}\N_B, \nonumber\\ &\Omega_{\rm eff} = -\sqrt{\chi_{AC}\chi_{CC}} d_C,\quad
 \tilde \omega_{20} \approx 2\omega_C -3\alpha -4|d_C|^2\alpha, \nonumber \\
&\tilde \omega_A = \omega_A-\chi_{AC}/2 -\chi_{AC}|d_C|^2.  \nonumber
\end{align}
We have restricted to the first and third levels of the eigenmode $C$ by defining $\sigma_+ = |2_C\rangle \langle 0_C|.$ The conditions to restrict to the two-level subspace are 
\begin{align}
\label{eq:TLS_condition_2}
|\delta_{\rm eff}|,|\Omega_{\rm eff}| &\ll \alpha.
\end{align}
In this rotating frame, the effective frequency $\hat \delta_{\rm eff}$ of the two-level system depends on the cavity photon numbers $\N_A,\N_B$ through the cross-Kerr interaction between modes $A,B$ and $C$. 

$H_{\vec\sigma}$ is the same as the Jaynes-Cummings Hamiltonian, except that the frequency of the two-level system depends on the cavity photon numbers. Diagonalizing $H_{\vec\sigma}$ leads to the following Hamiltonian:

\begin{align}
\label{eq:tilde_delta_d_m0_2}
& H_{\vec\sigma}/\hbar= \frac{\hat {\tilde \delta}_{\rm eff}(\N_A,\N_B)}{2}\sig_z ,  \quad \hat {\tilde \delta}_{\rm eff}(\N_A,\N_B) = \sgn(\hat \delta_{\rm eff}) \nonumber \\ &\times \left\{\hat \delta_{\rm eff}(\N_A,\N_B)^2  
+4|\Omega_{\rm eff}|^2 [\N_A + (\sig_z + 1)/2]\right\}^{1/2}.
\end{align}
To see how the cavity nonlinearities arise from the above Hamiltonian, we expand $\hat {\tilde \delta}_{\rm eff}(\N_A,\N_B)$ with respect to $\N_A,\N_B$. 
To third order in $\N_A,\N_B$, we found that $H_{\rm RWA}$ in Eq.~(\ref{eq:TLS_2}) becomes:

\begin{widetext}
\begin{align}
\frac{H_{\rm RWA}}{\hbar} = &\left(\frac{\delta_{\rm eff}}{2}-\sum_{X\in\{A,B\}}(\chi_{XC}+\Delta\chi_{XC})\N_X \right) \sig_z -\frac{1}{2}\sum_{X,X'\in\{A,B\}} (\chi_{XX'}-\Delta\chi_{XX'}\sig_z)\N_X \N_{X'} \nonumber \\ 
& + \left(\sum_{X\in\{A,B\}} \frac{\Delta\beta_X}{3!}\N_X^3 + \frac{\Delta\beta_{AB}}{2!}\N_A^2\N_B^2 +\frac{\Delta\beta_{BA}}{2!}\N_B^2\N_A\right) \sigma_z, 
\end{align}
\end{widetext}
where to lowest order in $\chi_{AC}/\delta_{\rm eff}$ and $\Omega_{\rm eff}/\delta_{\rm eff}$, we have
\begin{align*}
\frac{\Delta\chi_{AC}}{\chi_{AC}}& = -\epsilon_{\rm eff},\,\frac{\Delta\chi_{BC}}{\chi_{BC}} = -\epsilon_{\rm eff}\frac{\chi_{AC}}{\delta_{\rm eff}}(\sigma_z+1),\nonumber \\
\frac{\Delta\chi_{AA}}{\chi_{AA}} &= 16\epsilon_{\rm eff}\frac{\alpha  }{ \delta_{\rm eff}}\left(1- \frac{\epsilon_{\rm eff}}{2}\right), \,
\frac{\Delta\chi_{AB}}{\chi_{AB}} =4\epsilon_{\rm eff}\frac{\alpha}{ \delta_{\rm eff}}, \nonumber \\
\frac{\Delta\chi_{BB}}{\chi_{BB}} &=16\epsilon_{\rm eff}\frac{\alpha }{ \delta_{\rm eff}} \frac{\chi_{AC}}{\delta_{\rm eff}}(\sigma_z+1),
\end{align*}
\begin{align}
\label{eq:Delta_chi_TLS_2}
\frac{\Delta\beta_{A}}{\chi_{AA}} & = 48\epsilon_{\rm eff}\frac{ \chi_{AC}}{\delta_{\rm eff}} \frac{\alpha}{\delta_{\rm eff}}\left(\epsilon_{\rm eff}^2-3\epsilon_{\rm eff}+2 \right),\nonumber \\
\frac{\Delta\beta_B}{\chi_{BB}} & = 96\epsilon_{\rm eff}\frac{\chi_{BC}}{\delta_{\rm eff}}\frac{\alpha}{\delta_{\rm eff}}\frac{\chi_{AC}}{\delta_{\rm eff}}(\sigma_z+1), \nonumber \\
\frac{\Delta\beta_{AB}}{\chi_{AB}} & = 24\epsilon_{\rm eff}\frac{\chi_{AC}}{\delta_{\rm eff}}\frac{\alpha}{\delta_{\rm eff}}\left( 2-\epsilon_{\rm eff}\right),\nonumber \\
\frac{\Delta\beta_{BA}}{\chi_{AB}}& = 24\epsilon_{\rm eff}\frac{\chi_{BC}}{\delta_{\rm eff}}\frac{\alpha}{\delta_{\rm eff}},\nonumber \\
\epsilon_{\rm eff} &\equiv \alpha|d_C|^2/\delta_{\rm eff},\quad \delta_{\rm eff}\equiv \hat \delta_{\rm eff}(0,0). 
\end{align}
Although we have assumed  $|\Omega_{\rm eff}/\delta_{\rm eff}|\ll1$, parameter $\epsilon_{\rm eff}$ can be of order $\mathcal O(1)$. Interestingly, the fractional change in cavity-$A$ self-Kerr and its cross-Kerr with cavity-$B$ can be much larger than unity in the considered regime of $\alpha \gg |\delta_{\rm eff}|$. In addition to the change in cavity nonlinearities, strength of cavity cross-Kerr with the transmon is also modified by the drive.

Comparing the drive-induced change to sixth-order cavity nonlinearity strength with Kerr nonlinearity leads to the following condition for the expansion of $ \hat {\tilde \delta}_{\rm eff}(\N_A,\N_B)$ with respect to $\N_A,\N_B$ to converge:
\begin{align*}
\label{eq:convergence_2}
\frac{|\Delta\beta_A|}{|\Delta\chi_{AA}|}\sim \frac{\chi_{AC}}{|\delta_{\rm eff}|} \rm {max}(1,|\epsilon_{\rm eff}|) \ll 1.
\end{align*}
The above condition indicates that for the purpose of obtaining a large fractional change of cavity Kerr nonlinearity while keeping higher-order cavity nonlinearity small, the coupling between the cavity $a$ and the transmon needs to sufficiently weak so that $\chi_{AC}\ll \alpha$ and there is a large bandwidth to place $\delta_{\rm eff}$ to satisfy $\chi_{AC}\ll |\delta_{\rm eff}|\ll \alpha$.

\section{Non-perturbative corrections to the weak-coupling expressions of the cavity Kerr nonlinearities}
\label{app:beyond_weak_coupling}
For any finite cavity-transmon coupling strengths, there are higher-order corrections to the weak-coupling expressions for cavity Kerr nonlinearities in Eqs.~(\ref{eq:self_Kerr},\ref{eq:cross_Kerr}). In the absence of the drive, these corrections are small perturbations as long as $|g_{a(b)}/\delta_{a(b)}|\ll 1$, as can be seen from the limiting cases in Sec.~\ref{sec:theory_no_drive}. In the presence of the drive, however, the corrections can be non-perturbative, even when $|g_{a(b)}/\delta_{a(b)}|\ll 1$. In this section, we discuss these cases and show how to incorporate the non-perturbative corrections to the weak-coupling expressions in Eqs.~(\ref{eq:self_Kerr},\ref{eq:cross_Kerr}).


\subsection{Near a drive-induced cavity-transmon resonance}
The first situation where corrections to the weak-coupling expressions are important is when a cavity mode is in near-resonance with the driven transmon. As discussed in Sec.~\ref{sec:near_resonance}, the drive can induce resonant interaction between the cavities and the transmon. Near those resonances, cavity nonlinearities sensitively depend on the frequencies of the modes and the drive parameters, as seen from Fig.~\ref{fig:self_Kerr_spec}(a). When exactly on resonance, the perturbation theory that leads to Eqs.~(\ref{eq:self_Kerr},\ref{eq:cross_Kerr}) predicts the divergence of cavity Kerr nonlinearity indicating the breakdown of the theory.  

To understand at what distance to a resonance the perturbation theory becomes inaccurate, we can inspect the perturbative expressions in Eqs.~(\ref{eq:self_Kerr},\ref{eq:cross_Kerr}). The distance to a particular resonance is essentially given by the energy denominator of a relevant term in Eq.~(\ref{eq:self_Kerr}) or~(\ref{eq:cross_Kerr}). In evaluating these denominators, we used the unperturbed eigenenergies, neglecting the fact that the transmon-cavity coupling leads to shifts in these eigenenergies (or equivalently in the transition frequencies of the cavities and the transmon). When the resulting shift in the energy denominator is comparable to or larger than the size of the unperturbed denominator, the effect of the shift is non-perturbative.

\begin{figure}[ht]
\includegraphics[width=8.cm]{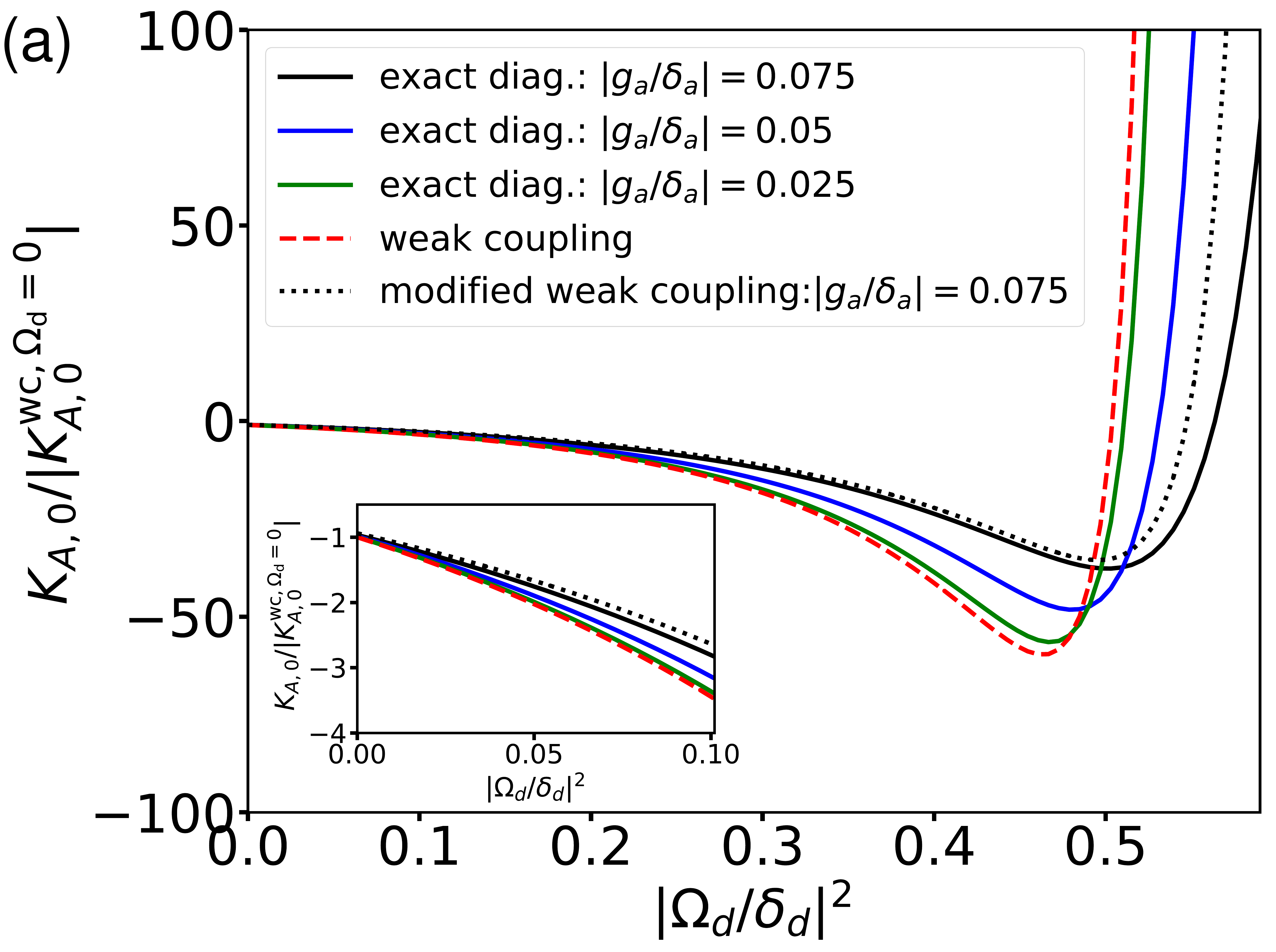} \\
\includegraphics[width=8 cm]{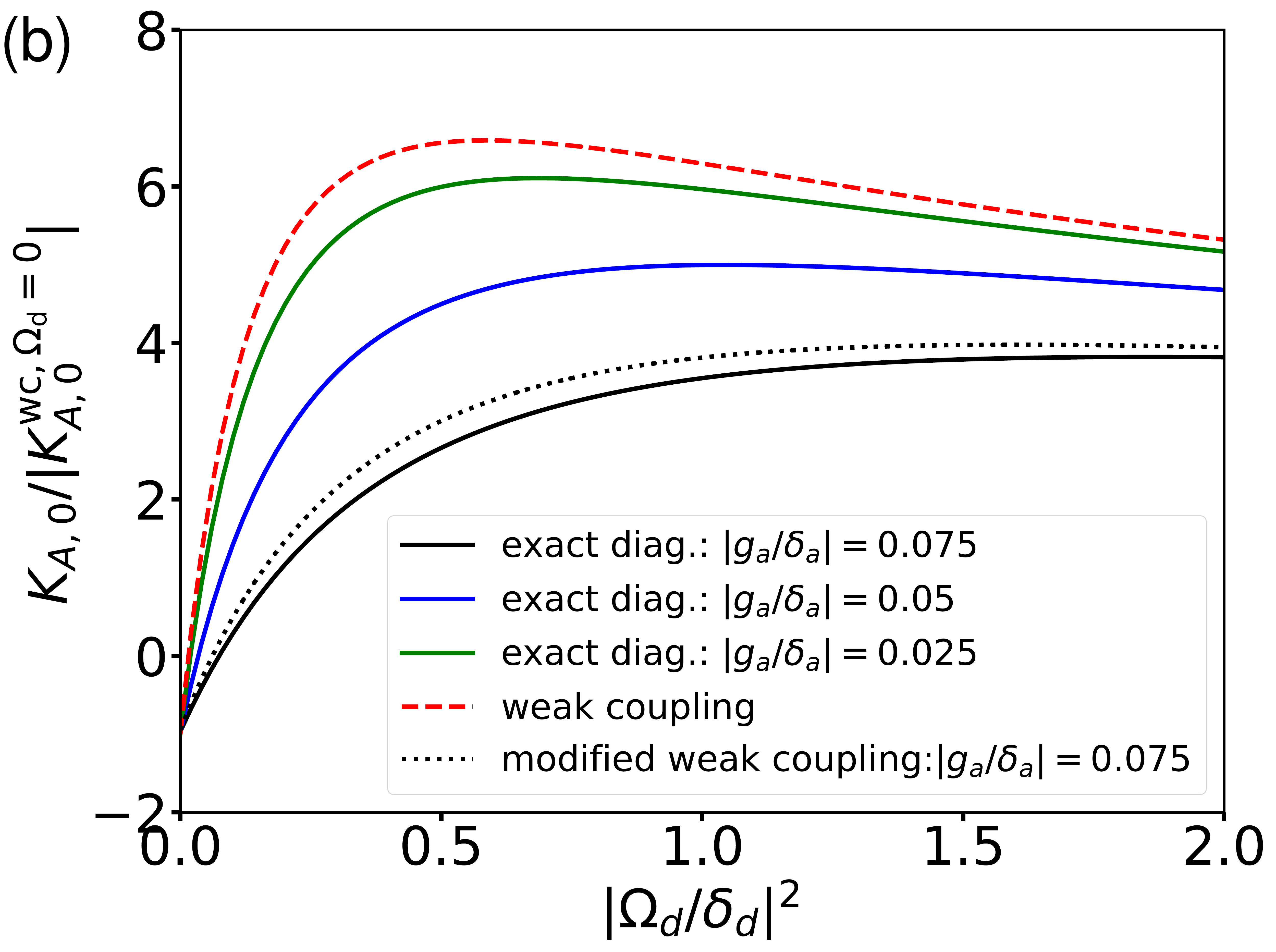} 
\caption{Comparison between the weak-coupling expression for the cavity self-Kerr in Eq.~(\ref{eq:self_Kerr}) (red dashed line) and the exact diagonalization (solid lines) of the full transmon-cavity Hamiltonian in Eq.~(\ref{eq:H_RWA}). The ordinate shows cavity self-Kerr scaled by its zero-drive value calculated using the weak-coupling expression. For this scaling, the red dashed line is independent of the scaled coupling strength $g_a/\delta_a$. The black dotted line refers to a modified version of Eq.~(\ref{eq:self_Kerr}) in which the coupling-induced cavity and transmon frequency shifts are taken into account; see the text for details. Panel (a)  Cavity-$a$ is near a drive-induced resonance, $\omega_a+\omega_d \approx \tilde \omega_{20}$. $\delta_a/\alpha = -5.1,\delta_d/\alpha = 3$. Panel (b) Cavity-$a$ is far-detuned from the transmon, but the drive is relatively close to $\omega_{10}$. For this panel, we have $\delta_a/\alpha=10$ and $\delta_d/\alpha=0.1$. }
\label{fig:self_Kerr_correction}
\end{figure}

To illustrate, we choose the frequency of cavity-$a$ to be near the resonance $\omega_a + \omega_d \rightarrow \tilde \omega_{20}$, and show the cavity self-Kerr $K_{A,0}$ as a function of the drive power in Fig.~\ref{fig:self_Kerr_correction}(a). Occurrence of the resonance can be seen as the sharp increase of the cavity Kerr as the drive approaches a certain power. As the figure shows, the Kerr as a function of the drive power calculated through the full diagonalization is shifted along the abscissa from the weak-coupling result using Eq.~(\ref{eq:self_Kerr}) and the shift is larger for stronger cavity-transmon coupling. This shift is precisely due to the coupling-induced frequency shifts of the cavity mode and the transmon. The deviation of the weak-coupling result from the full diagonalization becomes significant when the distance to the resonance becomes comparable to the shift. 

To leading order in the cavity-transmon coupling strengths, there are two types of coupling-induced frequency shifts. First, there are transmon-state-dependent cavity frequency shifts [corresponding to $c_{10,m}$ and $c_{01,m}$ term in Eq.~(\ref{eq:E_m_expansion})]. To leading order in the coupling strengths $g_a,g_b$ and neglecting the drive, they are equal to: \[c_{10(01),m}^{\Omega_d=0} = |g_{a(b)}|^2 \frac{\delta_{a(b)}-\alpha}{(\delta_{a(b)} + m\alpha)(\delta_{a(b)}+(m-1)\alpha)}.\] Second, there are cavity-photon-number-independent shifts in the transmon levels, corresponding to $c_{00,m}$ term in Eq.~(\ref{eq:E_m_expansion}). To leading order in $g_a,g_b$ and at zero drive strength, it is equal to: \[c_{00,m}^{\Omega_d=0} = -|g_a|^2 \frac{m}{\delta_a+(m-1)\alpha} - |g_b|^2\frac{m}{\delta_b+(m-1)\alpha}.\] 

The two types of frequency shifts can be taken into account by modifying Eqs.~(\ref{eq:self_Kerr},\ref{eq:cross_Kerr}) as follows. For the first type, one can simply replace the bare cavity frequencies with the shifted cavity frequencies in the energy denominators. For the second type, since the transmon transition frequencies also affect the matrix elements, we add a term $\sum_m c_{00,m}^{\Omega_d = 0}|m\rangle \langle m|$ to the ancilla Hamiltonian $H_{\rm anc}^{\rm RWA}$ in Eq.~(\ref{eq:H_RWA}) before diagonalizing it. The result of this procedure is shown as the ``modified weak-coupling" in Fig.~(\ref{fig:self_Kerr_correction}) and achieves a better agreement with the full diagonalization. Note that this modified weak coupling scheme is still numerically more efficient than the full diagonalization as it only requires diagonalization of the ancilla Hamiltonian, yet is able to capture non-perturbative effects beyond the weak coupling regime.

\subsection{Near a drive-transmon resonance}
A second situation in which the coupling-induced frequency shifts lead to non-perturbative corrections to the weak-coupling expression is when the drive frequency is close to certain transmon transition frequency while the cavity modes are far-detuned from any resonance with the transmon or the drive. For instance, the drive frequency can be close to transmon transition frequency $\omega_{(m_0+1)m_0}$. If their distance is small or comparable to the coupling-induced shift in $\omega_{(m_0+1)m_0}$ (i.e., $|\omega_d-\omega_{(m_0+1)m_0}|\lesssim |c_{00,m_0+1}^{\Omega_d=0}-c_{00,m_0}^{\Omega_d=0}|$), then the effect of the shift on the driven dynamics of the transmon is strongly non-perturbative. 
As explained in the previous section, to capture the non-perturbative effect of this frequency shift while still using the weak coupling expression, we add a term $\sum_m c_{00,m}^{\Omega_d=0}|m\rangle\langle m|$ to the Hamiltonian of the driven transmon before we diagonalize it. As shown in Fig.~\ref{fig:self_Kerr_correction}(b) for the case of $m_0=0$, this simple modification leads to a much better agreement with the full diagonalization.

\begin{figure}[ht]
\includegraphics[width=8. cm]{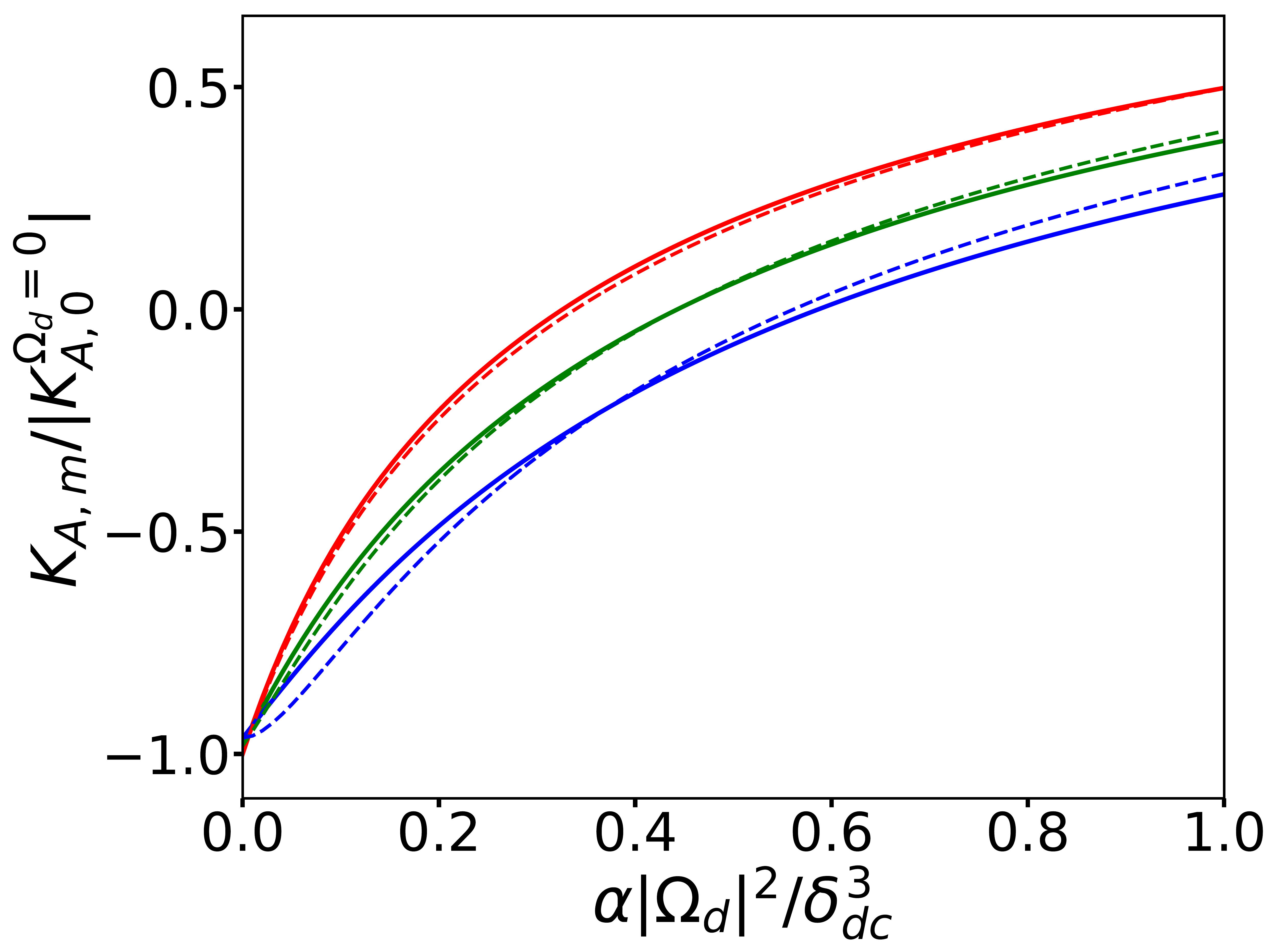} \hfill
\includegraphics[width=8. cm]{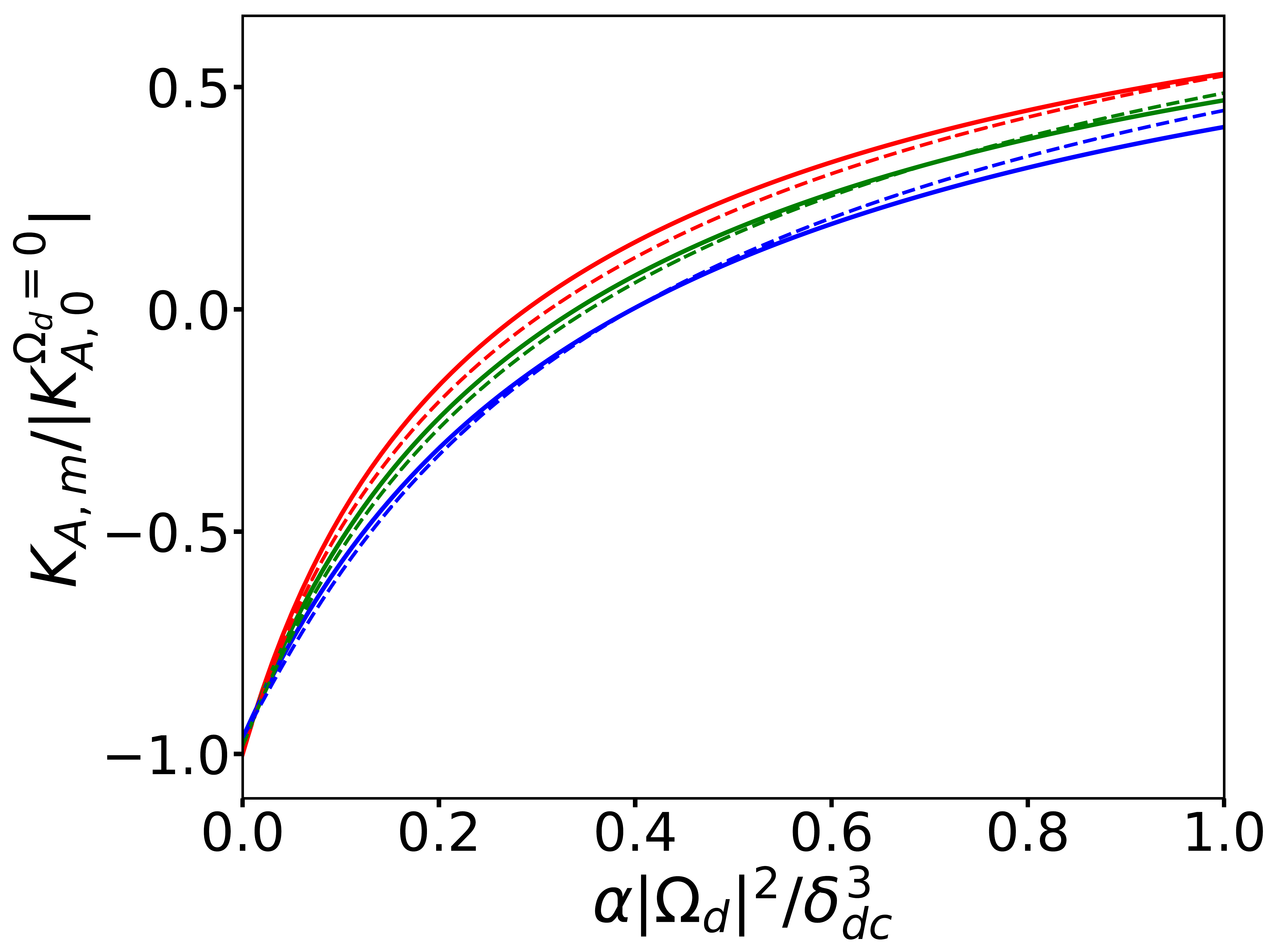}
\caption{Comparing the semiclassical result in Eq.~(\ref{eq:Delta_m_semiclassical})  (dashed lines) with the quantum mechanical calculation using Eq.~(\ref{eq:self_Kerr}) (solid lines) for cavity self-Kerr. Top: $\alpha/\delta_{dc} =1/15$. Bottom: $\alpha/\delta_{dc} =1/30 $. In both panels, $\alpha/\delta_a = 1/500.$}
\label{fig:self_Kerr_compare_semiclassical}
\end{figure}

\section{Comparing the semiclassical result with the quantum-mechanical calculation for cavity Kerr nonlinearity in the weak-coupling regime}
\label{app:semiclassical_compare_weak_coupling}
We compare in Fig.~\ref{fig:self_Kerr_compare_semiclassical} the drive-induced change of cavity self-Kerr obtained from Eq.~(\ref{eq:self_Kerr}) where the driven transmon is treated quantum mechanically and Eq.~(\ref{eq:Delta_m_semiclassical}) where it is treated semiclassically and the cavity is assumed to be far away from any drive-induced resonances.
As shown in the figure, in the regime where $|\delta_{a}|\gg \delta_{dc}\gg \alpha$, the semiclassical treatment agrees well with the quantum mechanical calculation. 

As shown by Eq.~(\ref{eq:Delta_m_semiclassical}) and Fig.~\ref{fig:self_Kerr_compare_semiclassical}, the behavior of $K_{A,m}$ is controlled by a dimensionless drive amplitude $\sqrt{\alpha}|\Omega_d|/\delta_{dc}^{3/2}$ and the effective Planck constant $\lambda = \alpha/2\delta_{dc}$. Finite $\lambda$ leads to a finite variation of $K_{A,m}$ with $m$. To leading order in $\lambda$, this variation is linear in $m\lambda$.

We note that in Fig.~\ref{fig:self_Kerr_dispersion}(b), although $|\delta_a|\gg \delta_{dc}$ is not strictly satisfied, the behavior of $K_{A,m}$ is already qualitatively captured by Eq.~(\ref{eq:Delta_m_semiclassical}). 


\section{Experimental details}
\label{app:experiment}

\begin{figure}[t]
\includegraphics[scale = 0.89]{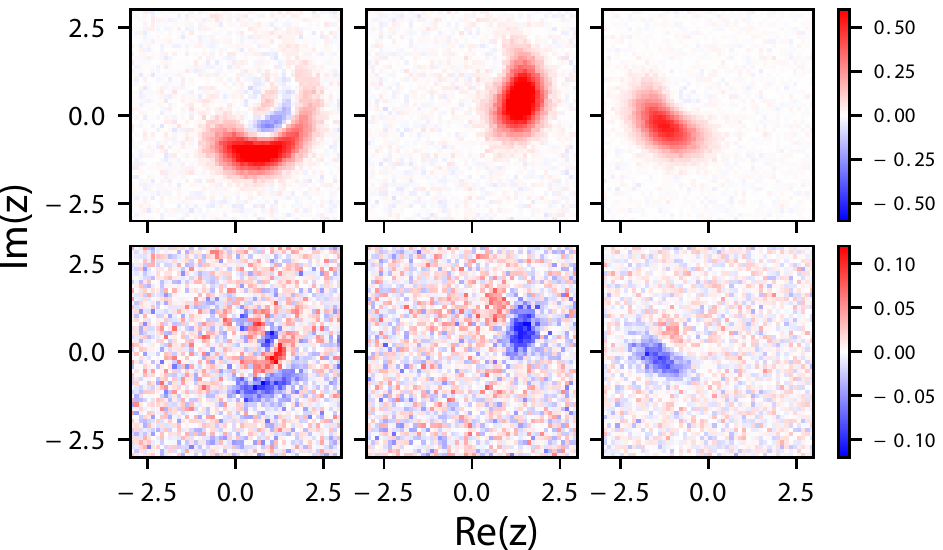}
\caption{The top row are measured cavity Wigner functions with the same transmon drive detuning and different drive amplitudes. From the left to right, they correspond to the same drive parameters as the third, fourth and fifth data point in Fig.~\ref{fig:convergence}(b) (counted from the lower amplitude side). We acquire these Wigner functions by measuring the displaced parity at each point $z$ shown in the plots after allowing a cavity coherent state of $\alpha=1.5$ to undergo evolution in the presence of the transmon drive for 10 $\mu$s. Next, we perform master equation simulations parametrized by $\delta\omega_{A,0},K_{A,0},\beta_{A,0}$ in Eq.~(\ref{eq:single_mode_expansion}) and a scale factor to account for Wigner normalization. From these we construct simulated Wigner functions, which we use to perform least-squares fits of the above parameters, excluding the normalization factor. The bottom row contains the fit residuals (experimental Wigners minus the simulated Wigners), with the same drive parameters as the plots above. The fits capture the main features of the experimental Wigners. The residuals are partly due to a $z$-dependent contrast reduction in the experimental Wigners which is currently under investigation.}
\label{fig:Wigners_exp}
\end{figure}

We determined the Hamiltonian parameters $K_{A,0}$ and $\beta_{A,0}$ in Eq. (\ref{eq:single_mode_expansion}) under different drive strengths and detunings by fitting simulated Wigner functions to measured Wigner functions. First, we actively cool the transmon to its ground state and prepare a coherent state $|\alpha=1.5\rangle$ in cavity mode-$A$ using a short resonant cavity drive of duration $72$ ns. Next, we apply an off-resonant transmon drive, with different detunings and strengths, of $10~\mathrm{\mu s}$ duration and $160$ ns risetime. The risetime is chosen to be longer than the inverse of the drive detuning so that the transmon adiabatically follows the drive and evolves from the vacuum state to the adiabatic Floquet state $\psi_0$. We then perform cavity Wigner tomography by measuring the displaced parity of cavity mode $A$~\cite{harocheb},
\begin{equation}
W(z)=\frac{2}{\pi}\text{Tr}(\hat{D}^\dagger(z)\hat{\rho}\hat{D}(z)\hat{P}),
\end{equation}
where $z$ is a complex variable that represents a point in the cavity phase space in the unit of $\sqrt{\hbar}$, $z = (Q_A +iP_A)/\sqrt{2\hbar}$; see Eq.~(\ref{eq:Wigner_def}). 
$\hat{D}$ is the displacement operator and $\hat{P}=e^{i\pi\hat{A}^\dagger\hat{A}}$ is the parity operator.

The Wigner function $W(z)$ contains full state information and allows extraction of the density matrix $\hat{\rho}$ via various methods including direct inversion, iterative methods, and maximum likelihood methods. For simplicity, we choose not to reconstruct $\hat{\rho}$, but rather perform a least-squares fit between a simulated and measured Wigner function. While this approach does not allow us to calculate the fidelity of the fitted and measured states, it does avoid possible systematic errors introduced in reconstructing $\hat{\rho}$. 

To extract the Hamiltonian parameters in Eq.~(\ref{eq:single_mode_expansion}), we fit a simulated Wigner function to the measured Wigner function via a least-squares cost function.  We obtain simulated Wigner functions through a Lindblad master equation simulation that excludes the $\sigma_{A, 0}$ term and includes single photon loss in cavity mode-$A$. For the cavity loss rate, we used the experimentally measured value of 1/330 $\mu s^{-1}$ at zero transmon drive. The simulated Wigners are only weakly influenced by this loss rate due to that the drive duration time (10 $\mu s$) is much shorter than 330 $\mu s$. We show in Fig.~\ref{fig:Wigners_exp} examples of experimentally-measured cavity Wigner functions and the residuals with respect to the simulated Wigners that best fit the measured ones. The simulation does not include loss, heating, or dephasing errors in the dressed transmon ancilla used for cavity parity measurement. These error channels reduce the contrast of the measured Wigner function, although not to the degree observed in the experiment. Work is ongoing to understand the $z$-dependence on Wigner function contrast reduction. 

Finally, note that our choice of small $\alpha=1.5$ is due to the small lifetime $T_1\approx330~\mathrm{\mu s}$ of cavity mode-$A$. This provides good signal-to-noise (SNR) for $K_{A,0}$, but reduced SNR for higher order nonlinearities due to the small fraction of population present in energy levels shifted by them. Attempts to fit $\sigma_{A, 0}$ failed for this reason.


\begin{thebibliography}{35}
\bibitem [{Note-nonlinear-resonance()}]{Note-nonlinear-resonance}%
  \BibitemOpen
  \bibinfo {note} {This statement holds also in the presence of nonlinear resonances between cavity and transmon excitations, because these resonances lead to a finite gap in the eigenspectrum which allows uniquely identifying state $|\overline{\psi_m,N_a,N_b}\rangle$ through adiabatic continuation.}\BibitemShut {Stop}%
\bibitem [{Note-scaling()}]{Note-scaling}%
  \BibitemOpen
  \bibinfo {note} {This scaling is motivated by the fact that $K_{A,0}$ is proportional to $|g_a|^4$ [see Eq.~\ref{eq:self_Kerr}], so we divide $K_{A,0}$ by $|g_a|^4$ and then multiply by $\alpha^3$ to make it dimensionless.}\BibitemShut {Stop}%
\end{thebibliography}

%

\end{document}